\documentclass{jpp}
\usepackage{graphicx,color}
\usepackage{amsmath}
\usepackage{amssymb}
\usepackage[normalem]{ulem}
\usepackage{cancel}

\newcommand{\pc}{\,{\rm pc}}
\newcommand{\kpc}{\,{\rm kpc}}

\newcommand{\cm}{\,{\rm cm}}

\def\vect#1{\ensuremath{\mathchoice{\mbox{\boldmath$\displaystyle#1$}}
		{\mbox{\boldmath$\textstyle#1$}}
		{\mbox{\boldmath$\scriptstyle#1$}}
		{\mbox{\boldmath$\scriptscriptstyle#1$}}}}

\shorttitle{Topological data analysis and compressible MHD turbulence}
\shortauthor{Makarenko et al.}

\title{Topological data analysis and diagnostics of compressible MHD turbulence}

\author{I.~Makarenko\aff{1}
  \corresp{\email{irina.makarenko@ncl.ac.uk}},
P.~Bushby\aff{1}, A.~Fletcher\aff{1}, R.~Henderson\aff{1}, 
N.~Makarenko\aff{2} \and A.~Shukurov\aff{1}}

\affiliation{\aff{1}School of Mathematics, Statistics and Physics, Newcastle University, Newcastle upon Tyne, NE1 7RU, UK 
\aff{2}Central Astronomical Observatory, Russian Academy of Sciences, 65 
Pulkovskoye Chaussee, Saint-Petersburg, 196140, Russia}

\begin{document}

\maketitle

\date{Submitted for publication in J. Plasma Phys.}
\pubyear{2018}
\pagerange{\pageref{firstpage}--\pageref{lastpage}}
\label{firstpage}

\begin{abstract}
The predictions of mean-field electrodynamics can now be probed using
direct numerical simulations of random flows and magnetic fields. When 
modelling astrophysical MHD, it is important to verify that such simulations 
are in agreement with observations. One of the main challenges in 
this area is to identify robust \emph{quantitative} measures to compare 
structures found in simulations with those inferred from astrophysical 
observations. A similar challenge is to compare quantitatively results from
different simulations. Topological data analysis offers a range of techniques, 
including the Betti numbers and persistence diagrams, that can be used to 
facilitate such a comparison. After describing these tools, we first apply them 
to synthetic random fields and demonstrate that, when the data are standardized 
in a straightforward manner, some topological measures are insensitive to 
either large-scale trends or the resolution of the data.  Focusing upon one 
particular astrophysical example, we apply topological data analysis to 
H\,\textsc{i} observations of the turbulent interstellar medium (ISM) in the 
Milky Way and to recent MHD simulations of the random, strongly compressible  
ISM. We stress that these topological techniques are generic and could be 
applied to any complex, multi-dimensional random field. 
\end{abstract}

\section{Introduction}
Mean-field electrodynamics is an integral part of the theory of turbulent 
flows in electrically-conducting fluids. Following its pioneering formulation 
using the second-order smoothing approximation (SOCA) \citep{SKR66,KR80}, the 
mean-field induction equation has been derived using a wide variety of 
approximations. Essentially the same equation is obtained when the assumptions 
of SOCA are relaxed, most importantly, the assumptions of scale separation 
between the mean and random magnetic fields and of a short velocity correlation 
time  \citep{DMSR84, ZeMoRuSo88, Molch91, BS05a}. The robustness of this theory 
to differing assumptions suggests its relevance and reliability. However,  
confidence in the theoretical basis of mean-field  electrodynamics has been 
shaken by direct numerical simulations of mean-field dynamo action in driven
and convective flows. Comparisons of such simulations with theory remain 
inconclusive \citep[e.g.,][and references therein]{FaBu13}. One of the possible
reasons for that is the complexity of the random flows that drive the dynamo:
they are often compressible, anisotropic and non-Gaussian, and 
the mean magnetic field may not be as symmetric as is often assumed 
(e.g., unidirectional or axially symmetric). However, analyses of 
the simulation results rarely allow for such complications. This is, perhaps, 
not surprising, since the known theory of random fields has been formulated 
almost exclusively for Gaussian random fields and their close relatives such 
as log-normal or $\chi^2$ fields. Techniques applicable to strongly 
non-Gaussian random fields are few and not widely known.

Motivated by the present difficulties experienced by comparisons between the 
theory and numerical simulations of mean-field dynamos, we focus upon one 
particular aspect of this problem: developing diagnostics of strongly 
non-Gaussian random fields that can be used to make quantitative comparisons 
between different data sets. The significance of such techniques extends 
beyond mean-field electrodynamics, as the need to compare sophisticated 
numerical simulations with experiments or observations of natural phenomena is 
common to many areas of physics. This is particularly true in astrophysics, 
where modern telescopes generate observational data at very high spatial 
resolution. With ever-increasing computer power, simulations of astrophysical 
processes are attaining greater levels of realism. Of particular interest is 
the notion of a topological comparison, which (in this context) assesses 
whether or not the connectivity of particular structures in simulations match 
those of the observations. It is clearly more difficult to quantify 
topology than it is to measure the standard statistics such as mean values, 
correlation functions of various orders, etc. However, a crucial advantage of 
topological methods is their generality and freedom from restrictive and often 
unjustifiable assumptions, such as the assumption of Gaussian statistics.

\begin{figure} 
\centerline{
\includegraphics[width=0.43\textwidth]{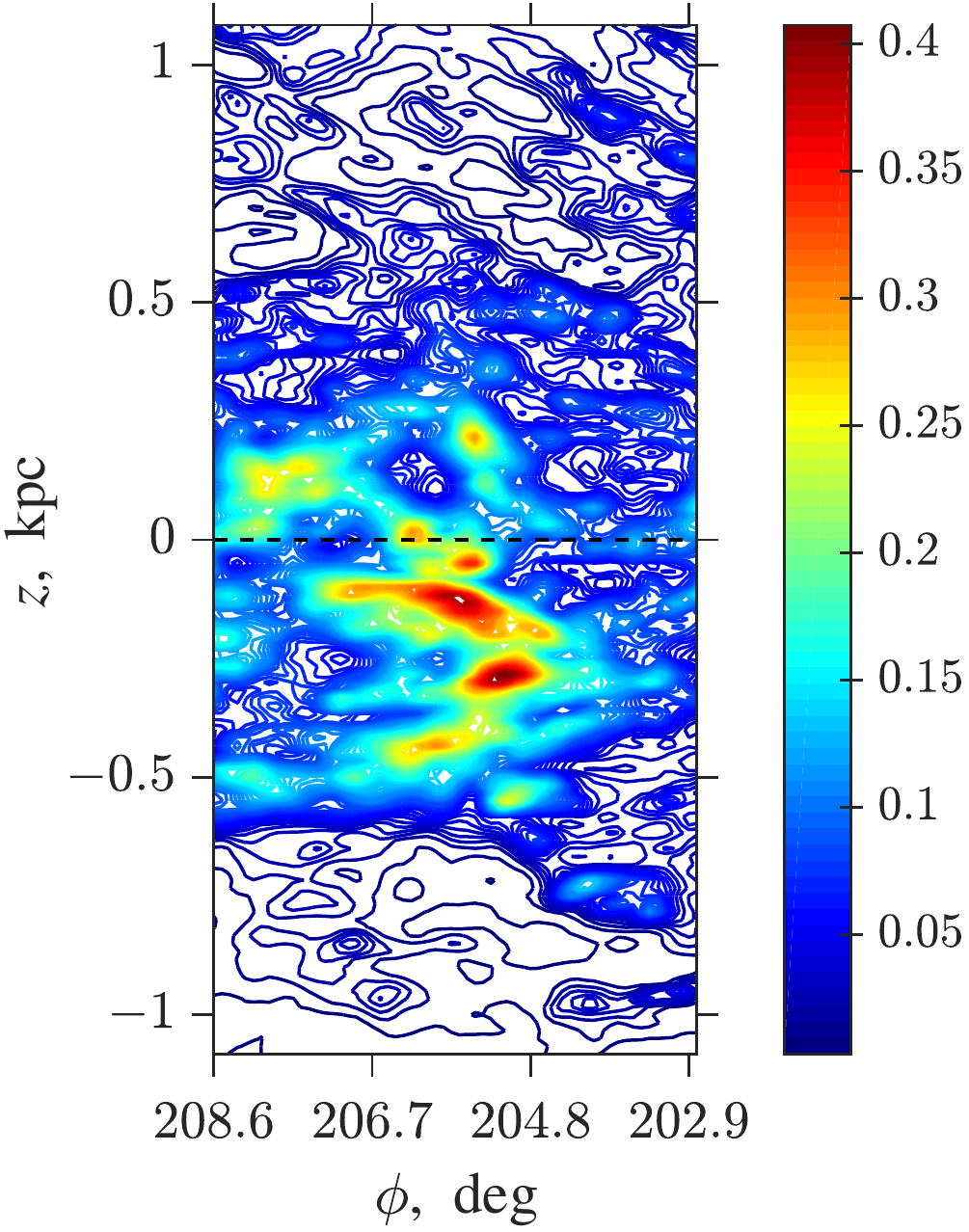} 
\hspace{4mm}
\includegraphics[width=0.40\textwidth]{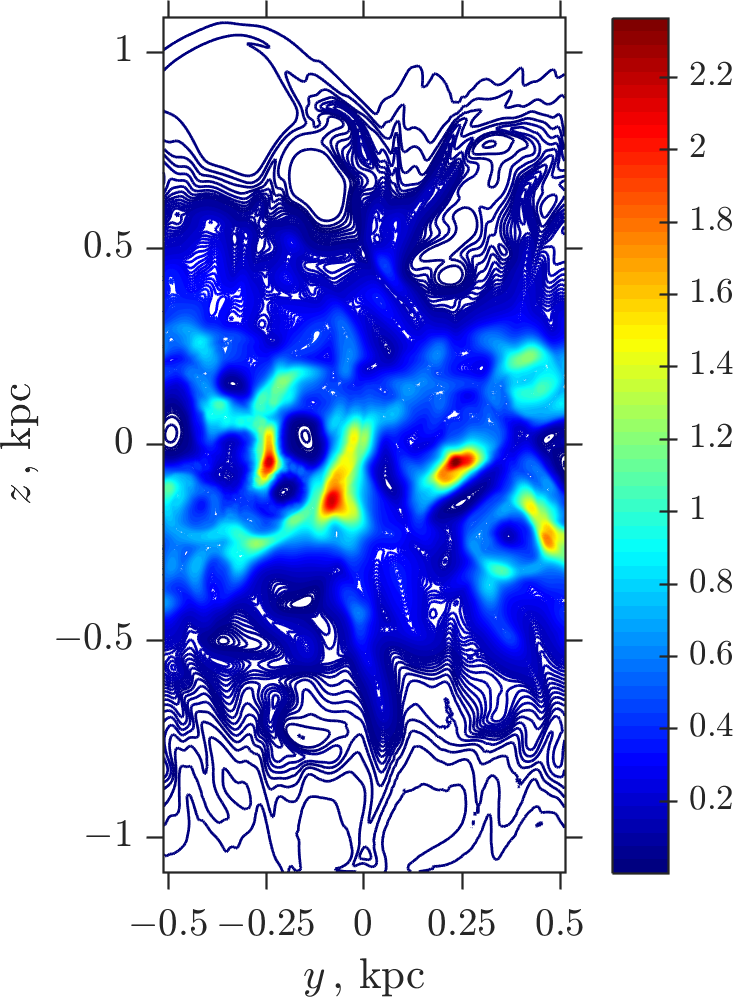} 
}
\caption{
\textbf{Left:} the isocontours of the H\,{\sc i} number density $n$ (in 
atom$\cm^{-3}$) from the Galactic All Sky Survey \citep{GASS10} at $R=10\kpc$, 
$202.9^\circ\leq\phi\leq208.6^\circ$, $|z|\leq1.1\kpc$, with $(R,\phi,z)$ the 
Galactocentric cylindrical coordinates with the Sun at $(R,\phi,z)=(8.5\kpc, 
180^\circ, 0$). The horizontal extent of the image is about $1\kpc$, its 
resolution is close to $17\pc$. \textbf{Right:} the isocontours of the gas 
number density $n$ (in atom$\cm^{-3}$) in a vertical plane, obtained from a 3D 
MHD simulation of \citet{Gent2013p1, Gent2013p2} at a resolution of $4\pc$. 
White areas in both panels correspond to low  density values, 
$n<3\times10^{-3}\cm^{-3}$ on the left and $5\times10^{-4}\cm^{-3}$ on the 
right.
} 
\label{fig:f1}
\end{figure}

We focus on the distribution of hydrogen atoms (H\,{\sc i}) in the Milky Way, 
obtained from recent observations \citep[see][]{GASS10}. The specific 
observable is the volume number density, $n$, measured in a Galactocentric 
cylindrical coordinate system $(R, \phi, z)$, where $R$ is the distance in the 
Galactic plane from the centre of the Galaxy, $\phi$ represents the azimuthal 
angle in the Galactic plane (with the Sun at $\phi = 180^\circ$) and $z$ is the 
vertical distance from the mid-plane of the Galactic disc. The left-hand panel 
of figure~\ref{fig:f1} represents a 2D cross-section of $n(\phi, z)$, at 
Galactocentric radius $R = 10\kpc$ (which is close to the Sun), in the region  
$202.9^\circ \leq \phi \leq 208.6^\circ$ and $|z| \leq 1.1\kpc$. The spatial 
extent of this region is approximately $1\times 2\kpc$, chosen for ease of 
comparison with existing simulations. Interstellar gas is involved in transonic 
MHD turbulence that supports the mean-field dynamo action \citep[e.g.,][]{S07} 
recently reproduced in direct numerical simulations \citep{GEZR09, Gent2013p1, 
Gent2013p2, BeGrEl15}.

The right-hand panel of figure~\ref{fig:f1} shows a snapshot of the gas density 
distribution in a numerical simulation of the multiphase interstellar medium 
in the solar neighbourhood. The model describes supernova-driven random flow 
in the interstellar gas, with gravity and differential rotation, including 
the effects of radiative cooling, photoelectric heating and various transport 
processes \citep[see][for more details]{Gent2013p1, Gent2013p2}. The spatial 
extent of the computational domain is $1\kpc\times1\kpc$ horizontally and 
$2\kpc$ vertically, symmetric about the Galactic mid-plane, so the simulation 
domain is comparable in size to the region shown on the left-hand panel of the 
figure. Note that the magnitude of the gas density is not directly comparable 
to the corresponding H\,{\sc i} density (which is only one component of the 
interstellar gas, although it does make up the majority of the gas). There are, 
however, some clear qualitative similarities between the simulation and the 
observations, notably the form of the higher density regions near the 
mid-plane. The question is: how well does the model describe reality? 

An initial inspection of the data shown in figure~\ref{fig:f1} immediately 
reveals three possible problems that may hinder meaningful comparisons. First, 
as mentioned above, the range of the gas densities in the observed and 
simulated fields in figure~\ref{fig:f1} differ; the mean and the standard 
deviation of the gas densities in the two fields are different too. The 
observed mean gas density is $\langle n \rangle=0.065\,\mathrm{atom\,cm}^{-3}$ 
and its standard deviation $\sigma_n=0.063\,\mathrm{atom\,cm}^{-3}$. In the 
simulation, $\langle n \rangle=0.220$ and 
$\sigma_n=0.349\,\mathrm{atom\,cm}^{-3}$. These values can also vary quite 
strongly from snapshot to snapshot in a simulation and from slice to slice in 
the real data. The data need to be standardized before their statistical 
properties are compared. Second, both the observed and simulated data have a 
strong trend in the vertical direction -- the gas density is higher at $z = 0$ 
and gets lower as $|z|$ increases. Does the presence of trends influence the 
comparison and, if so, how can this be mitigated? Given that it can often be 
difficult to separate trends from turbulent fluctuations, can we find methods 
of analysis that are insensitive to large-scale trends in the data? Third, the 
image resolutions are different -- the simulated data has four times higher 
resolution than the observation. The standard approach to this problem would be 
to reduce the resolution of the simulation to that of the observation. However, 
this may lose important information. A better approach would be to select 
methods of analysis that are independent of the image resolution.
We will show that topological data analysis meets these requirements. 

Over the last decade, topological approaches have been widely applied to the 
analysis of random fields \citep{CarGun2009, ABBSW2010, AdTay2011, Ed2014, 
YA2012}. Topological invariants, such as the Betti numbers, and the associated 
techniques of persistence diagrams and barcodes, rank functions and persistence 
landscapes, as well as persistence images for random fields, have been applied 
in cosmology \citep{Gay2010, Sousbie2011part1, Sousbie2011part2, Pranav2015, 
Pranav2017} and are beginning to be used in fluid dynamics \citep{Kram2016}, 
solar MHD \citep{MKN07, MMMKM14, KMU15, KMU15a, KMKD16, Knyaz2017} and 
astrophysics \citep{Li2016arXiv, Hend2017arXiv, Makar2018MNRAS}.

Complex topological diagnostics, such as persistence landscapes 
\citep{Buben2015, Buben2017pers}, rank functions \citep{Robins2016}, 
persistence images \citep{Adams2017}, are difficult to interpret
in physical terms; they are often not robust and work well with idealized image 
textures rather than more irregular patterns. The bottleneck distance 
\citep{CEH2007} as a measure of similarity between persistence diagrams, is 
often inefficient in astrophysical and other applications \citep[see, 
e.g.,][]{Makar2018MNRAS}. We found that simpler statistics can be more 
efficient when applied to real data \citep{Hend2017arXiv}. In this paper, we 
discuss topological data analysis (TDA) for random fields and identify simple, 
yet efficient topological statistics. We informally describe what topological 
filtration is, how to use it for comparison and classification of different 
scalar fields, and how stable the results are. \S\ref{sec:filt} describes the 
topological filtration of functions and its representation in the form of 
persistence diagrams. We show two different ways to represent a persistence 
diagram and introduce statistics that we have found useful for distinguishing 
between one-dimensional signals and fields in 2D. In \S\ref{sec:stand} we 
discuss a standardization of fields, which makes comparisons more reliable. 
\S\ref{sec:trends} focuses on topology in the presence of large-scale trends. 
In \S\ref{sec:res}, we investigate how the results of topological filtration 
depend on the image resolution. In \S\ref{sec:app} we apply some of the TDA 
methods to observations and numerical simulations of the ISM and summarize the 
key principles for the practical application of TDA in \S\ref{Con}. 

\section{Topological filtration and persistence diagrams}
\label{sec:filt}

In this section we present the basic ideas of topological filtration along 
with some useful  topological statistics. Our aim here is to develop these 
topological ideas at an intuitive level. Rigorous definitions can be found, for 
example, in \citet{Moroz2008, ABBSW2010, EH10, AdTay2011, Ed2014, 
toth2017handbook}, while \citet{Park2013}, \citet{Ghrist2008barcodes}, 
\citet{Pranav2017} and \citet{Motta2018} also provide useful and less formal 
expositions.

\subsection{Topological filtration}
Consider a random function $f(\vect{x})$ defined in a domain $M$. Suppose that 
$f(\vect{x})$ is smooth, so that its extrema are critical points, $\nabla 
f(\vect{x})=0$. (We note that continuity is sufficient in many applications.)  
We first introduce sublevel sets  \citep[e.g.,][]{EH2008survey}, those regions 
where the values of a random function $f(\vect{x})$ are below a certain 
threshold $h$. The sublevel set is defined as
\[
A_{h} = \left\{\vect{x} \in M: f(\vect{x}) \leq h \right\}. 
\]
A sequence of levels $h_1 \leq h_2 \leq \cdots$ generates a nested family of 
sublevel sets of $f$, $A_{1} \subseteq A_{2} \subseteq \cdots\,$, which is 
called the sublevel set filtration of $f(\vect{x})$. 

Each sublevel set has certain topological properties that we wish to 
characterize. When $\dim(A_h) = 1$ (i.e., $f$ is a function of one variable), 
the topology is characterized by the Betti number $\beta_0$ that counts the 
number of connected components. When $\dim(A_h) = 2$, the topological 
properties are the number of connected components, $\beta_0$, and the number of 
holes in them (more precisely, loops whose interiors are the holes), $\beta_1$. 
In 3D, in addition to components and tunnels (loops), voids can appear (i.e., 
hollow spaces bounded by closed surfaces), which are counted by $\beta_2$.
Connected components, loops or tunnels, and closed shells are also referred to 
as cycles of dimension 0, 1, and 2, respectively. The alternating sum of the 
Betti numbers is a wider known topological quantity, the Euler characteristic,
\[
\chi = \sum^{N-1}_{k = 0} (-1)^k \beta_k\,,
\]
where $N$ is the dimension of $M$. 

Equivalent results can be obtained by considering superlevel sets, defined as 
\[
A_{h} = \left\{\vect{x} \in M: f(\vect{x}) \geq h \right\}. 
\]
The transition from the sublevel to superlevel sets only interchanges the
Betti numbers $\beta_0$ and $\beta_1$ in 2D, and $\beta_0$ and $\beta_2$ in 3D. 
For the rest of the paper we will only discuss the sublevel set filtration. 

The Betti numbers are topological invariants, i.e., they are not affected by 
translations, rotations, and continuous deformations of the space. In a typical 
astrophysical application, where we deal with random fields, the components 
represent matter clumps or clouds, the loops are closed chains of matter and 
the shells surround regions of reduced density (voids).

\begin{figure} 
\centerline{
\includegraphics[width=0.28\textwidth]{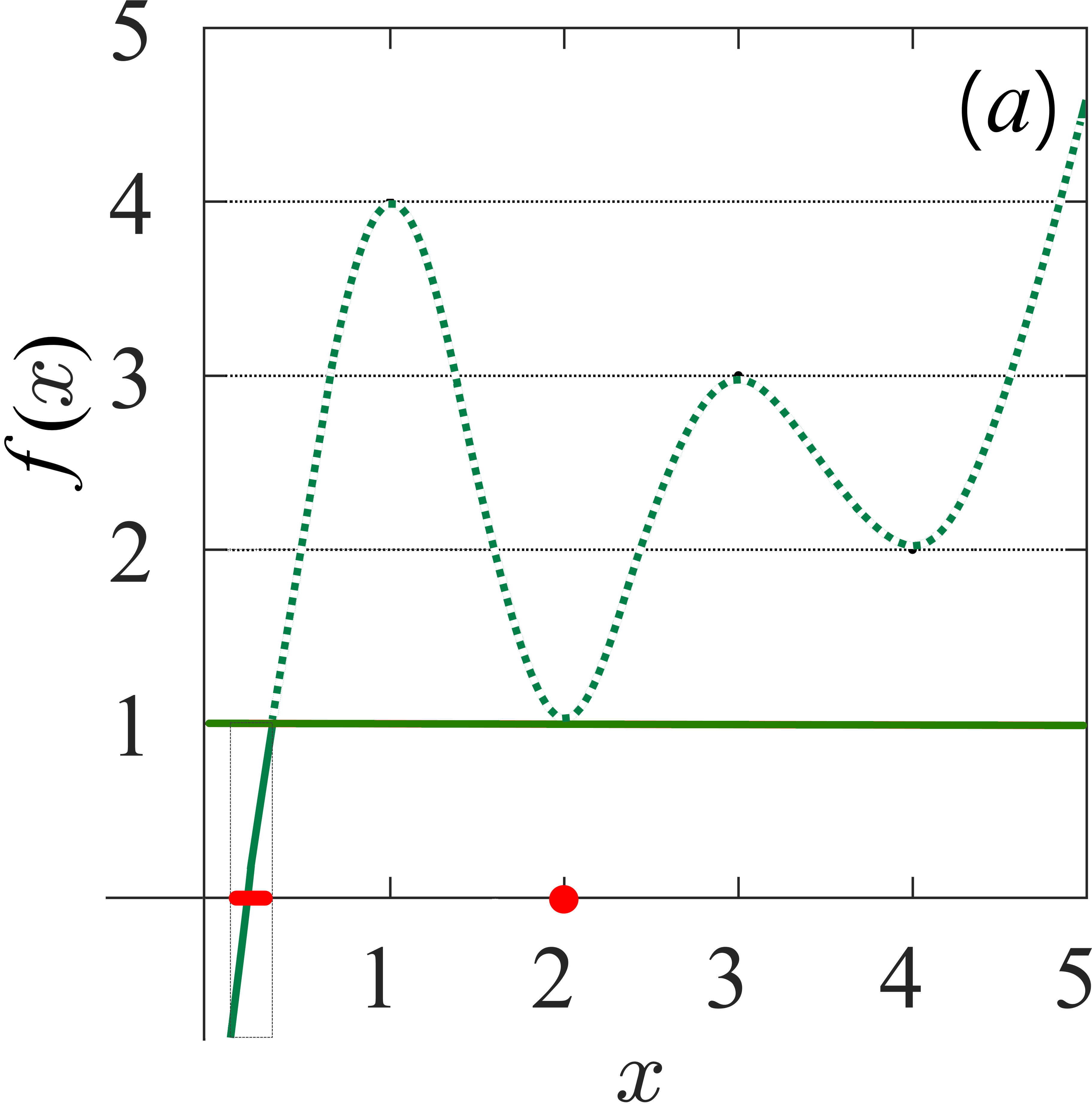} \hspace{3mm}
\includegraphics[width=0.28\textwidth]{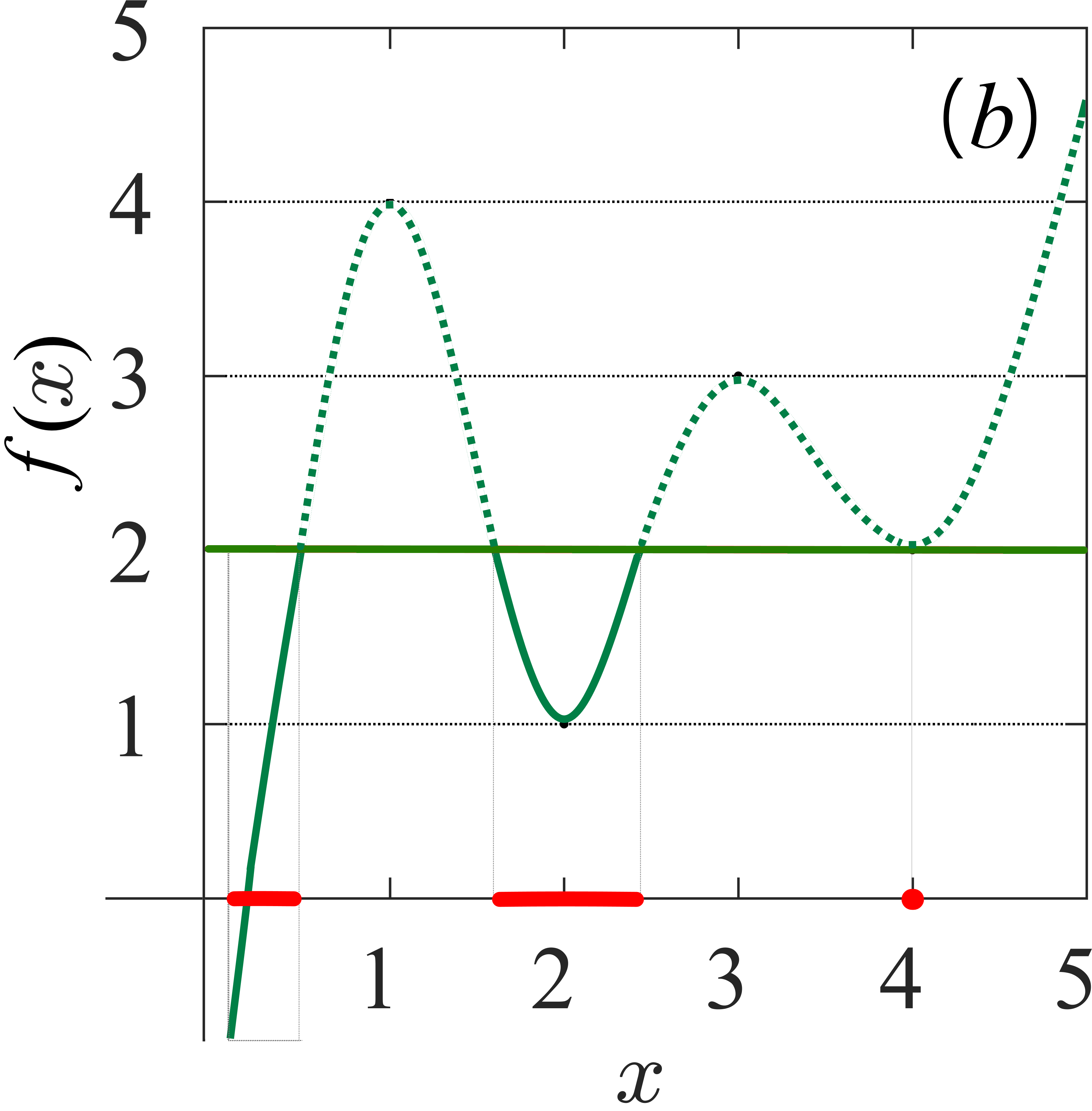} \hspace{3mm}
\includegraphics[width=0.28\textwidth]{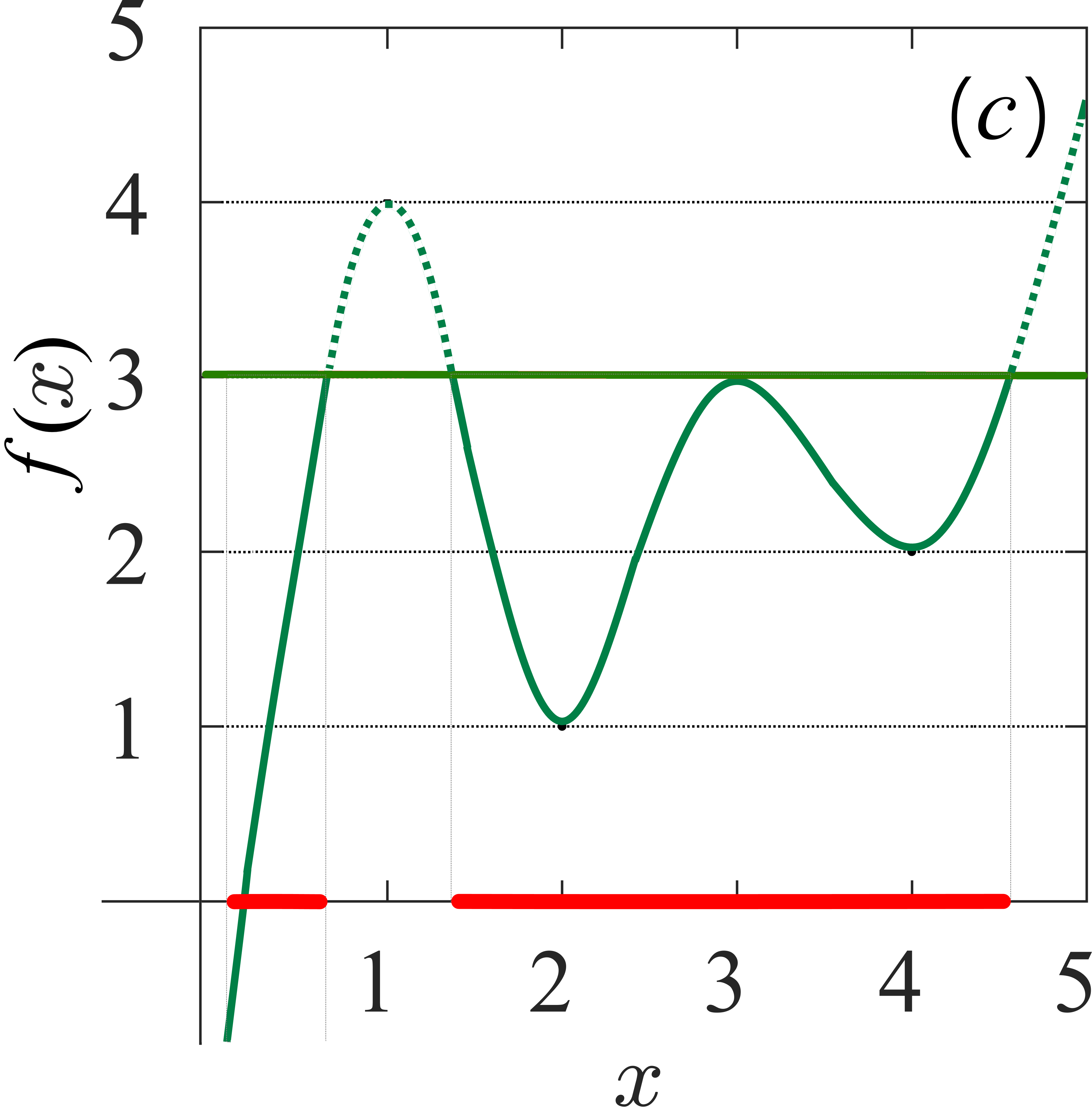}
}
\centerline{
\includegraphics[width=0.28\textwidth]{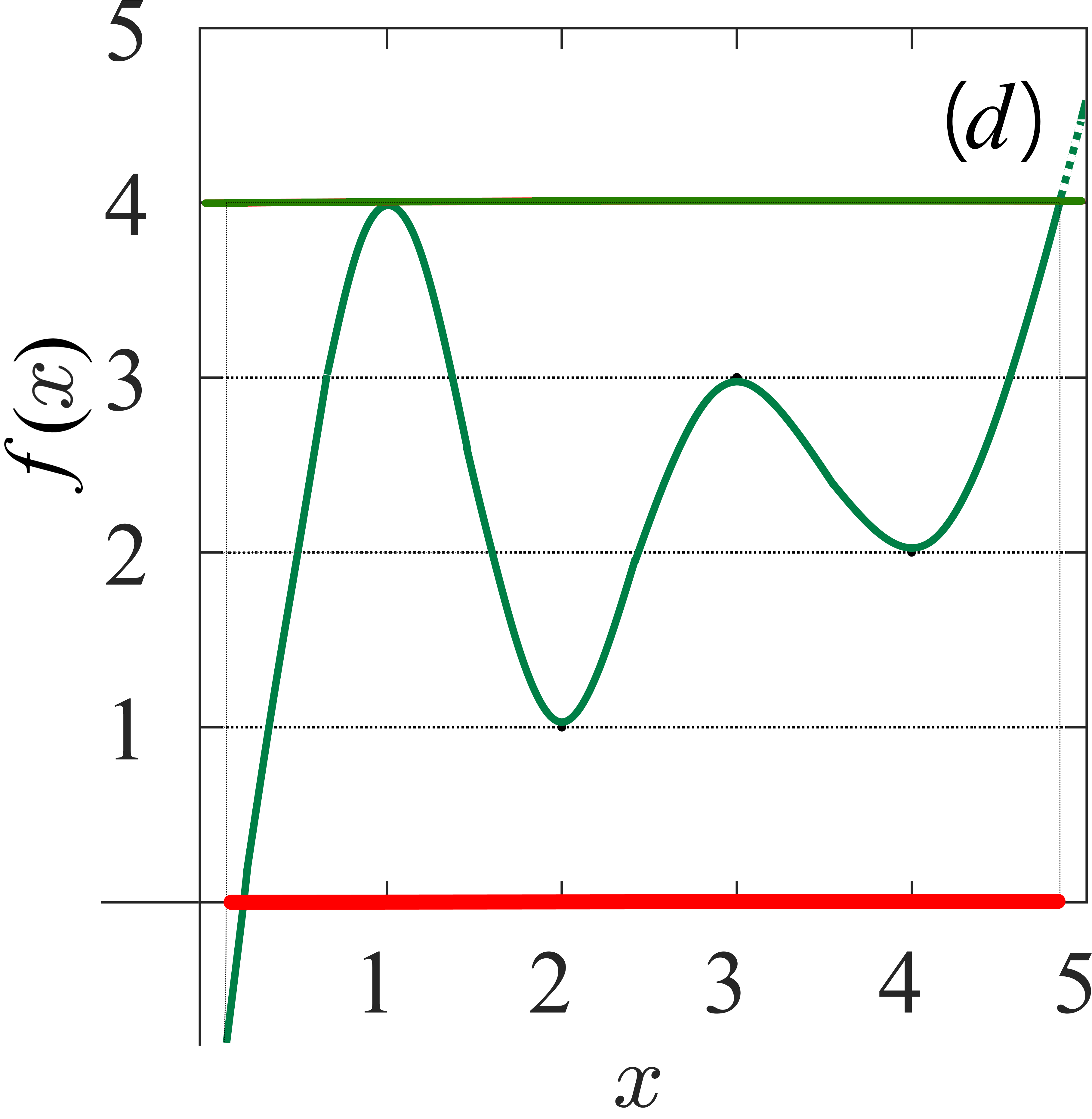} \hspace{3.5mm}
\includegraphics[width=0.605\textwidth]{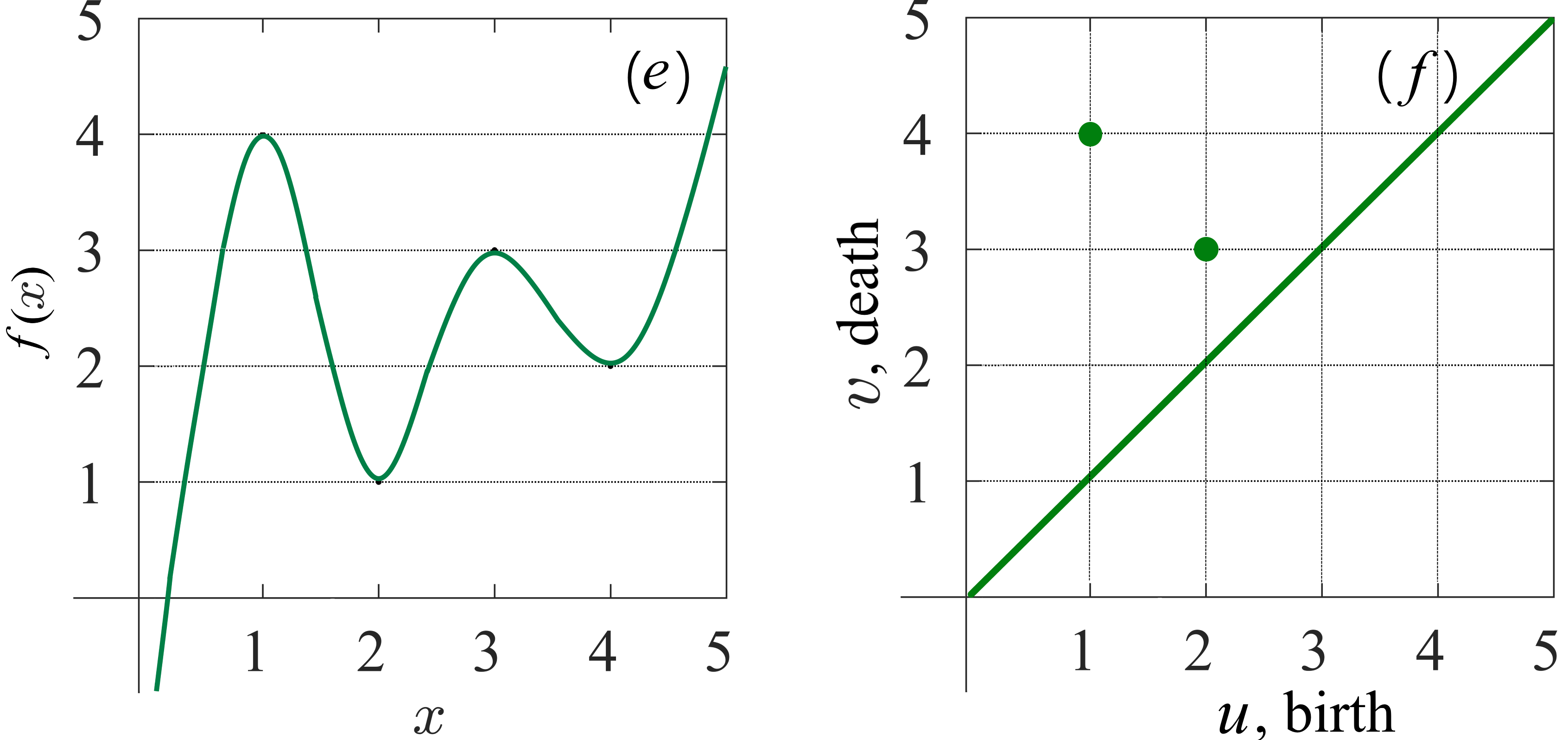}
}
\caption{ (\textit{a})--(\textit{d}) Sublevel set filtration of a function of 
one variable (\textit{e}) and the resulting persistence diagram (\textit{f}).
The horizontal green line indicates the increasing threshold $h$; the values 
of the function $f(x) \leq h$ are shown in a solid line, while $f(x) > h$ in a 
dotted line. On each panel the sublevel set at a fixed $h$ is shown as the red 
intervals on the $x$ axis. The number of red intervals corresponds to a number 
of connected components, $\beta_0$, at this level.
}
\label{fig:f3}
\end{figure}

\subsection{Topological filtration of a function of one variable} 
It is easiest to demonstrate a sublevel set filtration in 1D; this will allow 
us to develop some analogies and intuition about the method. We therefore 
consider the behaviour of a random function with reference to a threshold level 
$h$, considering the values of $x$ where $f(x)\leq h$. Figure~\ref{fig:f3} 
provides an example of this technique. As we vary the level $h$, the number of 
components in $A_h$, shown as red intervals on the $x$ axis, changes. When $h < 
1$ the corresponding sublevel set contains one component, an interval around $x 
= 0.3$, and so $\beta_0 = 1$. At $h = 1$ a new component appears, a point at 
$x=2$, which implies that $\beta_0 = 2$ at this level. When we reach $h = 2$ 
one more component is born, at $x = 4$, and so $\beta_0 = 3$. At $h = 3$, the 
second and the third components merge and $\beta_0 = 2$ again. According to the 
(so-called) \textit{Elder Rule}, the merger of two components results in the 
death of the younger one, i.e., the component that was born later. In this 
particular case, the third component, which appeared at $h = 2$, dies at $h = 
3$, merging with the second one. Finally, the first and the second components 
merge at $h = 4$, which results in $\beta_0 = 1$. We could continue to increase 
$h$, until this component also dies, if the function was defined in a wider 
range of $x$. Note that at each local minimum the sublevel set gains a new 
component, while at a local maximum, two components merge into one. Functions 
of one variable can also contain inflection points, but these do not change the 
configuration of the sublevel sets.

\begin{figure} 
\centerline{
\includegraphics[width=0.55\textwidth]{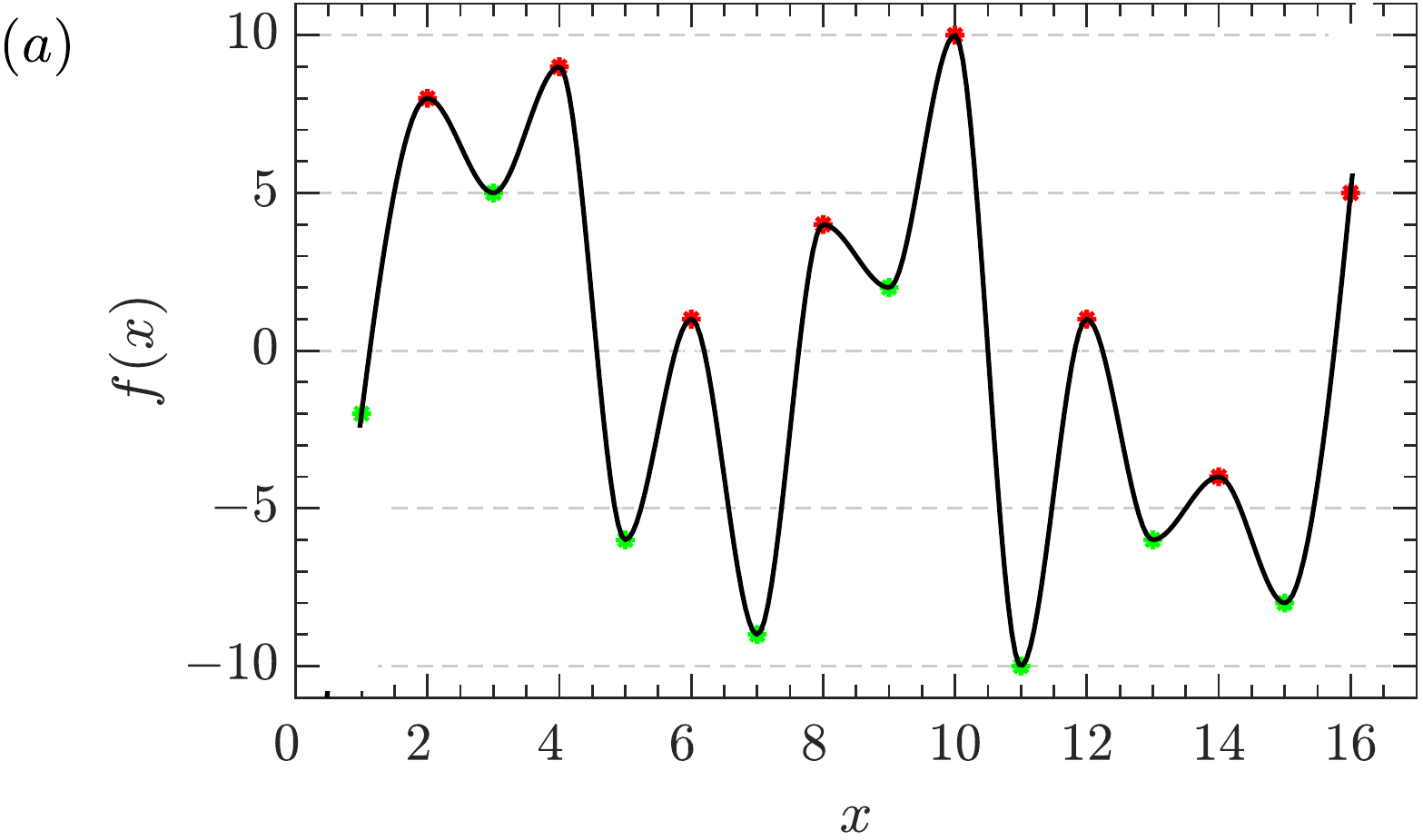}}\vspace{0.1in}
\centerline{
\includegraphics[width=0.38\textwidth]{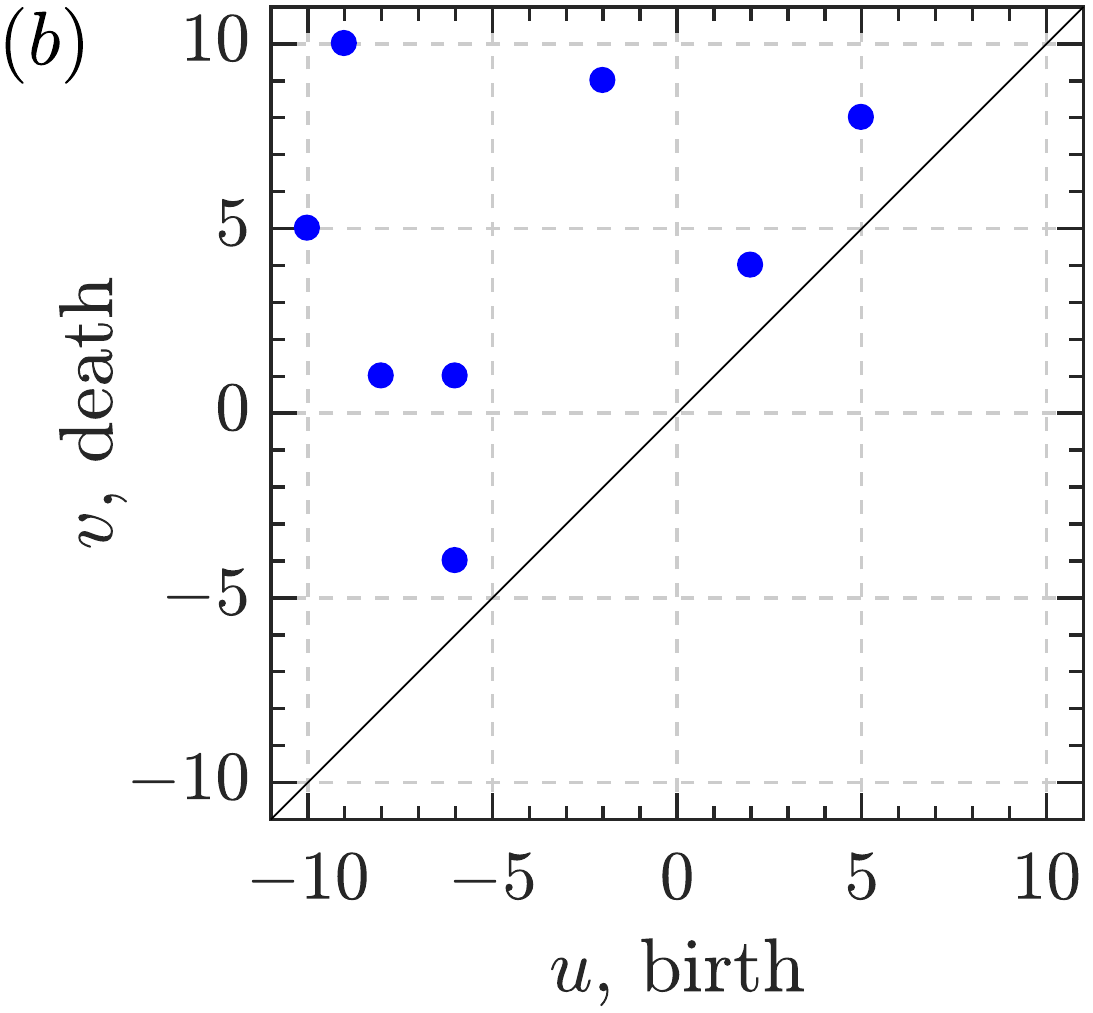}\hspace{0.1in}
\includegraphics[width=0.52\columnwidth]{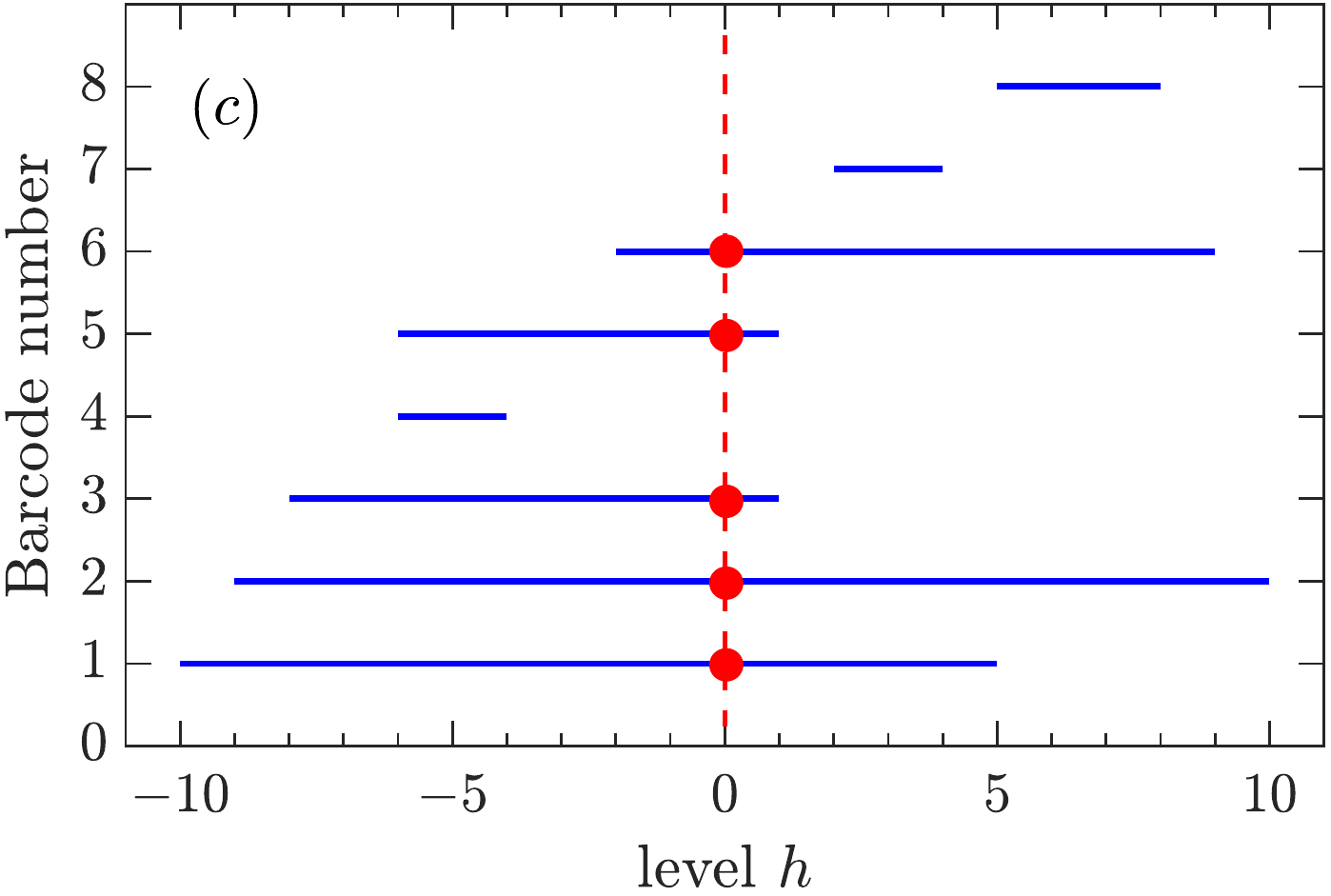}}
\caption{
(\textit{a}) A Gaussian random function $y = f(x)$, with the minima marked as 
green points, and the maxima as red ones. (\textit{b}) The persistence diagram 
obtained as the result of sublevel set filtration of this function. Every point 
corresponds to one persistent pair $(u_i, v_i)$, $i = 1,\dots,8$. (\textit{c}) 
One more way to illustrate the result of topological filtration of $f$ is to 
plot pairs $(u_i, v_i)$ as a sequence of barcodes, horizontal lines $u_i\leq 
h\leq v_i$, one for each component. The red vertical line in panel~(\textit{c}) 
corresponds to the threshold value $h = 0$. This line intercepts five out of 
the eight barcodes, which means there are five components at this level: 
$\beta_0 = 5$. The total length $L_T$ of these barcodes is equal to 61. As will 
be demonstrated below, $L_T(h)$ is a useful statistic for comparison of random 
functions.
}
\label{fig:F6}
\end{figure}

The filtration process illustrated in figure~\ref{fig:f3} captures and 
quantifies the topology of the sublevel sets. Tracing the varying sublevel sets 
as a function of threshold $h$, the topology remains static, or persists, 
between the critical points of $f(x)$, and changes only when $h$ reaches a 
critical point. The lifetime, or persistence of a component is defined as 
$v-u$, where $u$ is the level at which the component appears (its birth), and 
$v$ is the level at which it disappears (its death). This is therefore simply a 
measure of the range of levels $h$ over which this component exists. 
Topological filtration of a function results in a list of values $(u, v)$, one 
pair for each component, and these can be plotted in the $(u,v)$-plane to 
obtain what is called a \textit{persistence diagram}. For our example, the 
events of birth and death of the components are displayed on the persistence 
diagram in figure~\ref{fig:f3}(\textit{f}). Zero persistence corresponds to the 
diagonal line on the diagram; the most persistent components are those that are 
furthest from the diagonal.

Thus, topological filtration identifies critical points of a random function, 
pairs them and isolates those that correspond to its most significant 
(persistent) features. This represents topological properties of the function 
in terms of a relatively small number of quantifiable variables.

\subsection{Presenting the results of filtration: persistence diagrams and 
barcodes}

Figure~\ref{fig:F6}(\textit{a}) shows a small part of the realization of a
$\delta$-correlated Gaussian random process, $f(x)$, which has a mean of zero 
and a standard deviation of $6.6$. The result of the sublevel set filtration of 
$f(x)$ is a $2 \times m$ array where each of the $m$ pairs $(u_i, v_i)$ 
represents the birth, $u_i$, and the death, $v_i$, of the $i$-th component, $i 
= 1, \dots, m$. For the range of $x$ shown, the number of such pairs is $m = 
8$. Figure~\ref{fig:F6}(\textit{b}) is the persistence diagram obtained from 
the topological filtration of the function shown in 
figure~\ref{fig:F6}(\textit{a}) derived as described above. The persistence 
pairs $(u_i, v_i)$ can also be represented as a sequence of \textit{barcodes}, 
as shown in figure~\ref{fig:F6}(\textit{c}), where every point of the 
persistence diagram is shown as a horizontal bar extended between its birth and 
death levels. The bars are sorted, from the bottom up, according to when they 
first appear in the filtration process (which is also bottom up). 

Since the number of components depends on the size of the domain, it is useful 
to normalize the Betti numbers to unit length, area or volume as appropriate. 
We found that different random functions can be usefully compared by 
considering the dependence of the Betti numbers per unit length, area or volume 
on the level $h$, normalized to the unit area under the curve. The variation of 
the total length of the barcodes, 
\[
L_T=\sum_{i=1}^m(v_i-u_i),
\] 
also normalized to unit area under the curves, also turns out to be a useful 
diagnostic. The barcode length can be conveniently expressed in terms of the 
standard deviation of $f(x)$. These statistics make the result of topological 
filtration independent of the size of the dataset and some are also insensitive 
to the presence of a large-scale trend and the resolution of the data: we 
demonstrate these valuable properties in \S\ref{sec:trends} and \S\ref{sec:res}.

\begin{figure} 
\centerline{
\includegraphics[width=0.28\textwidth]{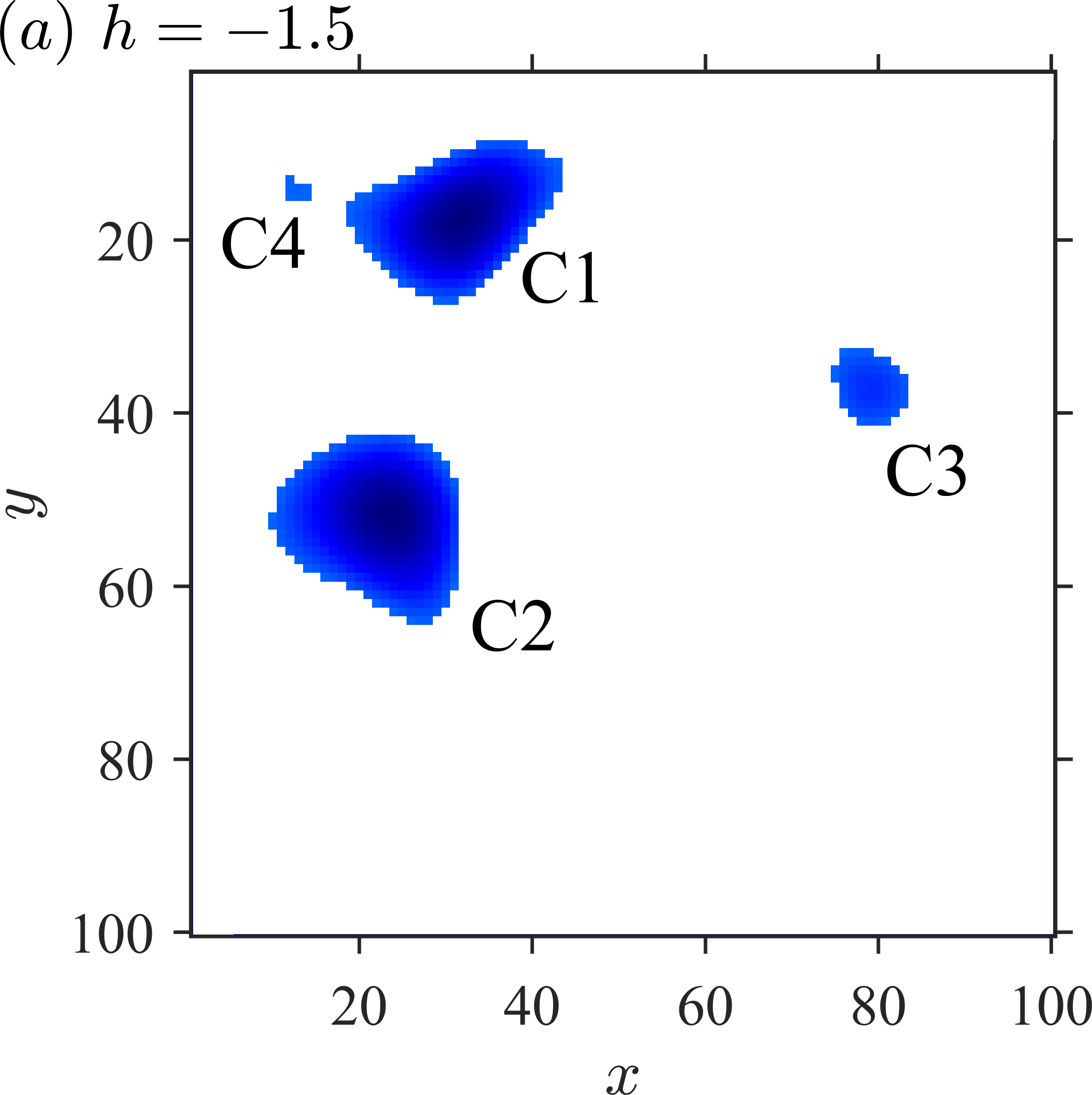} \hspace{2mm}
\includegraphics[width=0.28\textwidth]{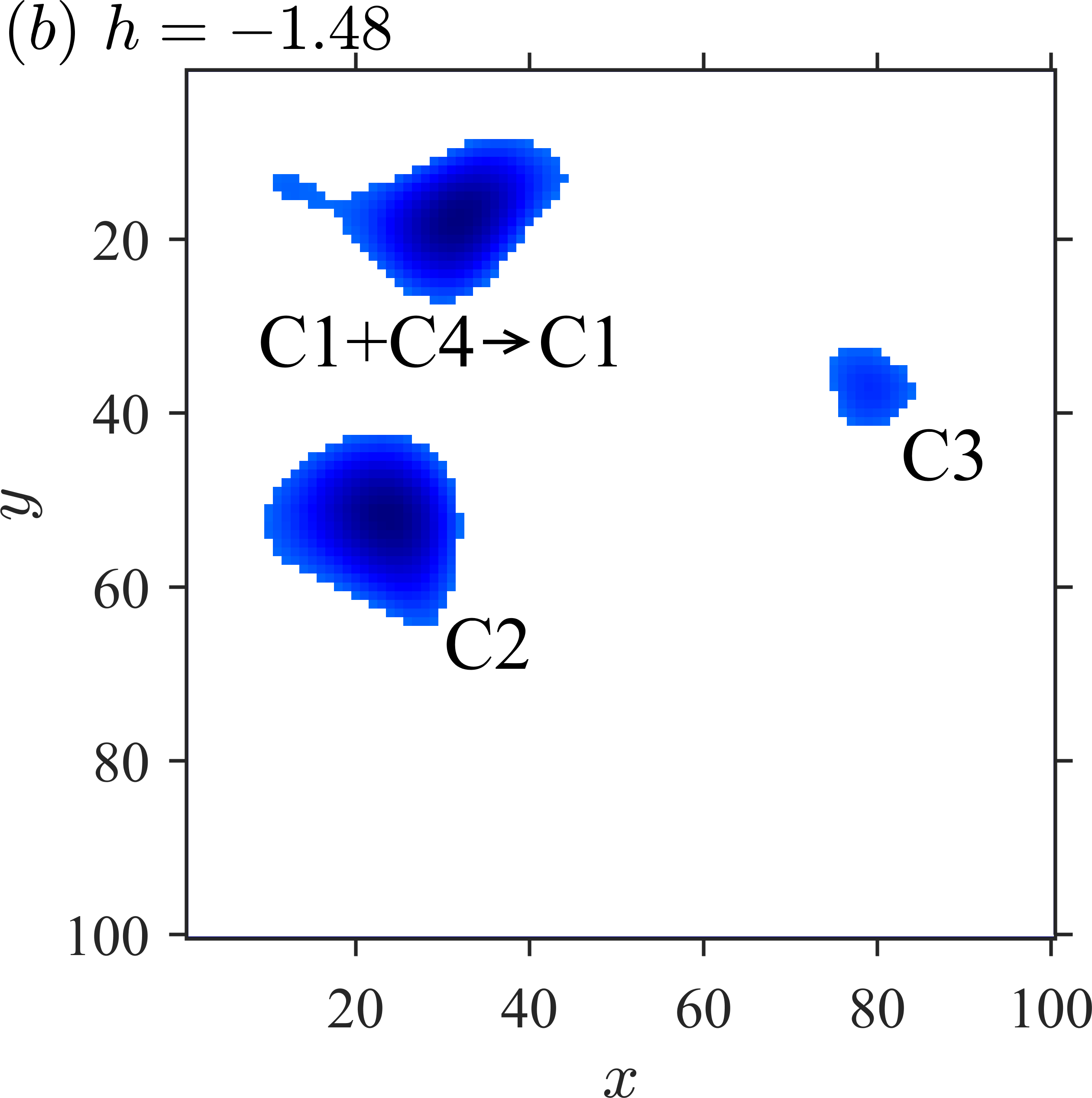} \hspace{2mm}
\vspace{-5mm}
\includegraphics[width=0.27\textwidth]{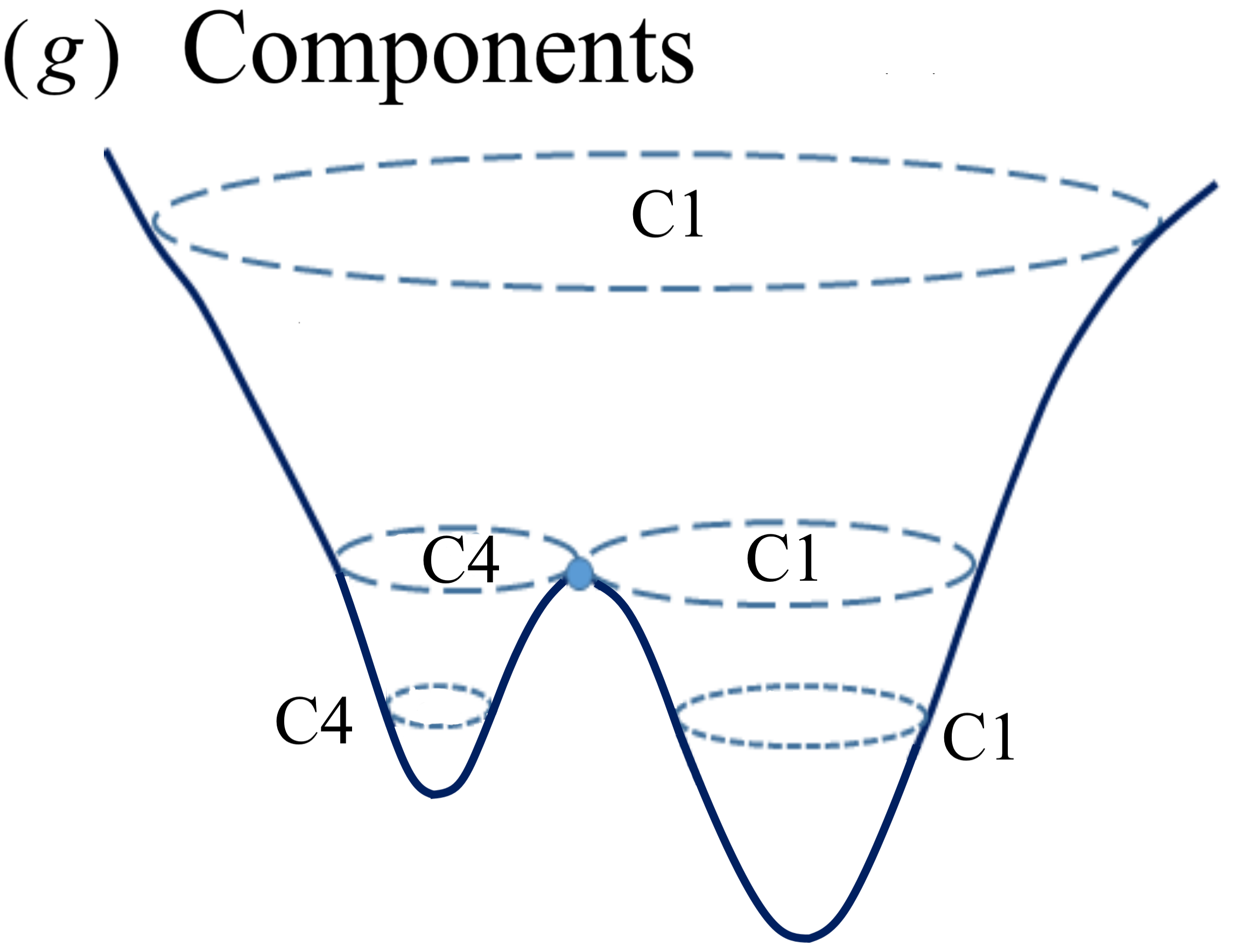}
}
\vspace{2mm}
\centerline{
\includegraphics[width=0.28\textwidth]{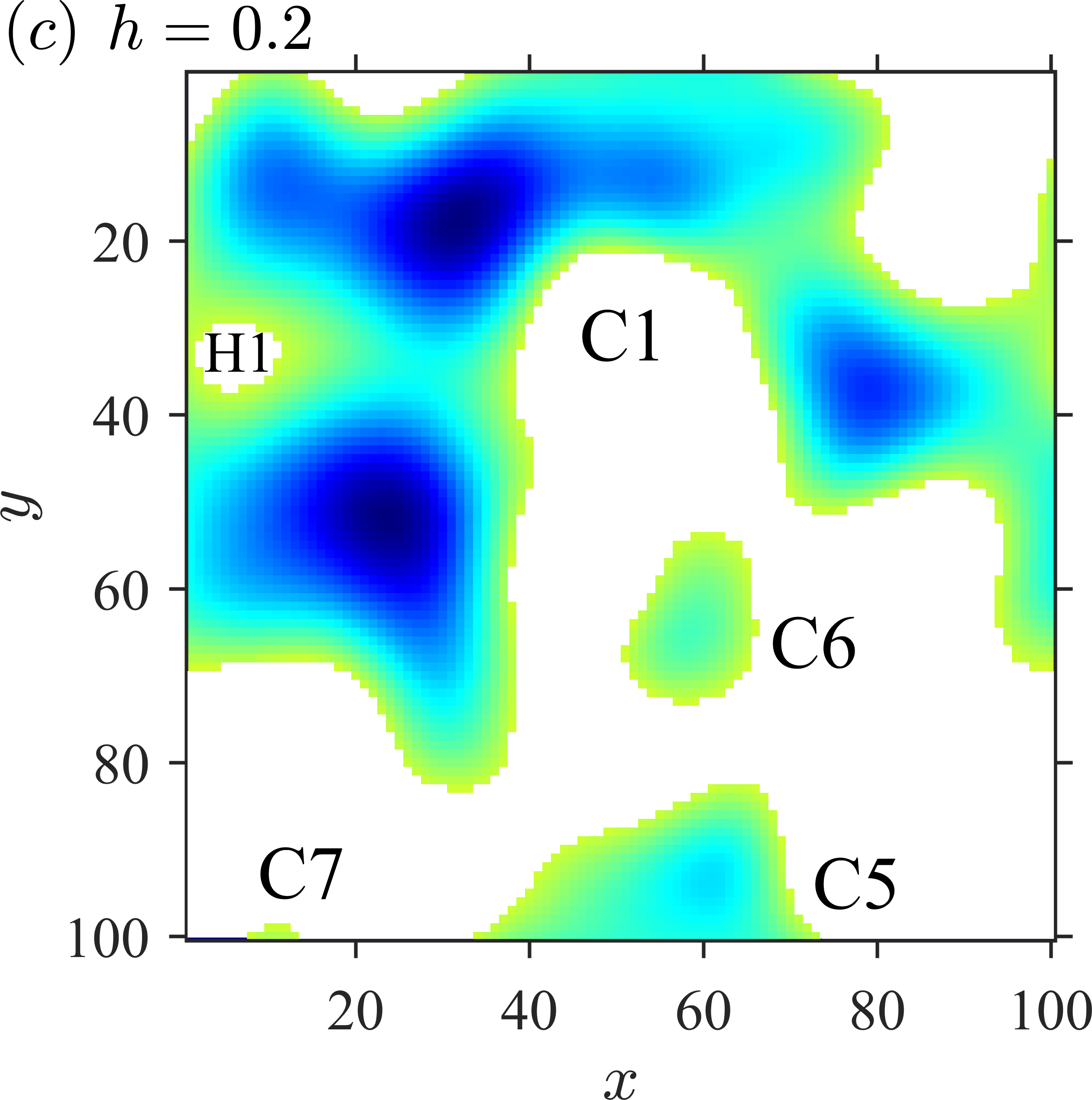} \hspace{2mm}
\includegraphics[width=0.28\textwidth]{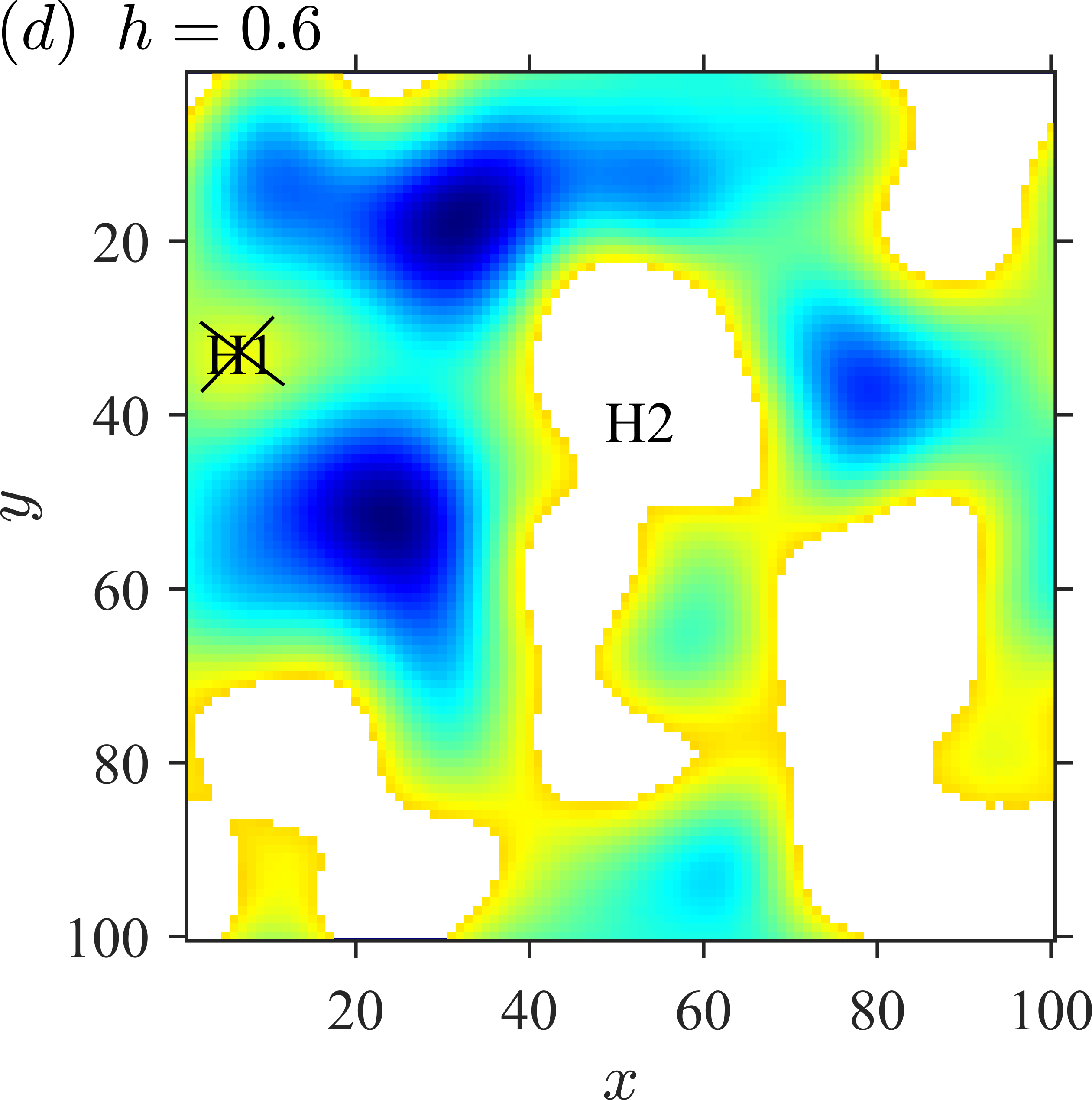} \hspace{2mm}
\includegraphics[width=0.27\textwidth]{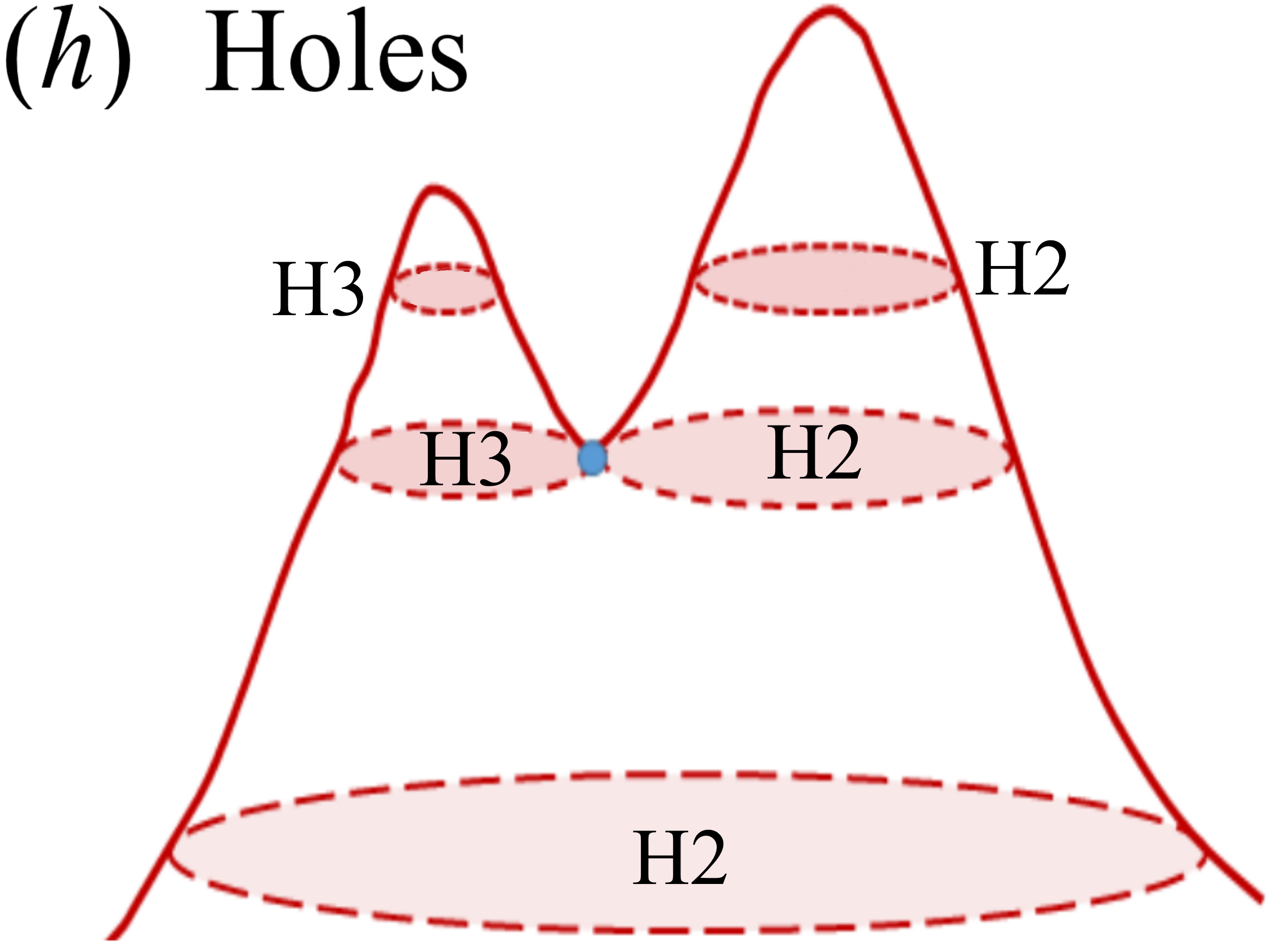}
}
\vspace{2mm}
\centerline{
\includegraphics[width=0.28\textwidth]{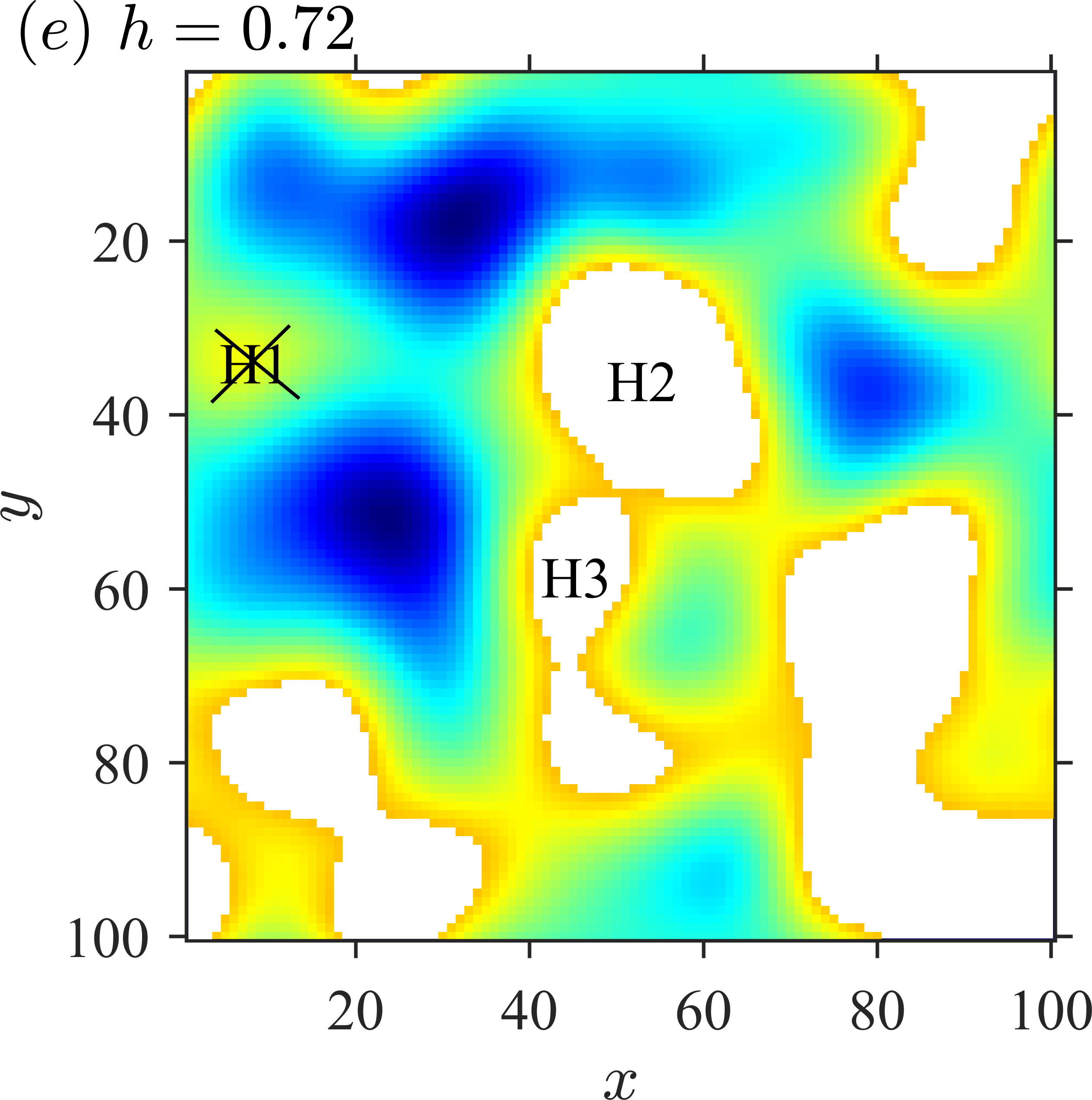} \hspace{2mm}
\includegraphics[width=0.315\textwidth]{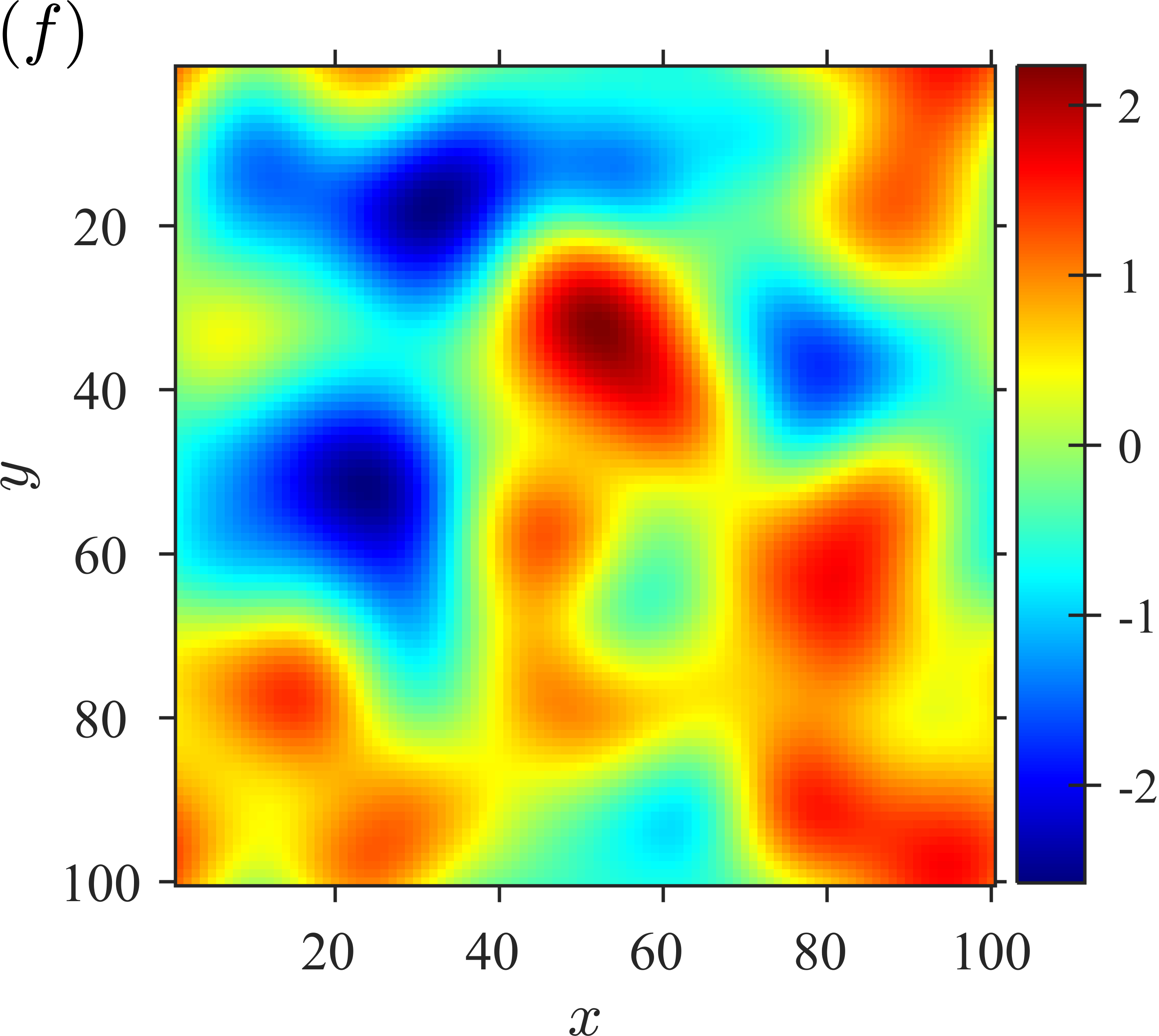}
\hspace{0.26\textwidth}}
\caption{Topological filtration of a continuous, smooth, 2D Gaussian scalar 
field, $f (x, y)$ (colour coded with the colour bar in panel (f)), which has a 
vanishing mean value, unit standard deviation, and an autocorrelation function 
$C(l) = \exp [-l^2/(2 L^2)]$, where $L \approx 15$. Sublevel sets, i.e., 
regions where $f(x, y) \leq h$, are shown for increasing values of $h$: 
(\textit{a}) $h = -1.5$, (\textit{b}) $h = -1.48$, (\textit{c}) $h = 0.2$, 
(\textit{d}) $h = 0.6$ and (\textit{e}) $h = 0.72$. The components (C) and 
holes (H) are labelled in panels (\textit{a}), (\textit{b}), (\textit{c}), 
(\textit{d}) and (\textit{e}) as described in the text. Panels (\textit{g}) and 
(\textit{h}) show how the components merge and the holes split as the level $h$ 
changes. Panel (\textit{f}) presents the full range of $f(x, y)$.}
\label{fig:fRff}
\end{figure}

\subsection{Topological filtration of a 2D field}
The topological filtration and related techniques can straightforwardly be 
generalized to multi-dimensional scalar random fields but they are easier
described at an intuitive level in two dimensions where the Betti numbers 
$\beta_0$ and $\beta_1$ are defined. Consider a continuous random function
$f(x,y)$ defined in a finite domain, and its isocontour at a level $h$, i.e., a 
curve in the $(x,y)$-plane where $f = h$. The set of points $(x,y)$ where $f 
\leq h$ is the sublevel set. We vary $h$ from smaller to larger values and 
record the number of components (clouds) and holes in the isocontours at each 
level together with the values of $h$ where the components and holes appear and 
disappear. As in 1D, the isocontours can also be scanned down from larger to 
smaller values of $h$: this does not affect the results. The transition from 
the sublevel sets to superlevel sets swaps the persistent diagrams of $\beta_0$ 
and $\beta_1$ (in 2D). Here, we use the sublevel sets, i.e., scan $f(x,y)$ up 
from smaller to larger values.

When the level $h$ reaches the lowest value of $f$ in the domain, the first 
component emerges as illustrated in figure~\ref{fig:fRff}. Each local minimum 
of $f$ adds a new component when it is reached by the increasing isocontour 
level $h$. Two (or more) components can merge, when $h$ increases, if they are 
connected by a valley (that necessarily contains a saddle point); then the 
labelling convention is that the younger component (i.e., that formed at a 
larger $h$) dies whereas the older one continues to exist. Two components merge 
when the isocontour contains a saddle point, see 
figure~\ref{fig:fRff}(\textit{g}); three components merge, when $h$ passes 
through a monkey saddle (a degenerate critical point with a local minimum along 
one direction and inflection point in another, as opposed to the ordinary saddle
with a minimum in one direction and maximum in another). Three (or more) 
components can merge, when there are three (or more) saddle points at the same 
level $h$. The components can merge to form a loop whose interior is a hole; 
each hole at a level $h$ surrounds a local maximum of $f$ that is reached at 
some higher level. Holes are born when a loop is formed and die when $h$ passes 
through the corresponding local maximum of $f$. Holes can split up at a saddle 
point, in which case the labelling convention is to ascribe the original birth 
time to the hole with the larger eventual local maximum, and deem the current 
level to be the birth time of the second hole.  
Figure~\ref{fig:fRff}(\textit{h}) illustrates this.

It is then clear that the birth and death of components and holes are 
intrinsically related to the nature of, and connections between, the stationary 
points of the random field (its extrema and saddles) and to the values of $f$ 
at those points. Betti numbers contain rich information about the random 
function. Eventually, as $h$ approaches the absolute maximum of $f$ in the 
domain, only one, most persistent component remains (and no holes). Therefore, 
$\beta_0 = 1$ and $\beta_1 = 0$ at levels $h$ exceeding the absolute maximum of 
$f$ in the domain. The ‘lifetime’, or persistence of a component or a hole is 
characterized by the range of $h$ where it exists. Selecting only those 
features that are more persistent, one distils a simplified (and therefore, 
better manageable) topological portrait of the random field.

Figure~\ref{fig:fRff} illustrates the topological filtration of a 
two-dimensional random field using sublevel sets $f(x,y) \leq h$, where $h$ is 
the level altitude. Figure~\ref{fig:fRff}f shows a realization a continuous, 
smooth, 2D Gaussian scalar field, $f (x, y)$, of in a domain of $100^2$ pixels 
in size, which has a vanishing mean value, unit standard deviation, and an
autocorrelation function $C(l) = \exp [-l^2/(2 L^2)]$ with $L \approx 15$.

The absolute minimum of the field, $f = -2.57$, occurs at $(x, y) = (31, 18)$.
Thus, the first component C1 is born at $h = -2.57$ and there are no holes at 
this level: we have $\beta_0 = 1$ and $\beta_1 = 0$ at $h = -2.57$. As $h$ 
increases, more components are born. At $h = -1.5$, panel (\textit{a}), there 
are four components C1--C4 and no holes: $\beta_0 = 4$ and $\beta_1 = 0$. At $h 
= -1.48$, panel (\textit{b}), components C1 and C4 have merged via a saddle 
point between them, passed through at a smaller $h$; the surviving component is 
labelled C1 whereas C4 has died as it was born later than C1. There are no 
holes at $h = -1.48$: $\beta_0 = 3$ and $\beta_1 = 0$. At a higher level $h = 
0.2$, shown in panel (\textit{c}), most of the components are merged into a big 
island, and there are three smaller islands, C5, C6 and C7, in the bottom half 
of the panel; moreover, one hole H1 appeared: $\beta_0 = 4$ and $\beta_1 = 1$.  
At a level $h = 0.6$, panel (\textit{d}), the number of components is 2, 
$\beta_0 = 2$; and one more hole H2 has appeared, $\beta_1 = 1$. At a higher 
level $h = 0.72$, panel (\textit{e}), all components have merged into a single 
island, $\beta_0 = 1$. There are two holes labelled H2 and H3, each around a 
local maximum of $f$, so $\beta_1 = 2$. Note that some holes (bordering on the 
field frame) have not yet completely formed and are not taken into account as 
holes. There are six such cases in panel (\textit{e}). The holes H1 and H2, 
which surround the maximum lower than $f = 0.72$, have already died 
(contracted).
 
\section{Advantages of data standardization in topological filtration}
\label{sec:stand}

In this section we describe a procedure for standardizing 1D signals and 2D 
fields. This preprocessing simplifies the comparison of the results of 
topological filtration that have been applied to different data. For example, 
one of our main motivations is to compare the topology of astronomical 
observations and numerical simulations: standardizing the data ensures a 
natural non-dimensionalization of the data and allows us to focus on their more 
subtle properties. We will use 1D examples to show how data standardization 
affects persistence diagrams.

\begin{figure} 
\centerline{
\includegraphics[width=0.55\textwidth]{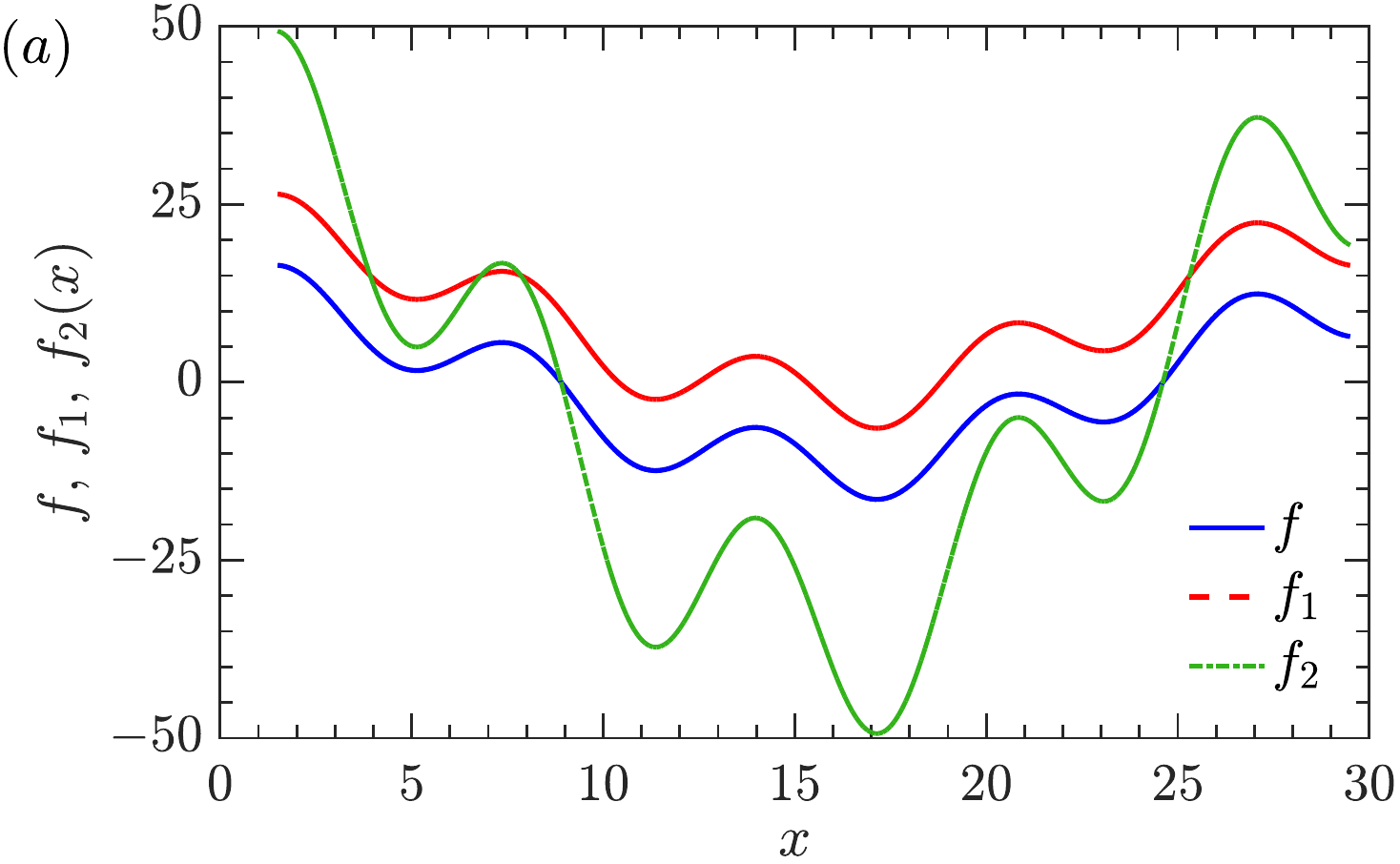} 
\includegraphics[width=0.37\textwidth]{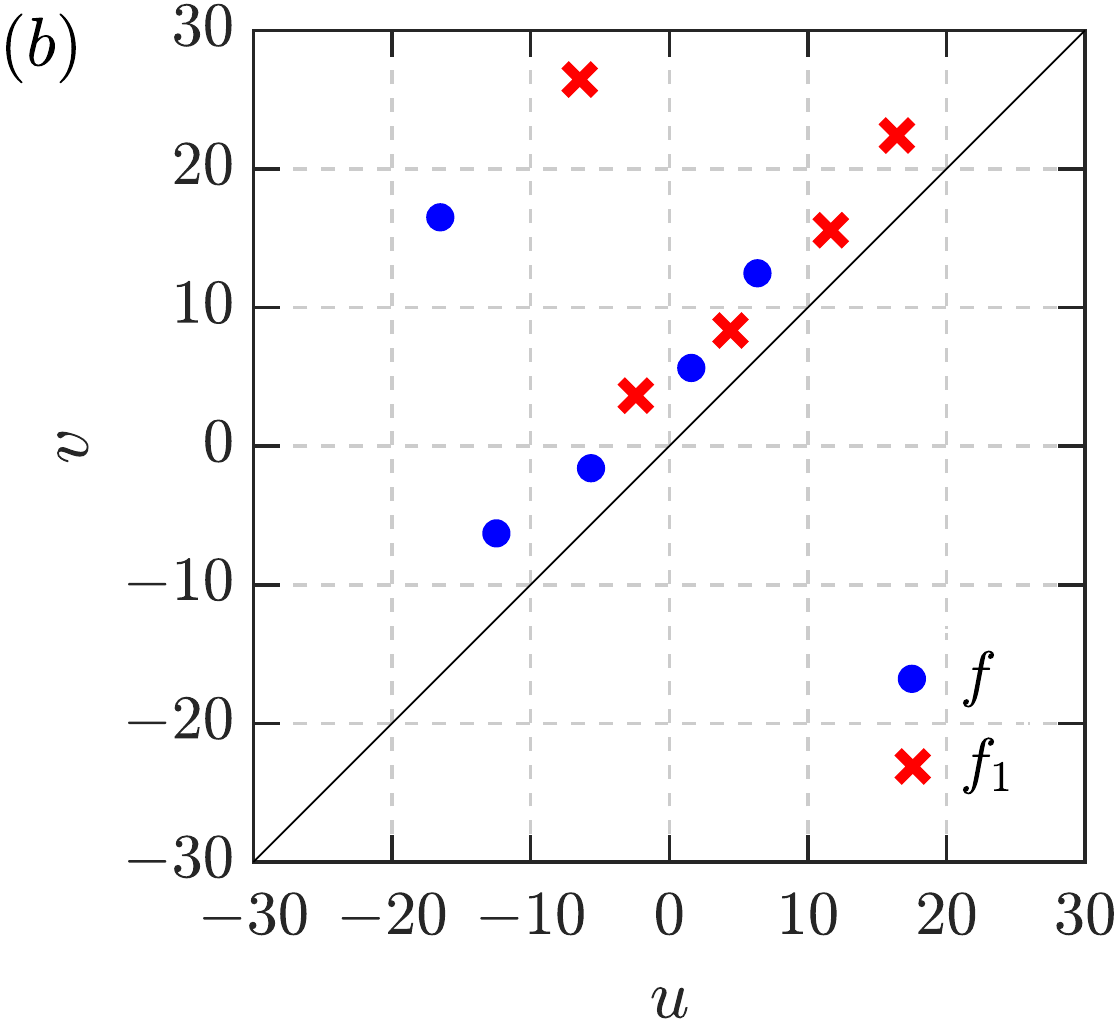}}
\vspace{2mm}
\centerline{
\includegraphics[width=0.37\textwidth]{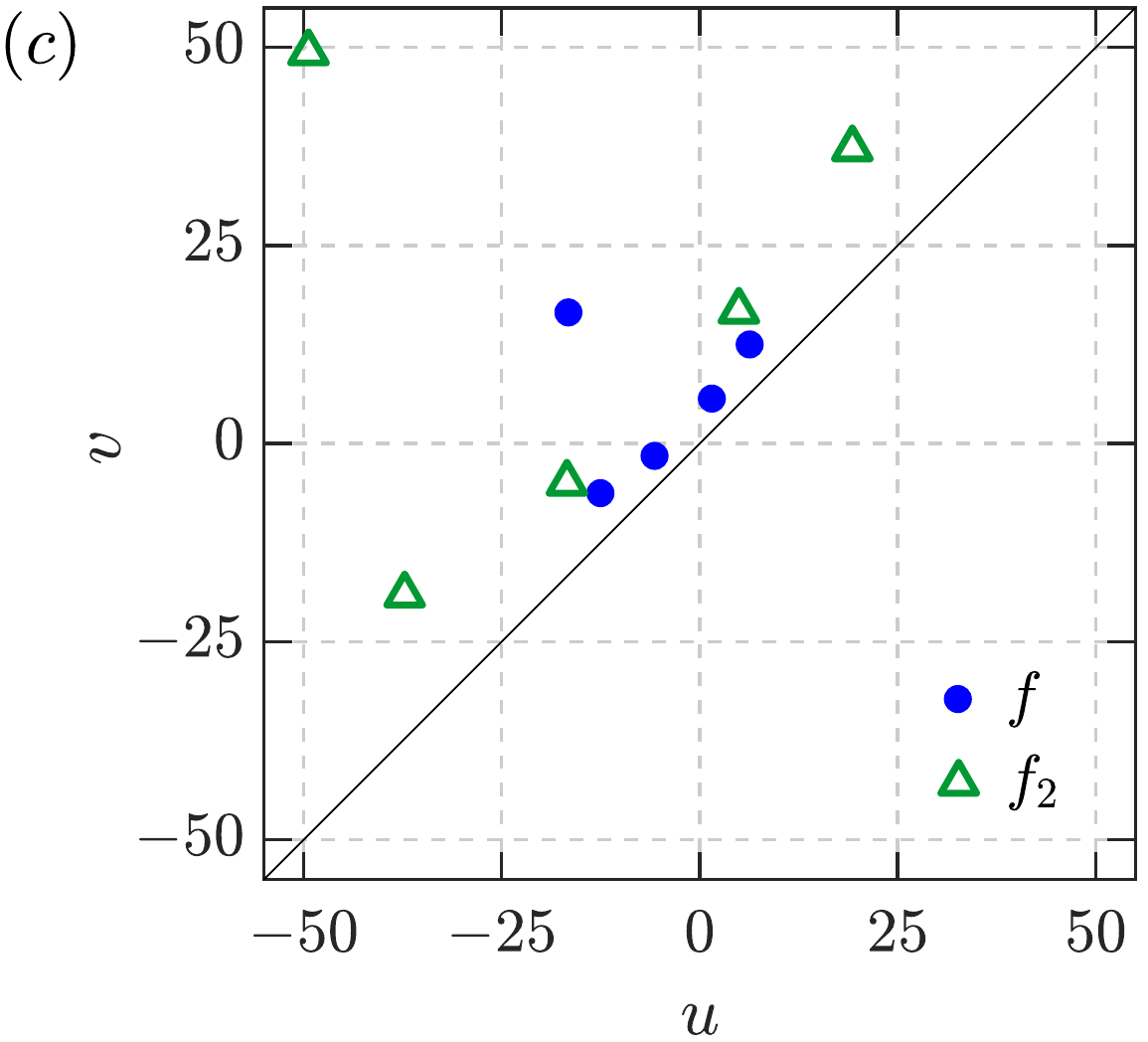} \hspace{5mm}
\includegraphics[width=0.365\textwidth]{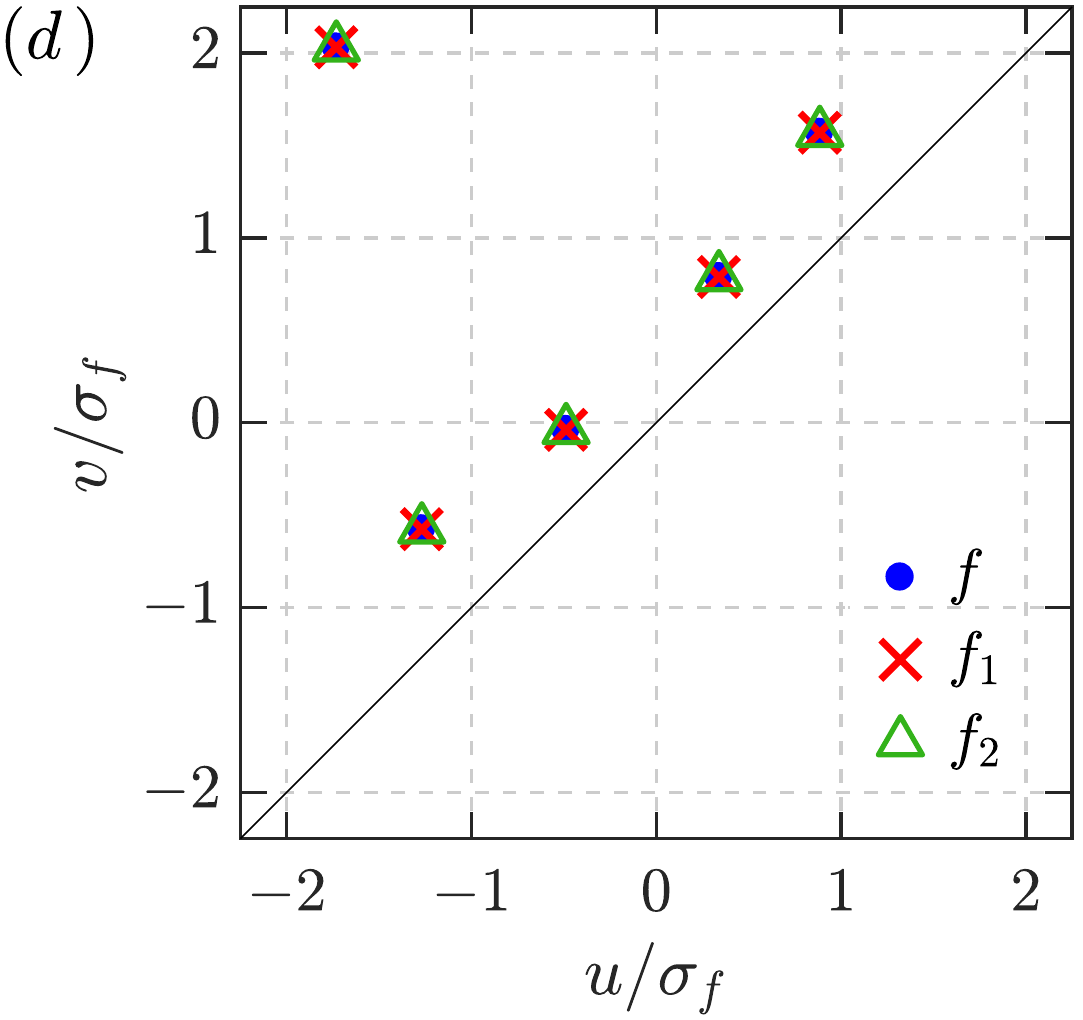}} 
\caption{Signal standardization affects persistence diagrams. 
(\textit{a}) A signal $f(x)$ of equation~(\ref{fxsin}) (the solid blue line), 
and its shifted $f_1 = f + 10$ (dashed red) and scaled $f_2 = 3 f$ versions 
(dash-dotted green). The persistence diagrams of (\textit{b})  $f$ (circles) 
and $f_1$ (crosses), and (\textit{c})  $f$ (circles) and $f_2$ (triangles).
(\textit{d}) The persistence diagrams of the signals $f$, $f_1$ 
and $f_2$ coincide when they are standardized.}
\label{fig:F7}
\end{figure}

For a random function $f(x)$, its mean value and standard deviation are denoted 
$\langle f \rangle$ and $\sigma_f$, respectively. We introduce a standardized 
version of $f(x)$ as
\begin{equation}
\tilde{f} = \frac{f - \langle f \rangle}{\sigma_f}\,,
\label{eq:stand}
\end{equation}
so that $\langle \tilde{f} \rangle = 0$ and $\sigma_{\tilde{f}} = 1$.
 
Figure~\ref{fig:F7}(\textit{a}) shows the graphs of three related functions.
The first one, shown in solid blue, is 
\begin{equation}\label{fxsin}
f(x) = 5 \sin(x) + 12 \cos(0.2 x)\,,
\end{equation}
with  $\langle f \rangle = -1.32$ and $\sigma_f = 8.75$. The first term 
represents a higher-frequency fluctuation superimposed on a lower-frequency 
trend. The second one is $f_1=f+10$ (dashed red), and the third is $f_2 = 3f$ 
(dash-dotted green). The blue dots in figure~\ref{fig:F7}(\textit{b}) is the 
persistence diagram of $f$, while red crosses show the results of filtration of 
$f_1$. Both diagrams have the same spatial configuration, but the increased 
magnitude of $f_1$ results in an upwards shift of its components along the 
diagonal (red crosses). However, the persistence diagrams of $f$ and $f_2$, 
shown in figure~\ref{fig:F7}(\textit{c}), differ in a non-trivial way: the 
increased amplitude results in the components of $f_2$ being spread-out in the 
persistence diagram, while retaining the same relative spacing as those of $y$. 
If we now standardize the signals using equation~(\ref{eq:stand}), the 
persistence diagrams of $f$, $f_1$ and $f_2$ become identical as shown in 
figure~\ref{fig:F7}(\textit{d}). Their \emph{topologies} are thus easily seen 
to be identical, as expected. Standardization of 2D and 3D fields has the same 
effect on the persistence diagrams.

\section{Large-scale trends}\label{sec:trends}
In practice, it is often necessary to work with data in which small-scale 
random fluctuations are superimposed on large-scale trends. For example, 
figure~\ref{fig:f1}({\it a}) shows the gas density $n$ observed in  the Milky 
Way. It contains large-scale trends along and across the Galactic mid-plane at 
$z = 0$, mainly caused by vertical stratification and azimuthal variations due 
to the spiral arms. The best method to isolate the trend from the fluctuations 
is not obvious though -- we could use horizontal averages, apply 2D smoothing, 
fit polynomials or splines, employ wavelet filtering -- and the results 
obtained will often be dependent on the choice that is made.  

In this section we examine how the presence of a trend influences the 
topological data analysis and we will show that some measures are relatively 
insensitive to the presence or absence of a large-scale trend in the data.

\subsection{Trends in 1D signals}

\begin{figure} 
\centerline{
\includegraphics[width=0.43\textwidth]{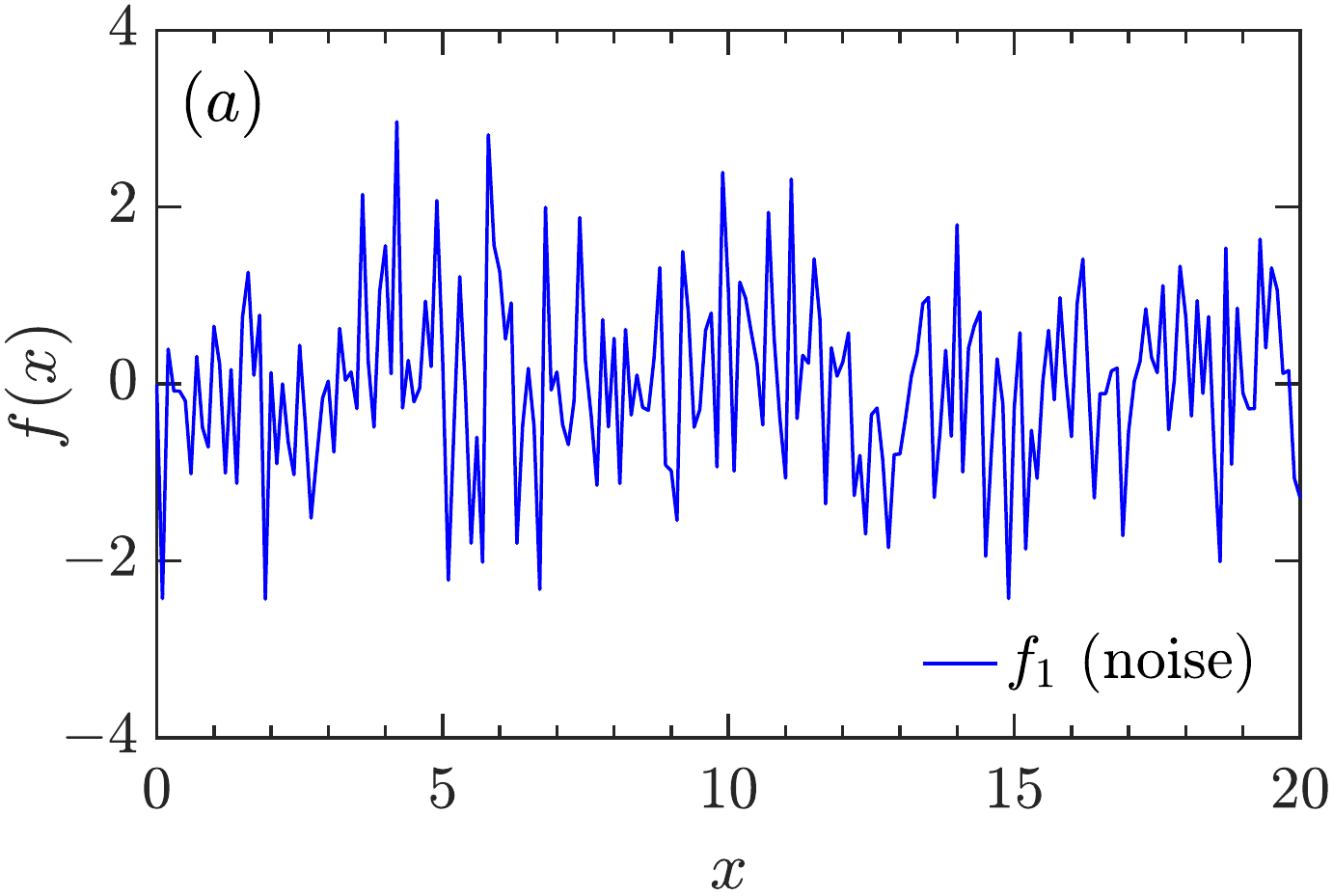}
\includegraphics[width=0.43\textwidth]{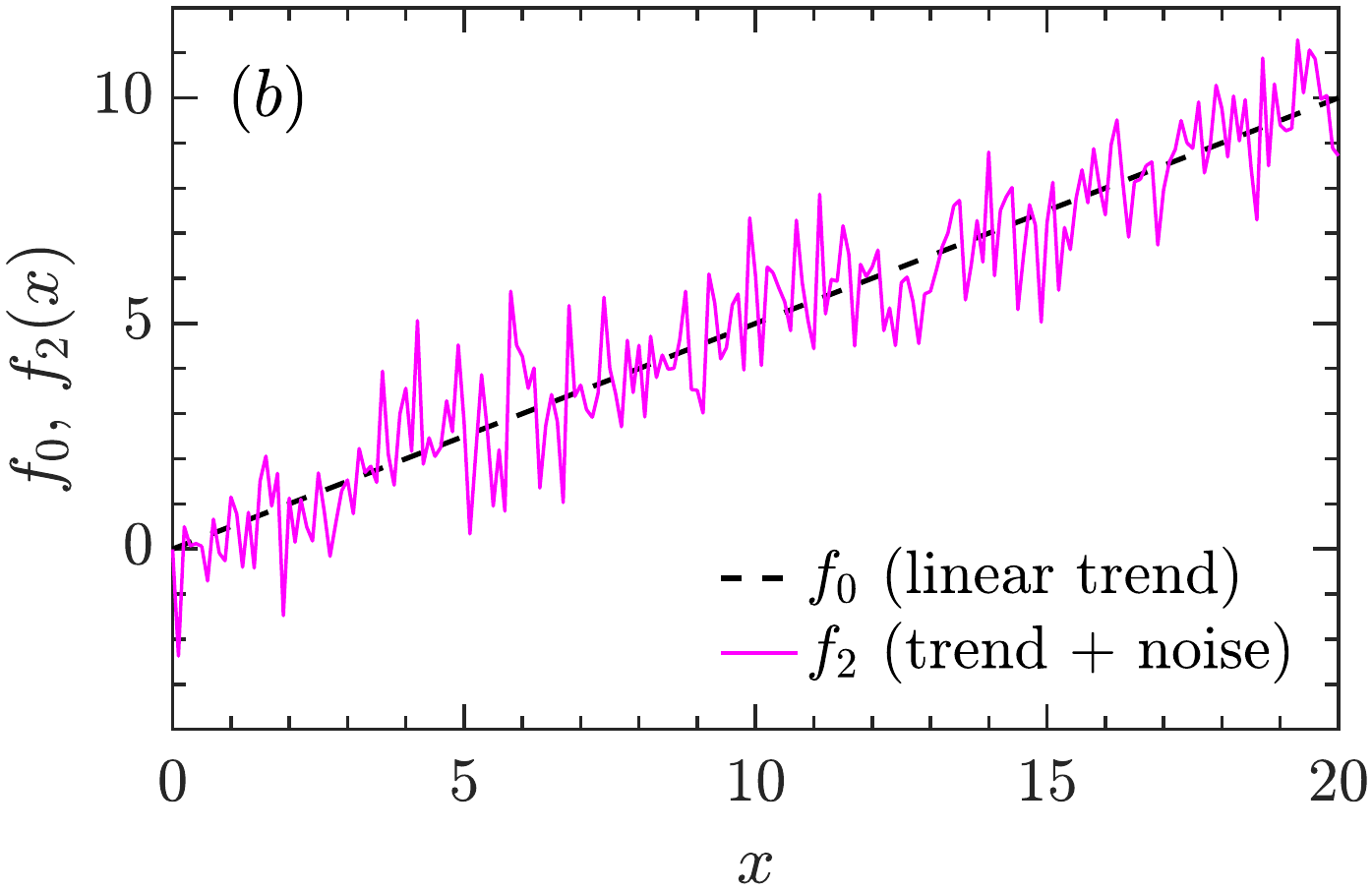}}
\vspace{2mm}
\centerline{
\includegraphics[width=0.38\textwidth]{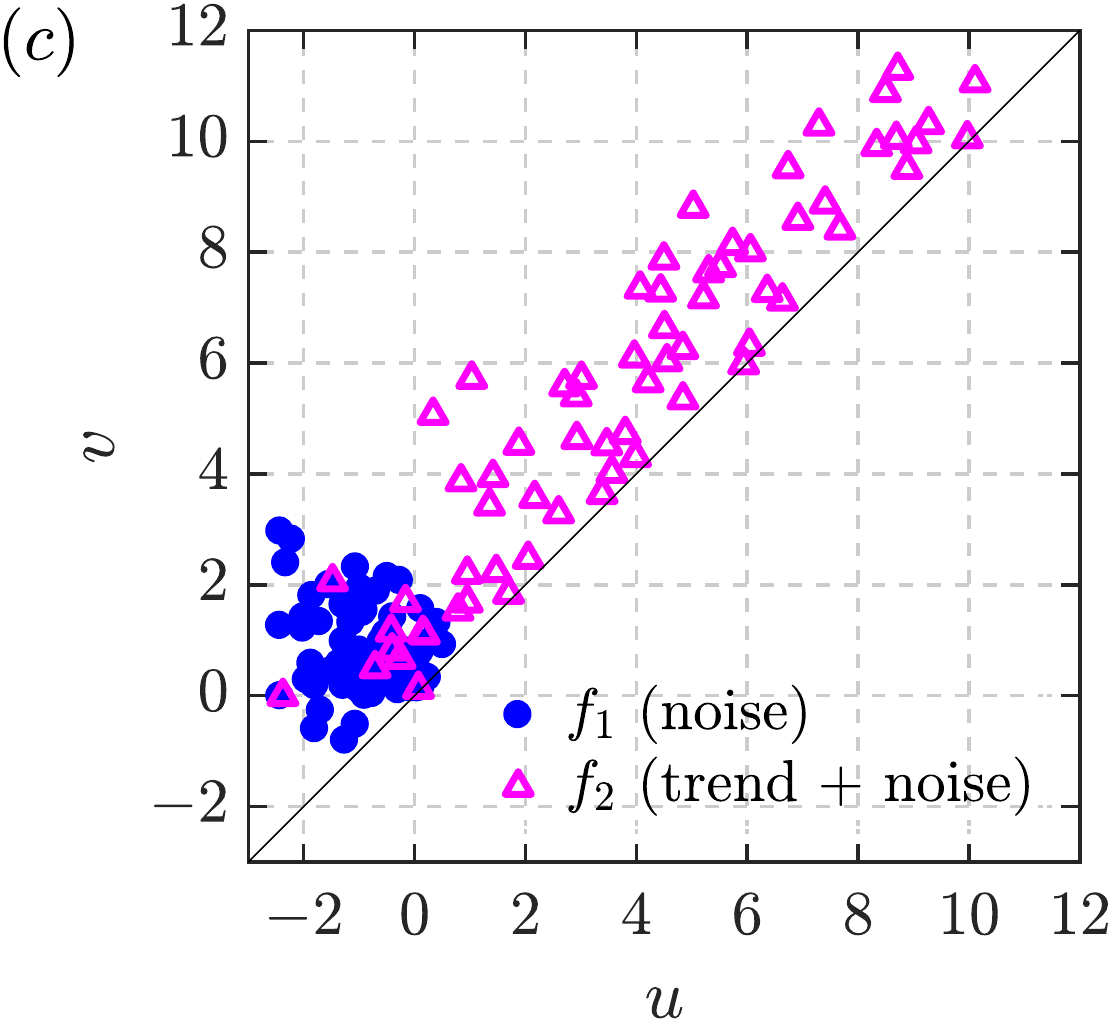}
\hspace{5mm}
\includegraphics[width=0.35\textwidth]{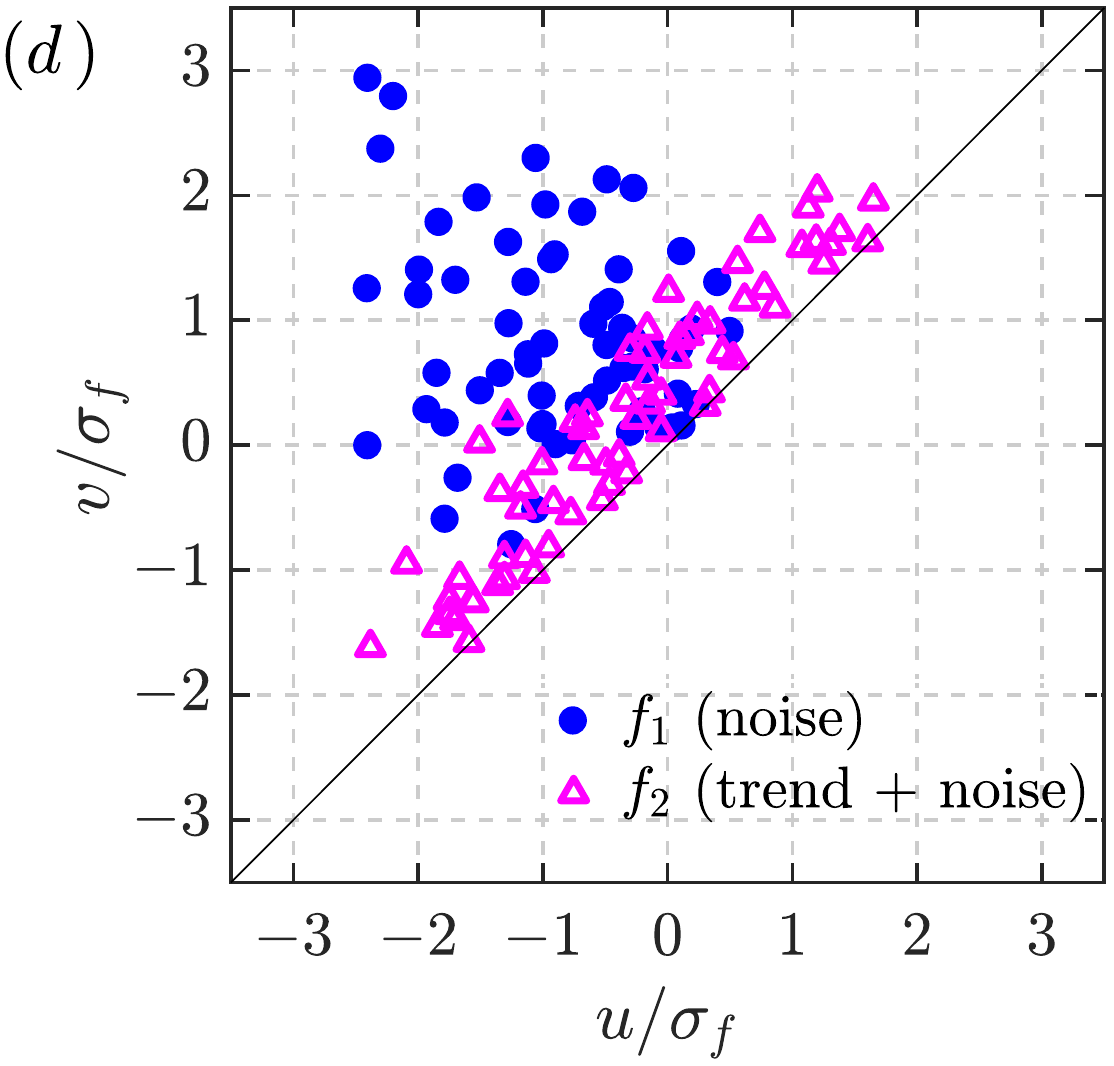}}
\caption{Influence of a linear trend on persistence diagrams. (\textit{a}) 
Gaussian $\delta$-correlated random fluctuations with zero mean and unit 
standard deviation. (\textit{b}) A new signal $f_2  = f_0 + f_1$ (magenta), 
formed by combining the Gaussian fluctuations $f_1$  and a linear trend $f_0 = 
x/2$ (dashed black line). (\textit{c}) The persistence diagram for the 
functions $f_1$ and $f_2$ obtained without standardization. (\textit{d}) The 
persistence diagram obtained after standardization of both signals, using 
equation~(\ref{eq:stand}). In the persistence diagrams, components arising 
solely from the fluctuations are shown as blue dots and those from the 
fluctuations plus trend are shown as magenta triangles.}  
\label{fig:linTrend}
\end{figure} 

\begin{figure} 
\includegraphics[width=0.45\textwidth]{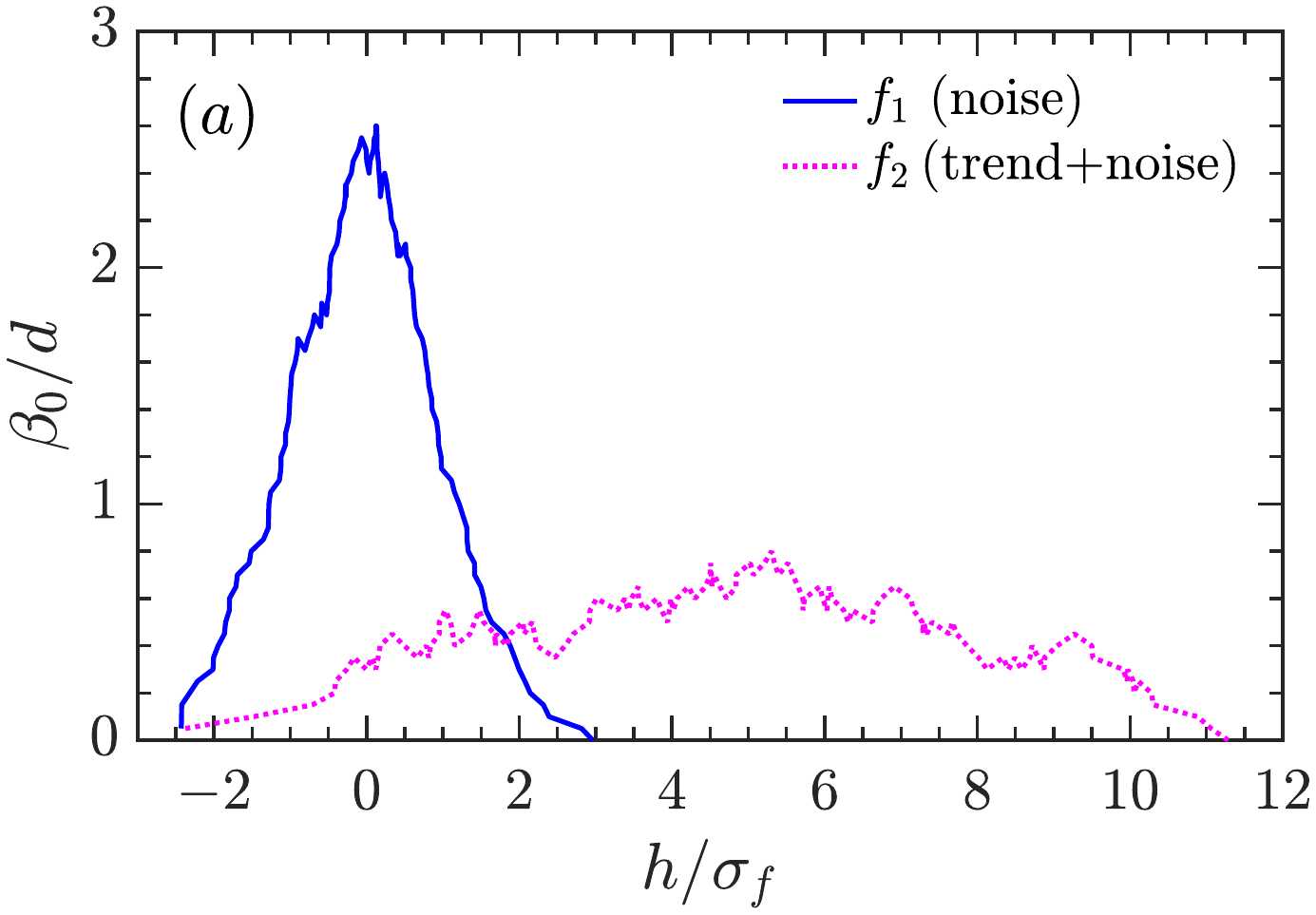} 
\hfill
\includegraphics[width=0.45\textwidth]{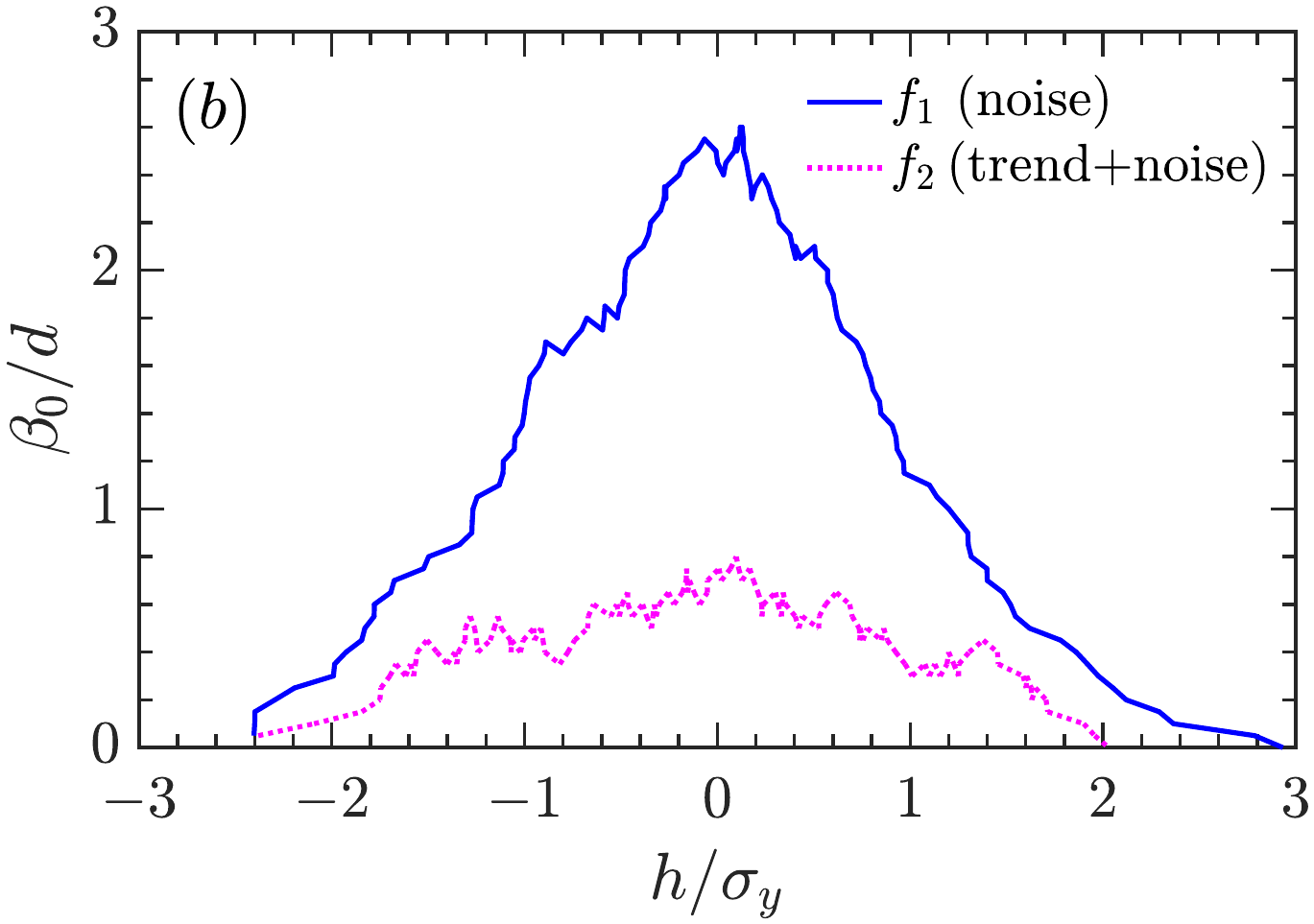} 
\vspace{2mm}
\includegraphics[width=0.45\textwidth]{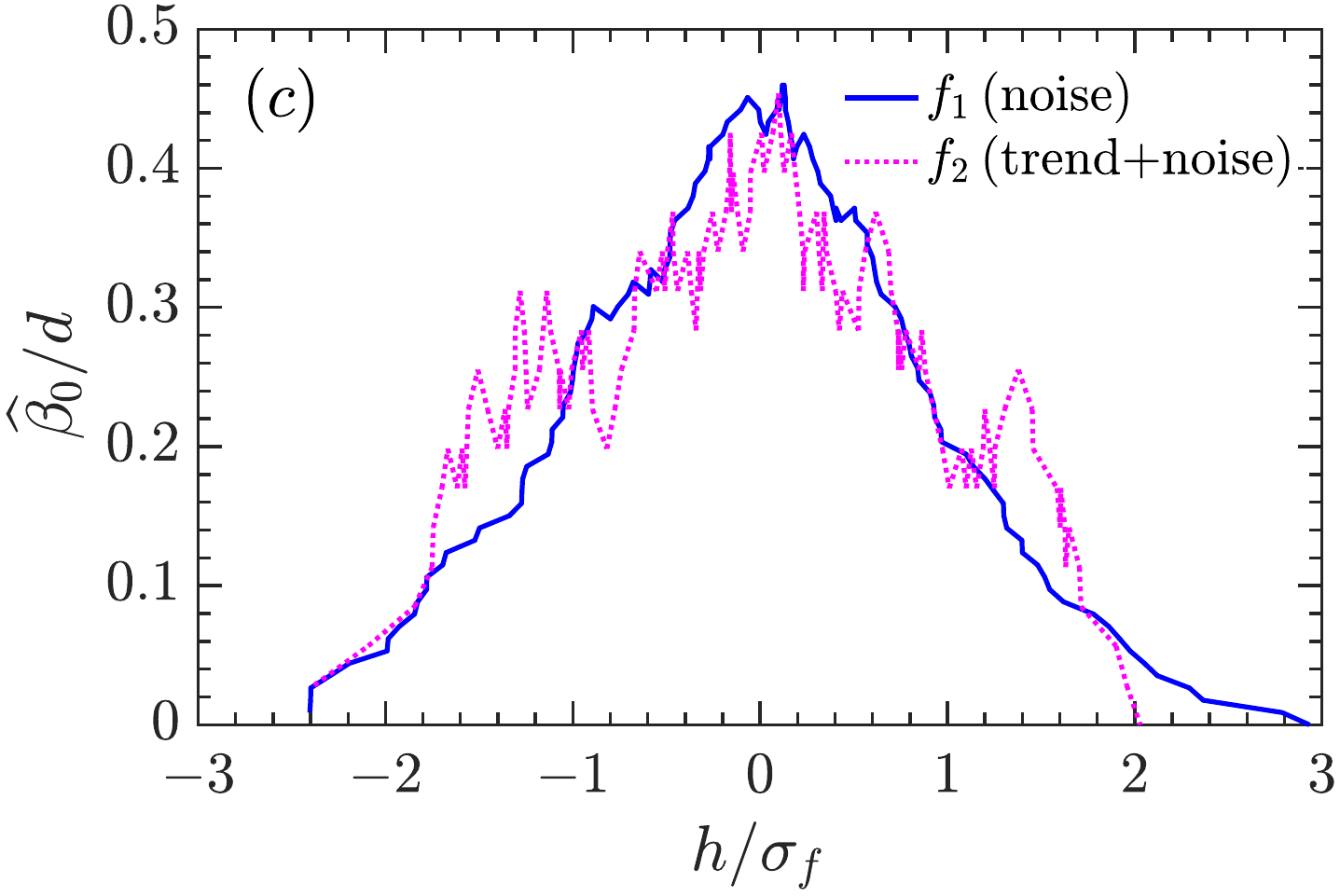} 
\hfill
\includegraphics[width=0.45\textwidth]{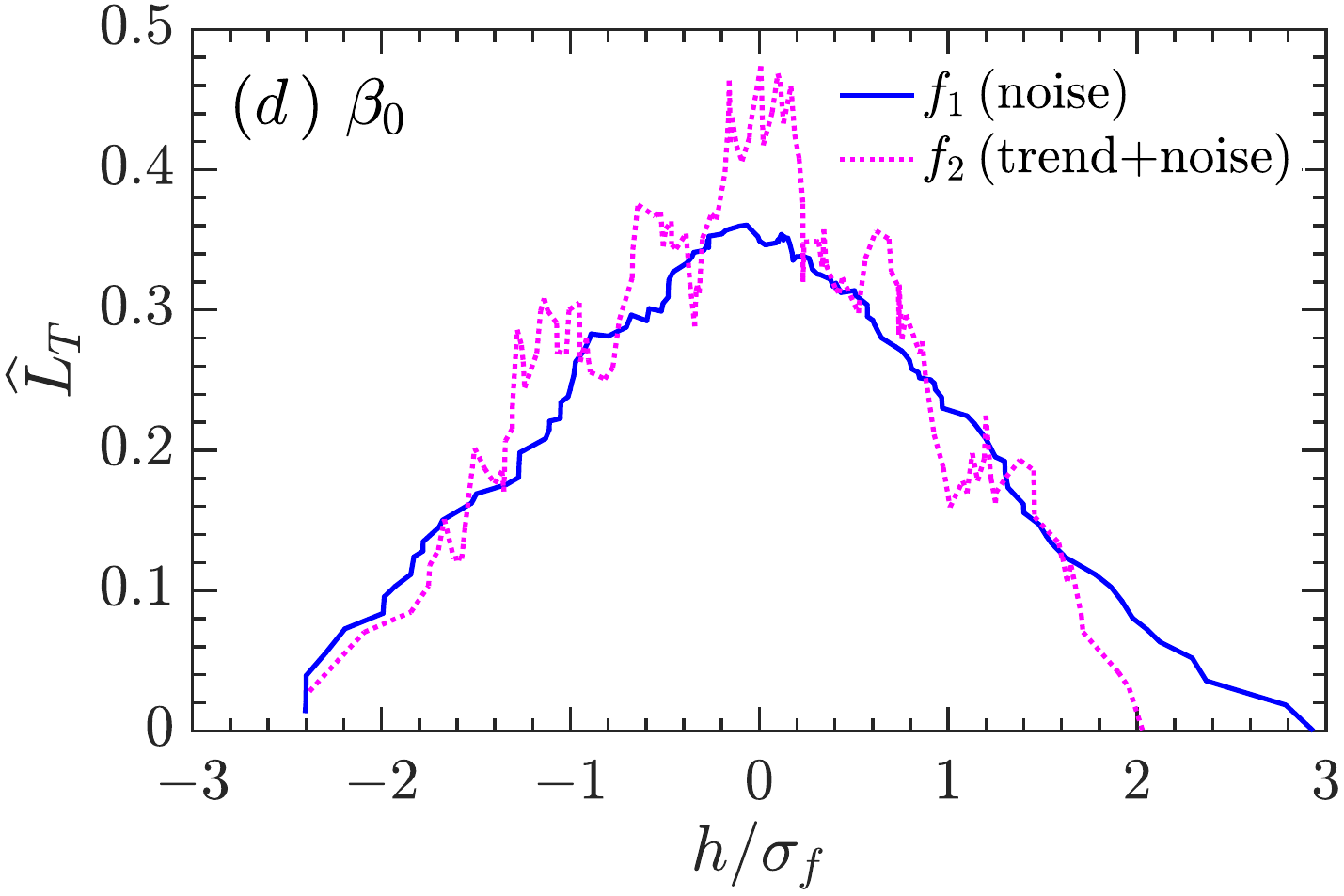} 
\caption{Influence of a linear trend and standardization on the Betti numbers 
($\beta_0$ in 1D case), and the total barcode length $L_T$ at levels. 
(\textit{a}) The number of components $\beta_0/d$ at each level $h$, calculated 
per unit length of the graph (here $d$ is the length along the $x$-axis). The 
result is shown for the Gaussian $\delta$-correlated noise $f_1$ (solid blue 
line) and for the function $f_2$ (in the dotted magenta line), that is sum of a 
linear trend $f_0 = x/2$ and the noise $f_1$. (\textit{b}) The normalized Betti 
number $\beta_0/d$ after standardization of the functions $f_1$ and $f_2$, 
using equation~(\ref{eq:stand}). (\textit{c}) The variation of $\beta_0/d$  
with $h$ normalized to unit area under each curve and  (\textit{d}) the 
corresponding normalized total barcode length $L_T(h)$.}
\label{fig:linTrendPDF}
\end{figure} 

The signal $f_1$ in figure~\ref{fig:linTrend}({\it a}) is Gaussian 
$\delta$-correlated random noise, with zero mean and a unit standard deviation. 
Figure~\ref{fig:linTrend}({\it b}) shows a signal $f_2$ (as solid magenta line),
which is obtained by adding $f_1$ to a linear trend $f_0 = x/2$ (dashed black 
line). Figure~\ref{fig:linTrend}(\textit{c}) demonstrates how the addition of 
the trend alters the appearance of the resulting persistence diagram: the 
compact cluster of components produced by the fluctuations (shown as blue dots) 
are spread out along the diagonal by the linear trend (magenta triangles). 
Applying the standardization of equation~\ref{eq:stand} removes the diagonal 
shift of the components but also produces a substantial difference in the 
lifetime of the components (see figure~\ref{fig:linTrend}(\textit{d})) that was 
not present in the non-standardized diagram. On the basis of this, we suggest 
that the persistence diagram is poorly suited to the analysis of fluctuations
in a signal in the presence of a large-scale trend.

Figure~\ref{fig:linTrendPDF}(\textit{a},\textit{b}) shows that this problem 
remains if we consider the Betti number $\beta_0$. However, panel (\textit{c}) 
of figure~\ref{fig:linTrendPDF} demonstrates that if we normalize the plot of 
$\beta_0/d$ against $h$ so that it has unit area under it, then the two curves 
are similar. This is true for similarly normalised curves for $L_T$ 
(figure~\ref{fig:linTrendPDF}(\textit{d}). Whilst the trend clearly introduces 
stronger local variations in these variables, their overall shape is the same. 

\begin{figure} 
\centerline{
\includegraphics[width=0.50\textwidth]{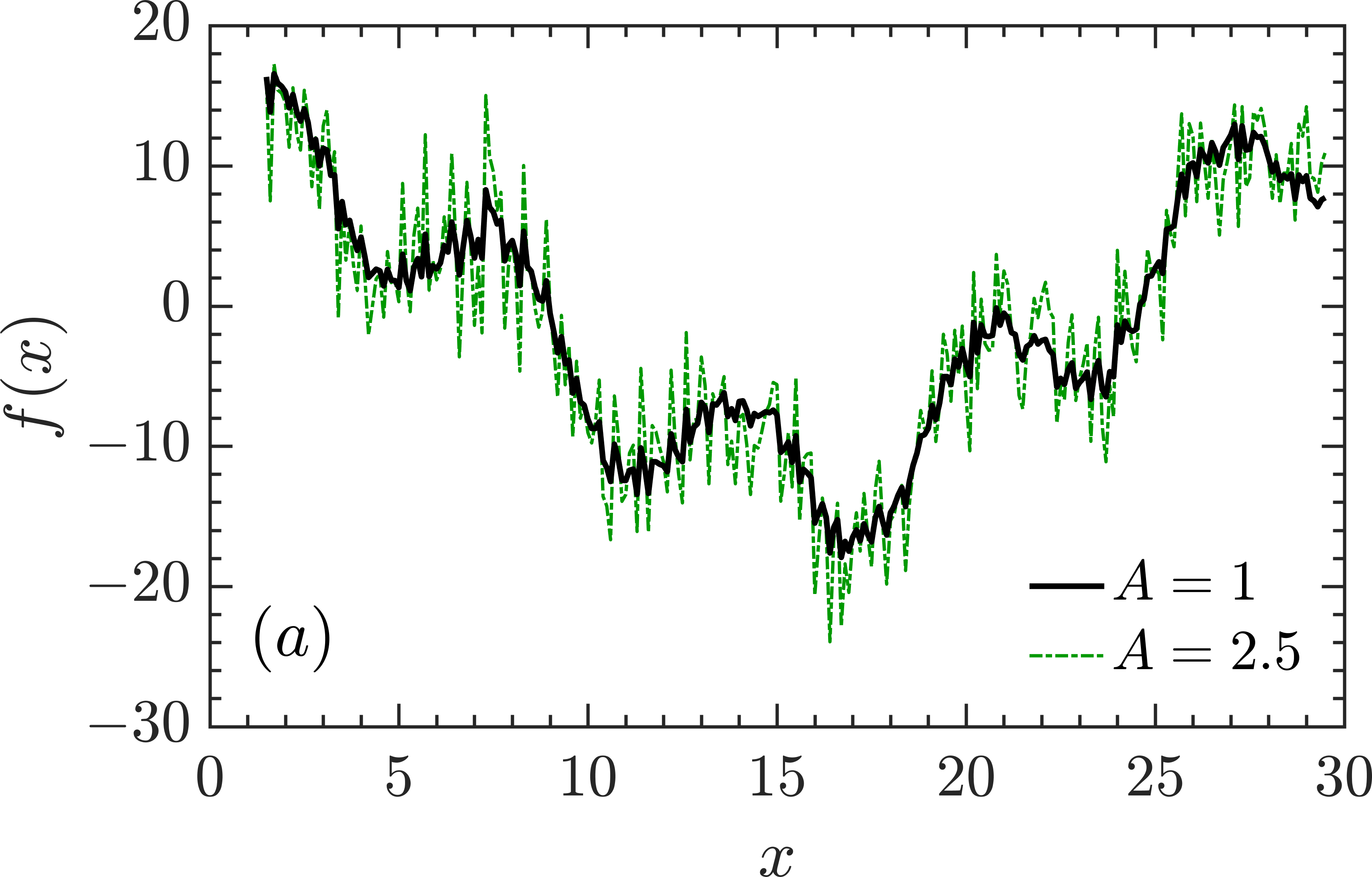}} 
\vspace{2mm}
\centerline{
\includegraphics[width=0.35\textwidth]{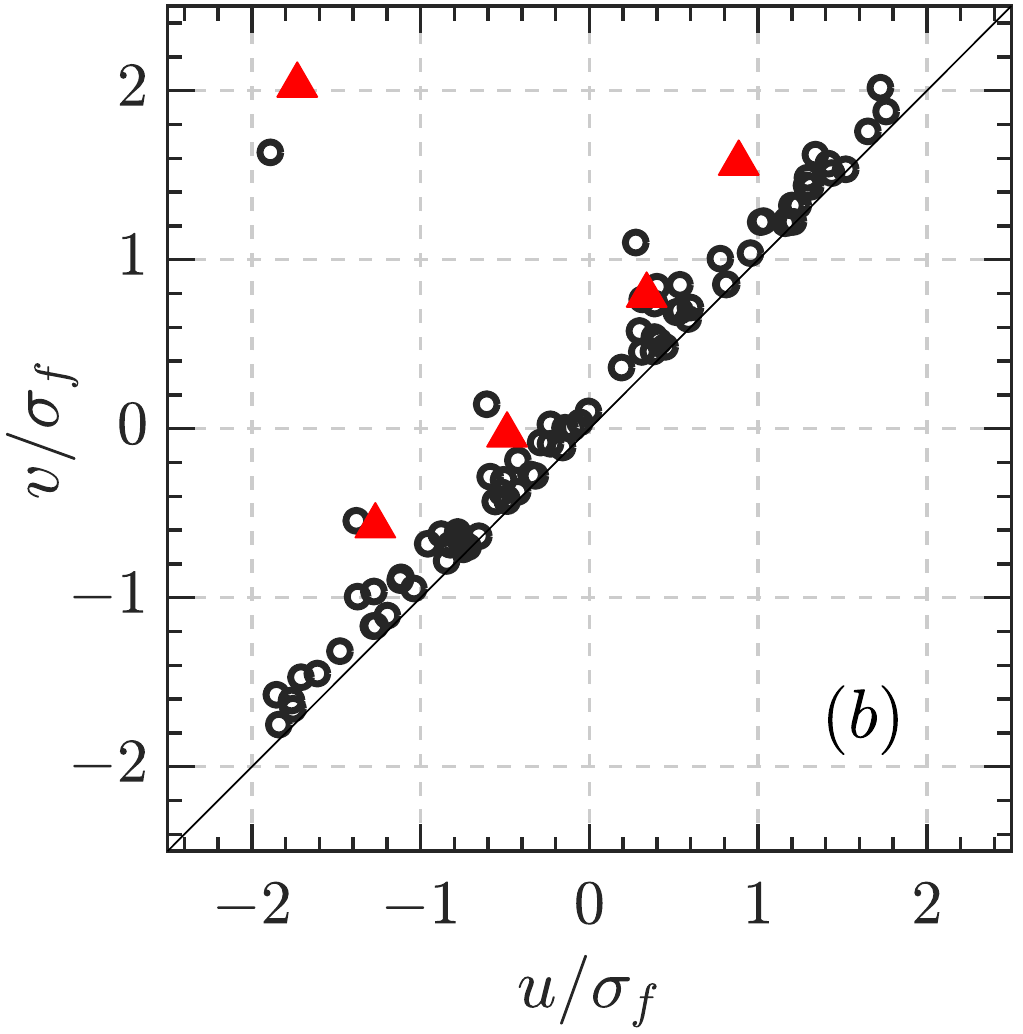} 
\hspace{5mm} 
\includegraphics[width=0.35\textwidth]{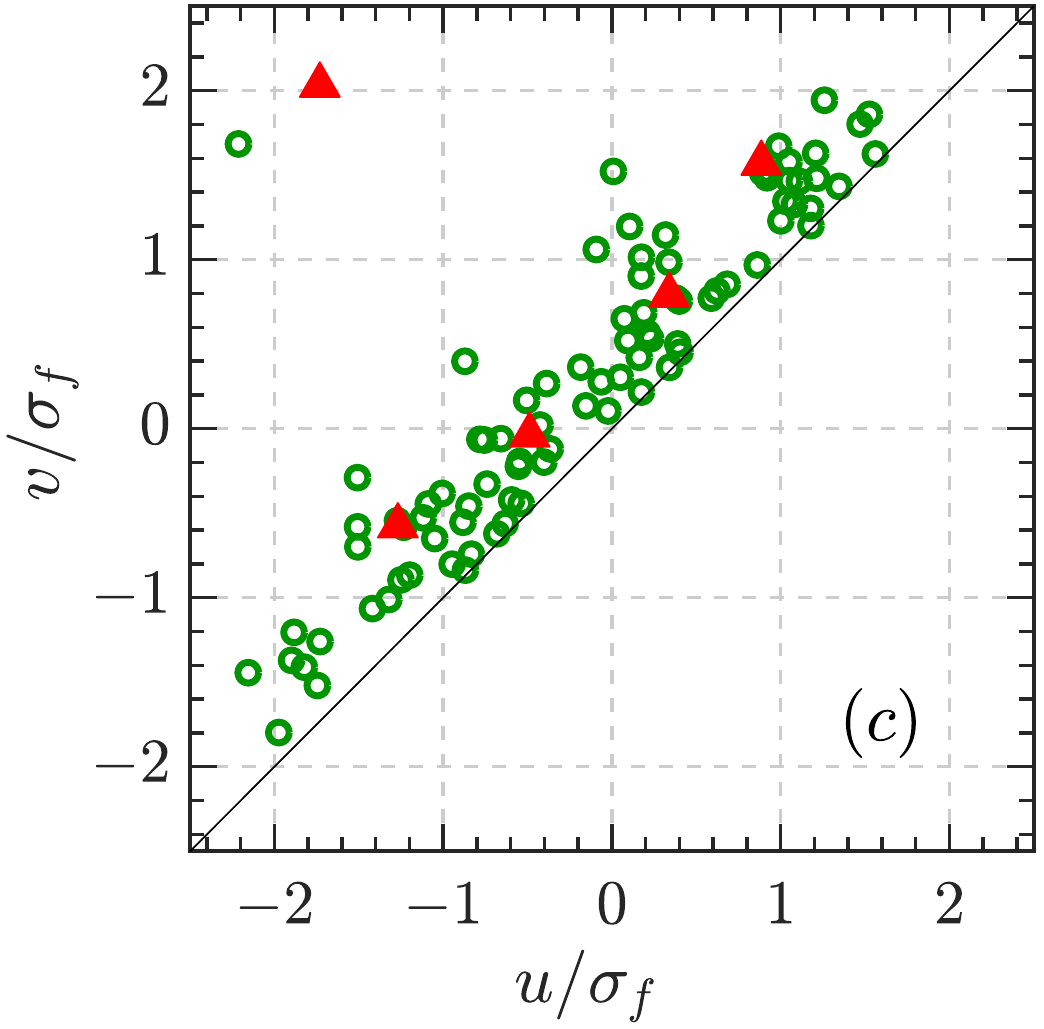}} 
\caption{
The influence of a trend on a persistence diagram. 
(\textit{a}) Solid black line: the 1D signal $f(x)$ of equation~(\ref{fxsin}), 
as shown in Figure~\ref{fig:F7}(\textit{a}), with added Gaussian 
$\delta$-correlated random noise with standard deviation $A = 1$. Dash-dotted 
green line: the Gaussian random noise added with $A = 2.5$. The persistence 
diagrams for the signals shown in the panel (\textit{a}): (\textit{b}) $A=1$, 
(\textit{c}) $A=2.5$. The red triangular markers on the persistence diagrams 
correspond to $f(x)$ in the absence of noise (i.e., $A=0$).
}
\label{fig:trend}
\end{figure}

\begin{figure} 
\includegraphics[width=0.45\textwidth]{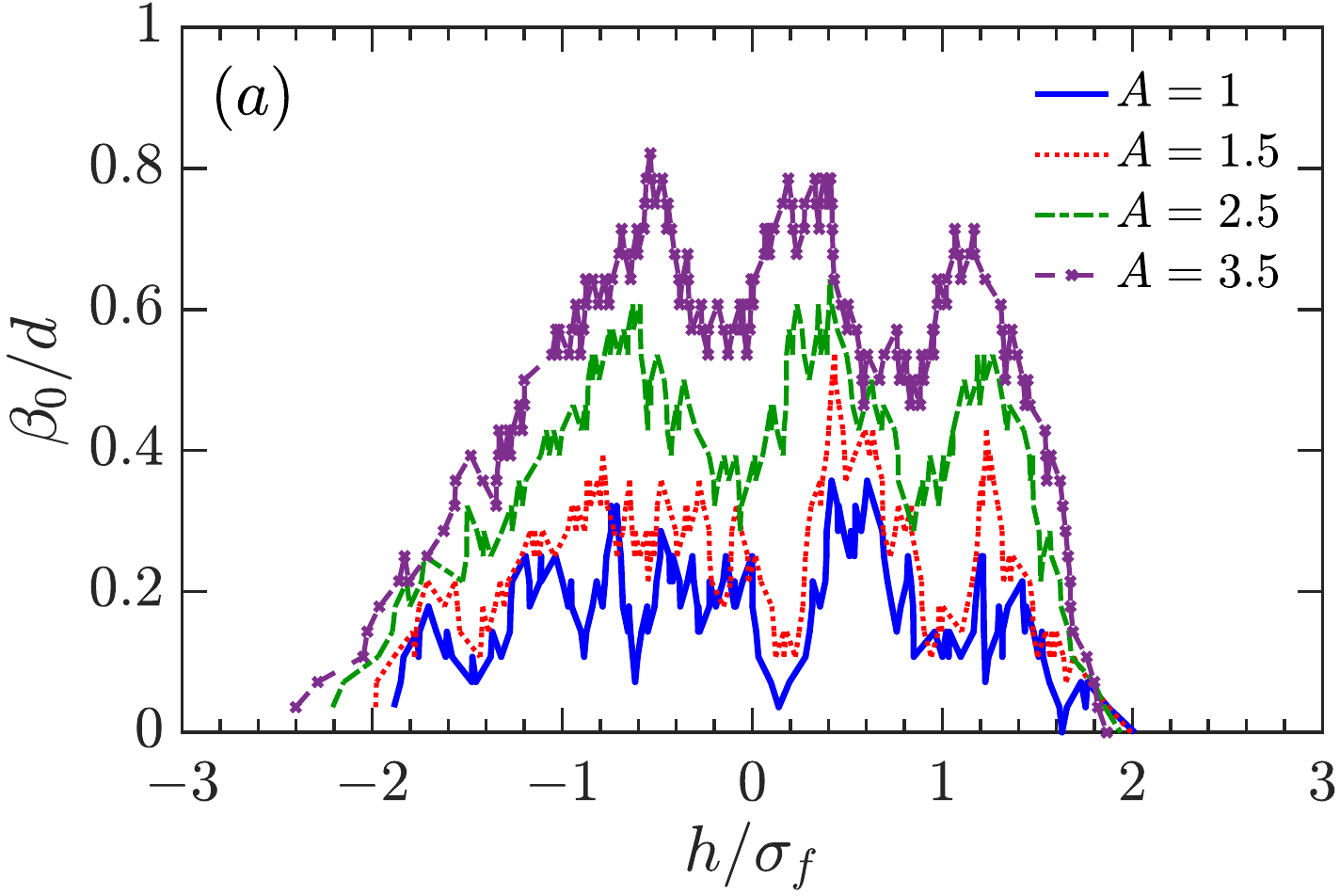}
\hfill
\includegraphics[width=0.45\textwidth]{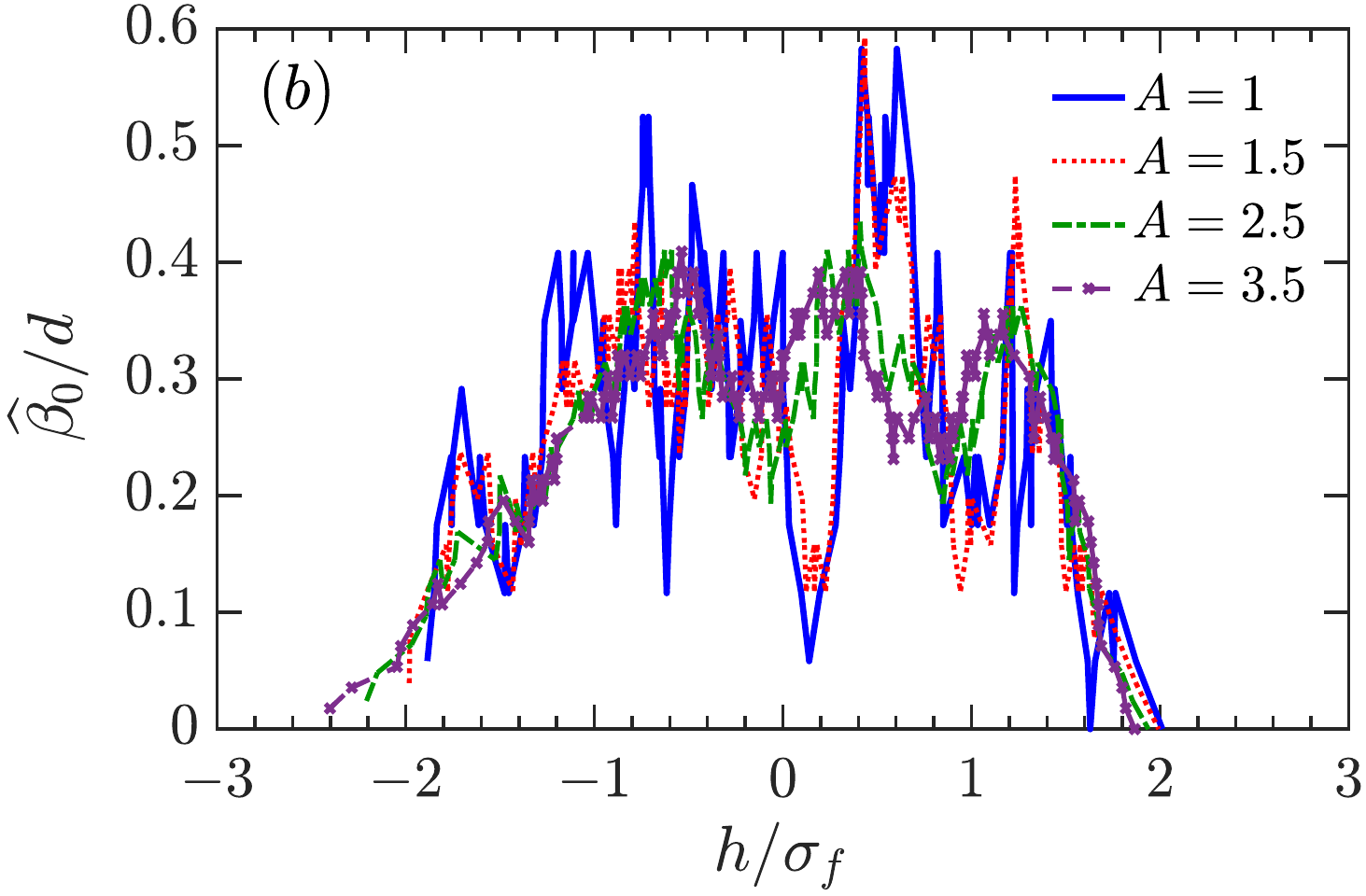}
\vspace{2mm}
\includegraphics[width=0.45\textwidth]{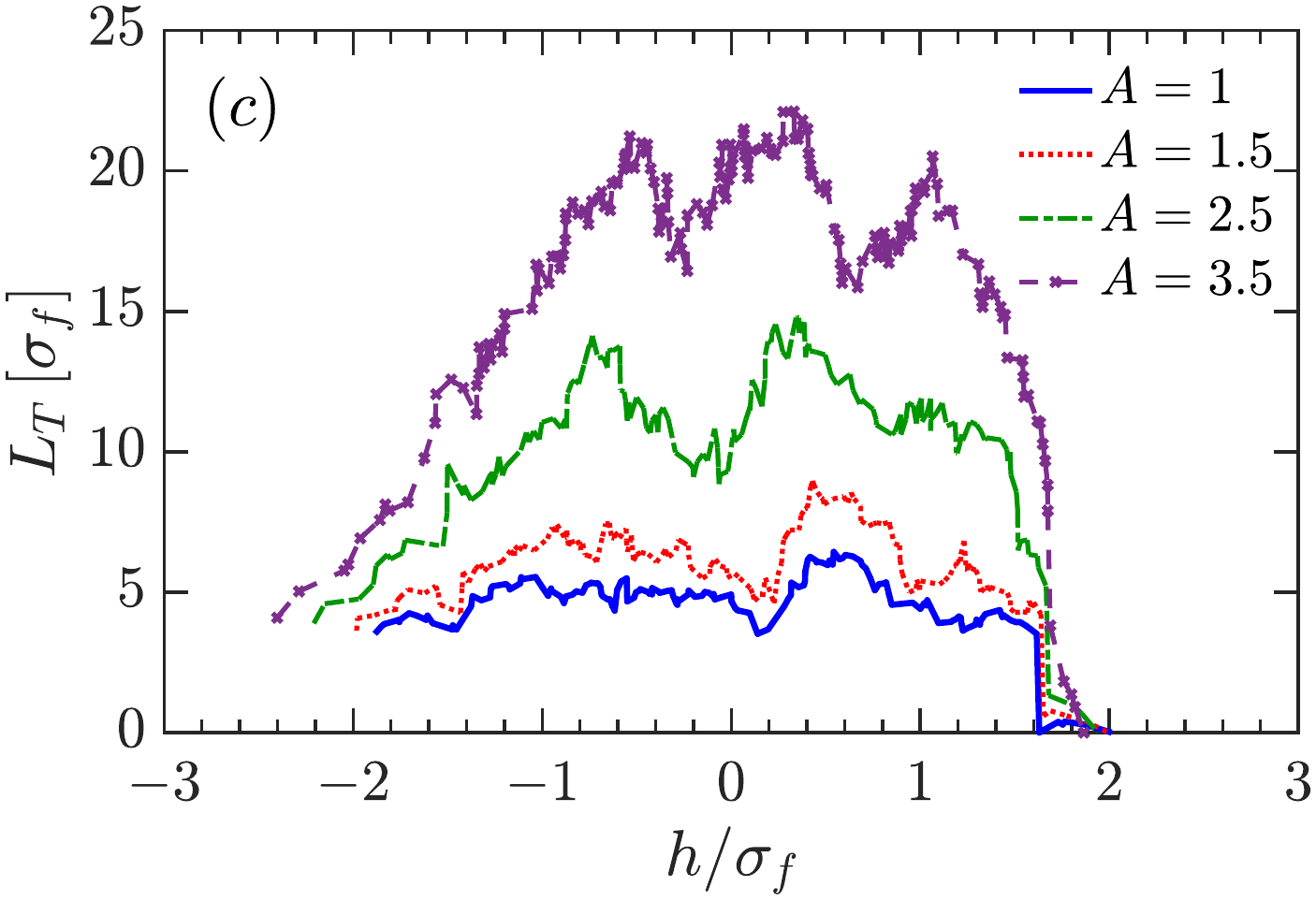}
\hfill
\includegraphics[width=0.45\textwidth]{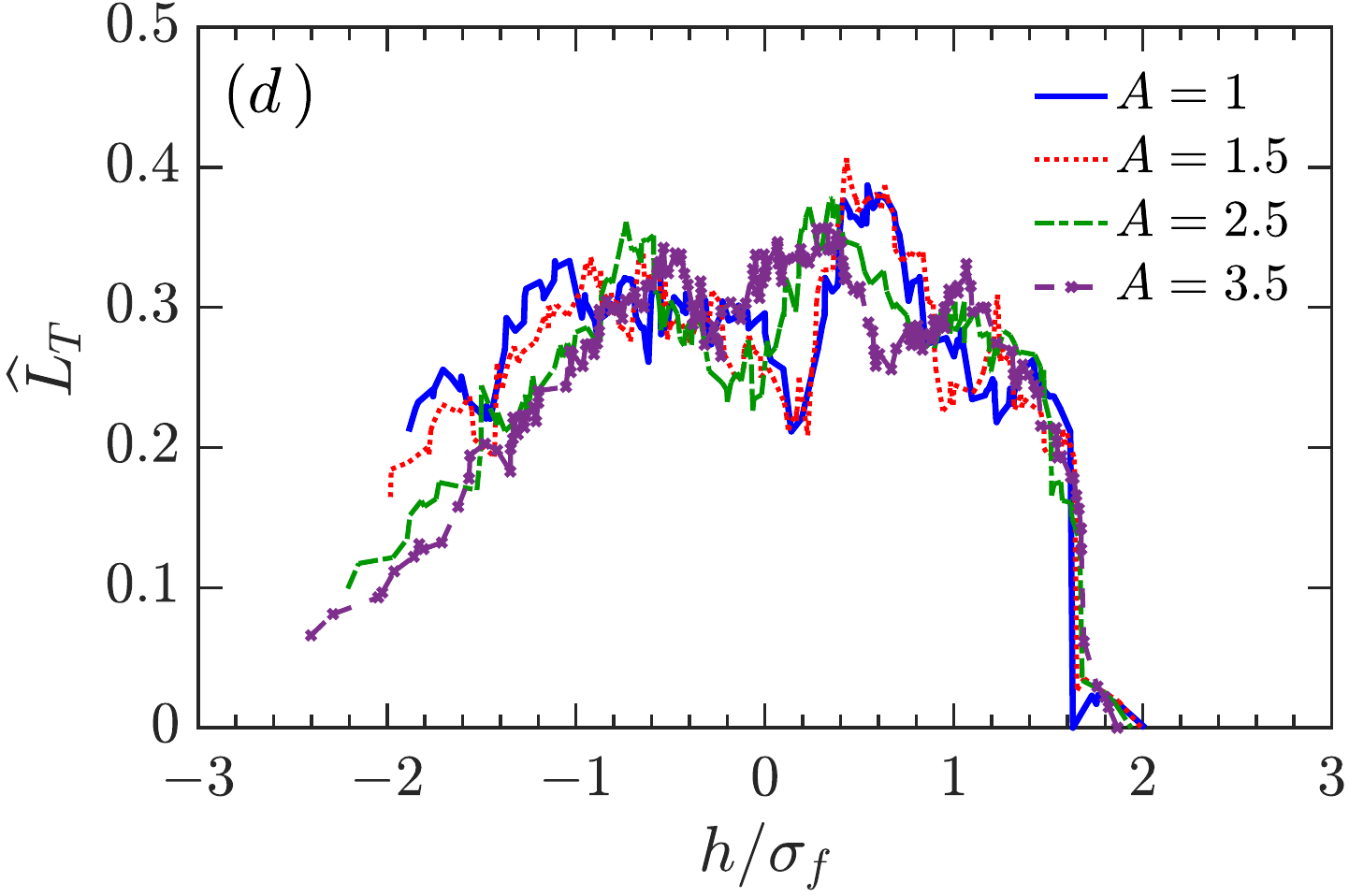}
\caption{Original $(\textit{a, c})$ and normalized  $(\textit{b, d})$ plots 
of $\beta_0/d$ and $L_T$ for function (\ref{fxsin}) with added noise of various 
standard deviations $A$.	
}
\label{fig:trendPDF}
\end{figure}

We may have identified a potentially useful way to compare the topology of 
fluctuations in different data sets without having to first separate the large 
and small-scale signals. This is worth investigating further, with another 
example.

Consider a more complicated case of $f(x)$ given in equation~(\ref{fxsin}), as 
plotted in solid blue line in figure~\ref{fig:F7}(\textit{a}), with added 
$\delta$-correlated Gaussian random noise with the zero mean and increasing 
standard deviations of $A = 1$, $1.5$, $2.5$ and $3.5$, shown in 
figure~\ref{fig:linTrend}(\textit{a}).

The resulting signals are shown in figure~\ref{fig:trend}(\textit{a}) by the 
solid black line for $A = 1$ and dash-dotted green line for $A = 2.5$, and 
panels (\textit{b}) and (\textit{c}) show the persistence diagrams for them.
Red triangles in the persistence diagram are for the trend $f$. The 
fluctuations produce a cloud of points in the persistence diagrams aligned with 
the diagonal. At the low noise amplitude, these components die fast and are 
separated from those arising due to the trend, which are more persistent. 
Increasing the amplitude of the fluctuations spreads this cloud out, mixing the 
trend and the fluctuations. The interpretation of the persistence diagram is no 
longer so clear.

Figure~\ref{fig:trendPDF} shows the original and normalised plots of 
$\beta_0/d$ and $L_T$ for the function (\ref{fxsin}) with all four choices of 
added noise, $A= 1, 1.5, 2.5, 3.5$. The normalised versions are not sensitive 
to the magnitude of $A$.

\begin{figure} 
\centerline{\includegraphics[width=0.75\textwidth]{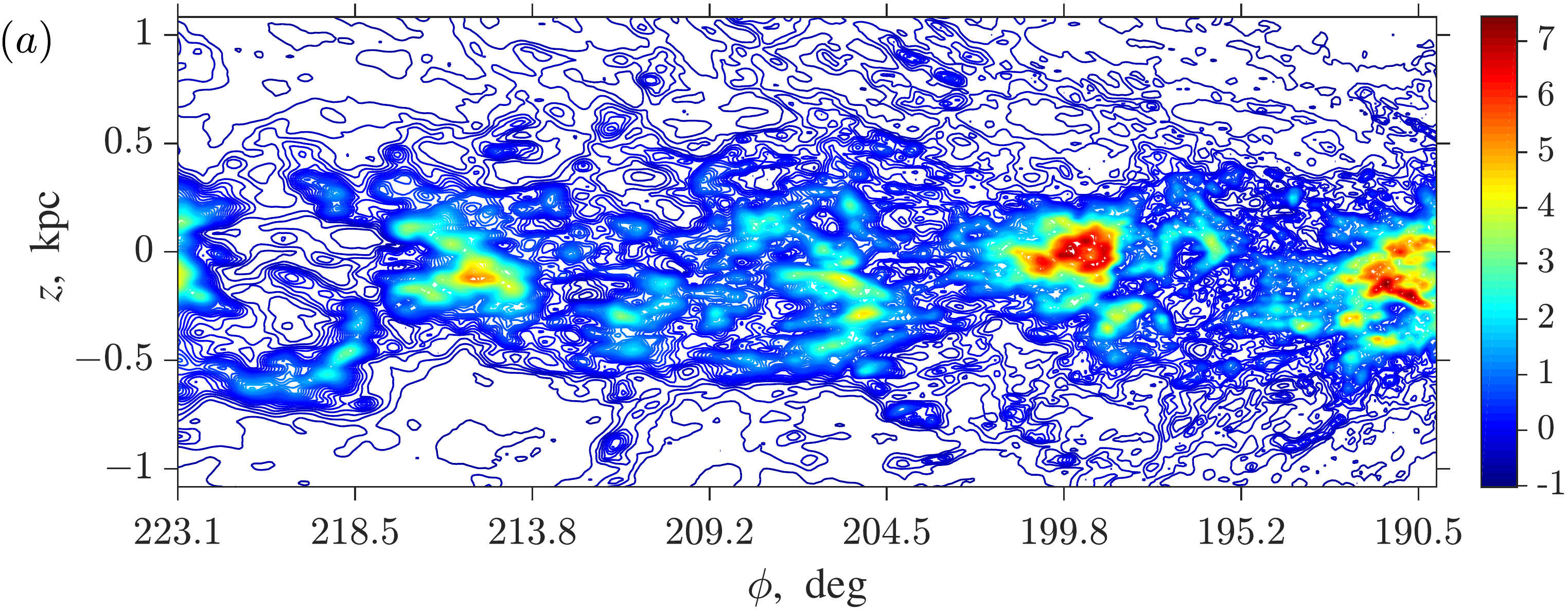}}
\centerline{\includegraphics[width=0.75\textwidth]{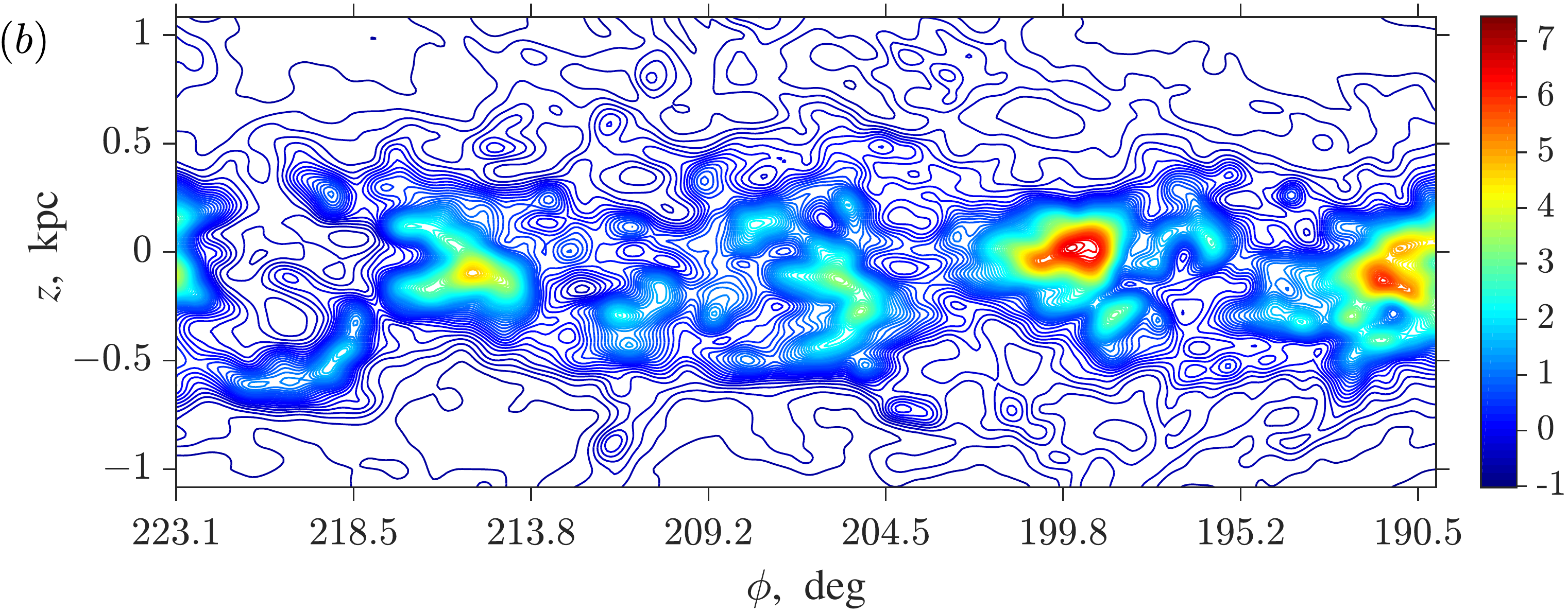}}
\centerline{\includegraphics[width=0.75\textwidth]{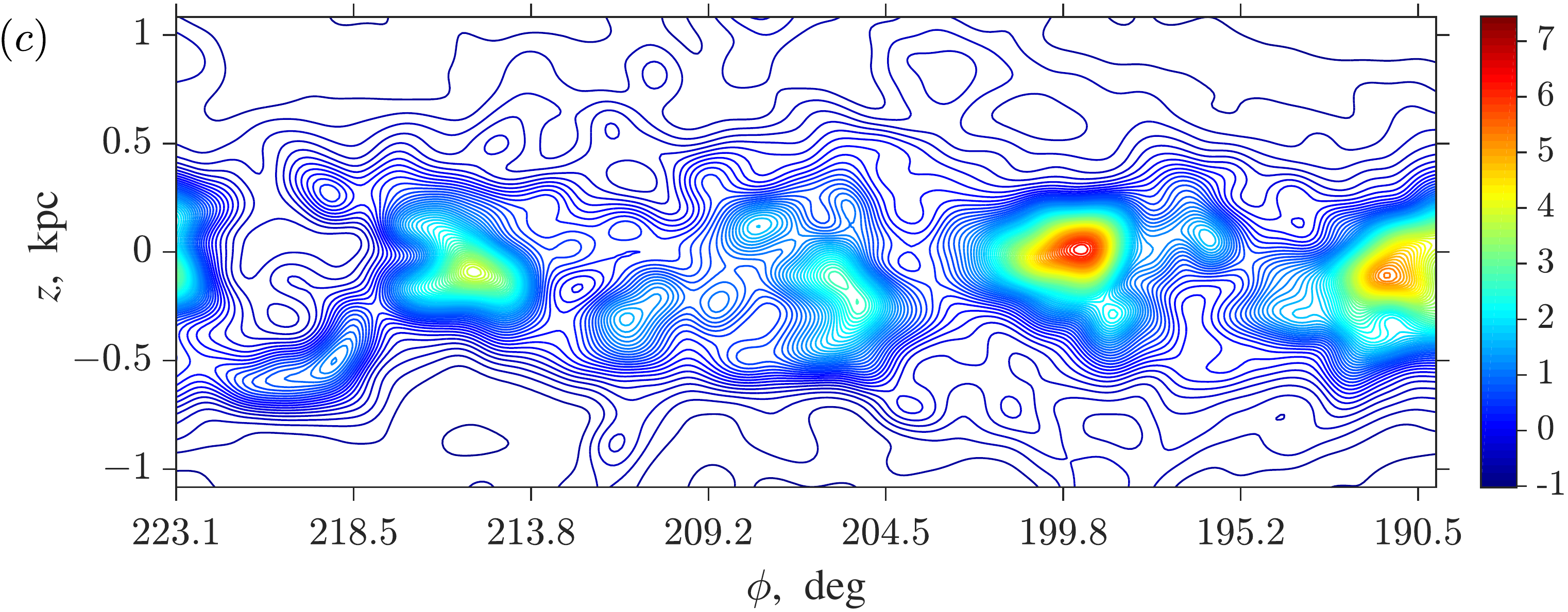}}
\centerline{\includegraphics[width=0.75\textwidth]{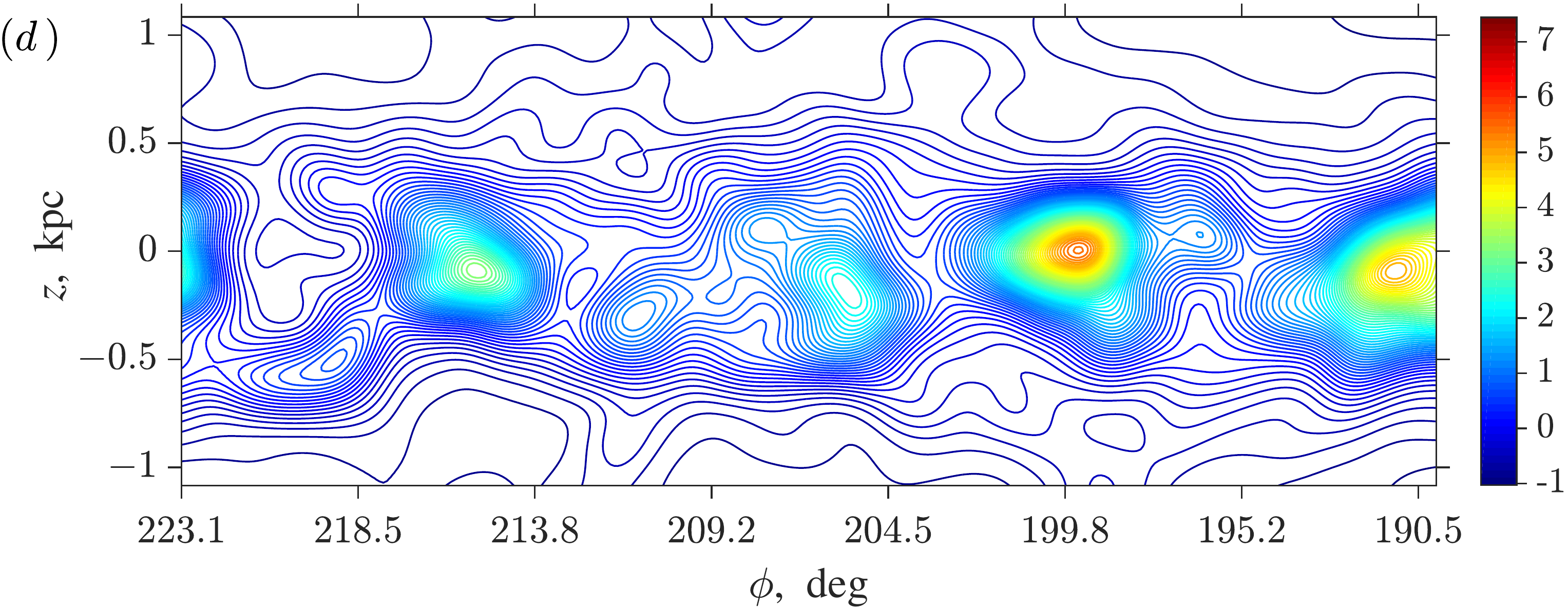}}
\caption{
(\textit{a}) Gaussian smoothing of H\,{\sc i} number density in the Milky Way
from \citet{GASS10} at $R = 10\kpc$. Panel (\textit{a}) represent a 
cross-section of the gas density distribution, $n(\phi, z)$. Here the field is 
shown after standardization, so that negative $n$ values occur. The results of 
blurring this image by a Gaussian function with a kernel width $l = 2$, $4$ and 
$6$ are shown in panels (\textit{b}), (\textit{c}) and (\textit{d}), 
respectively. We applied the Gaussian smoothing to the original 
(non-standardized) field, and then standardized the obtained 2D matrix. The 
colour bars are identical in all panels, the gas density values are in 
$\sigma_n$ units, the number of contours on each map equals to 80. 
}
\label{fig:gs1}
\end{figure}

\begin{figure} 
\centerline{
\includegraphics[width=0.45\textwidth]{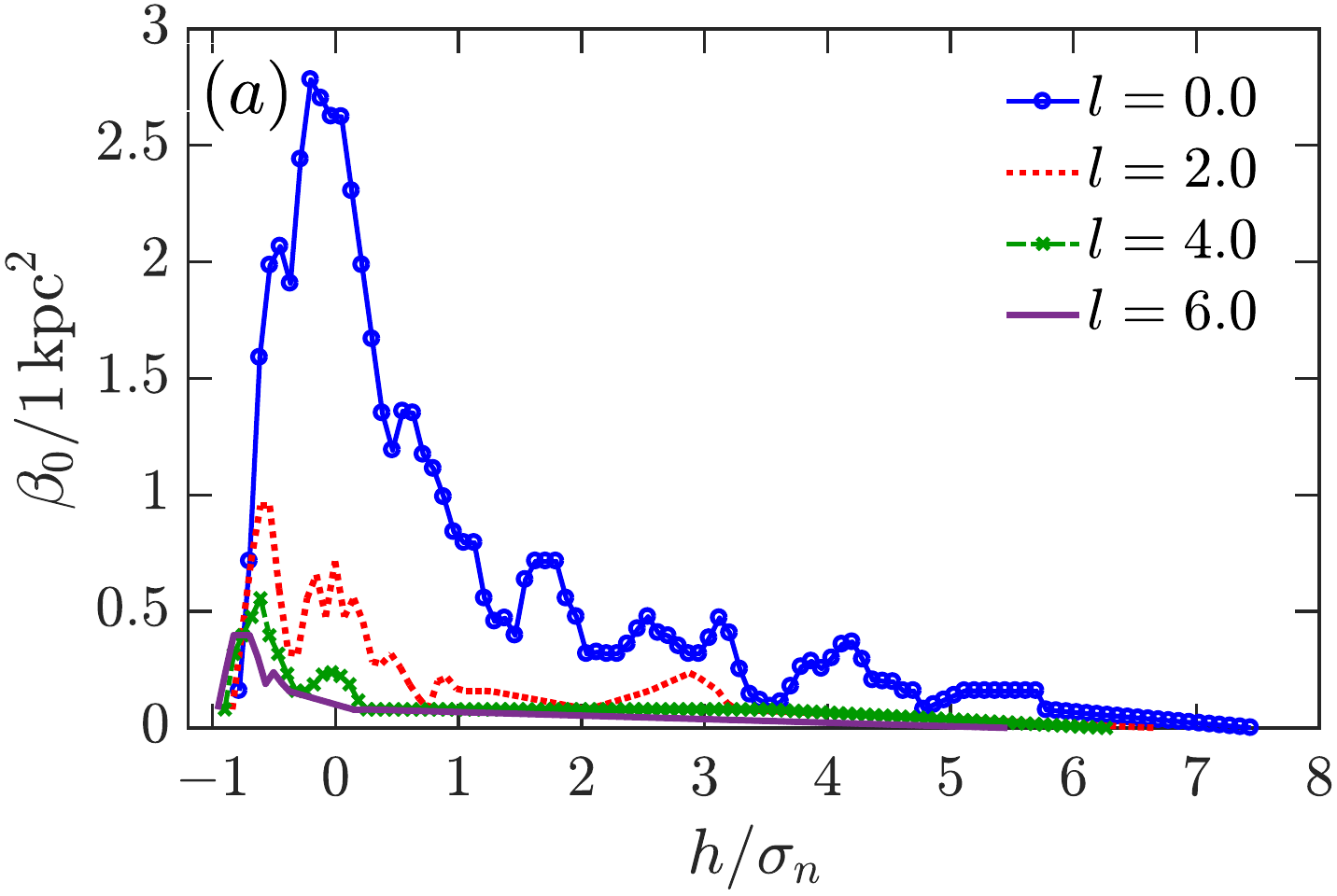} \hspace{2mm} 
\includegraphics[width=0.45\textwidth]{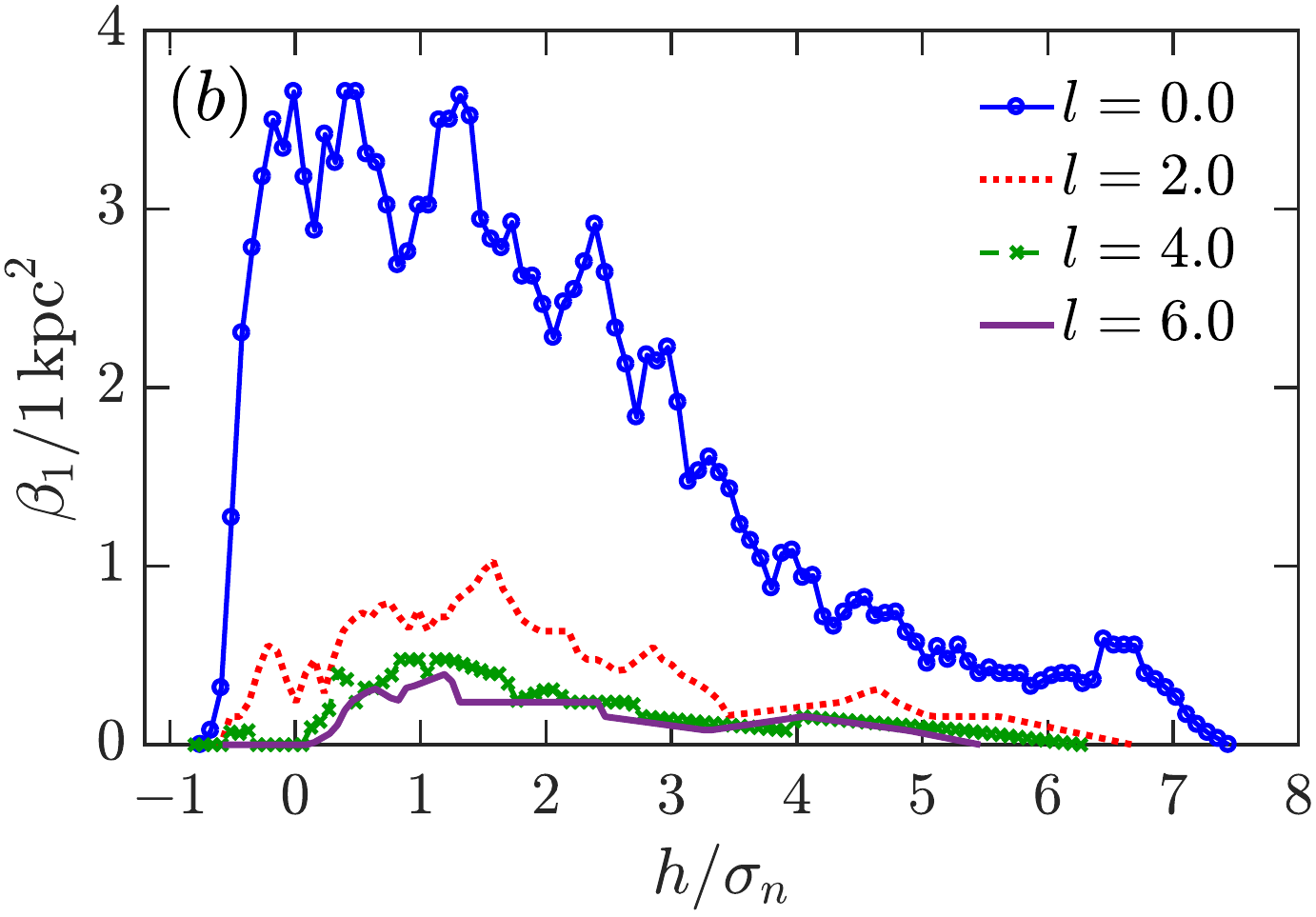}}  
\vspace{2mm}
\centerline{
\includegraphics[width=0.45\textwidth]{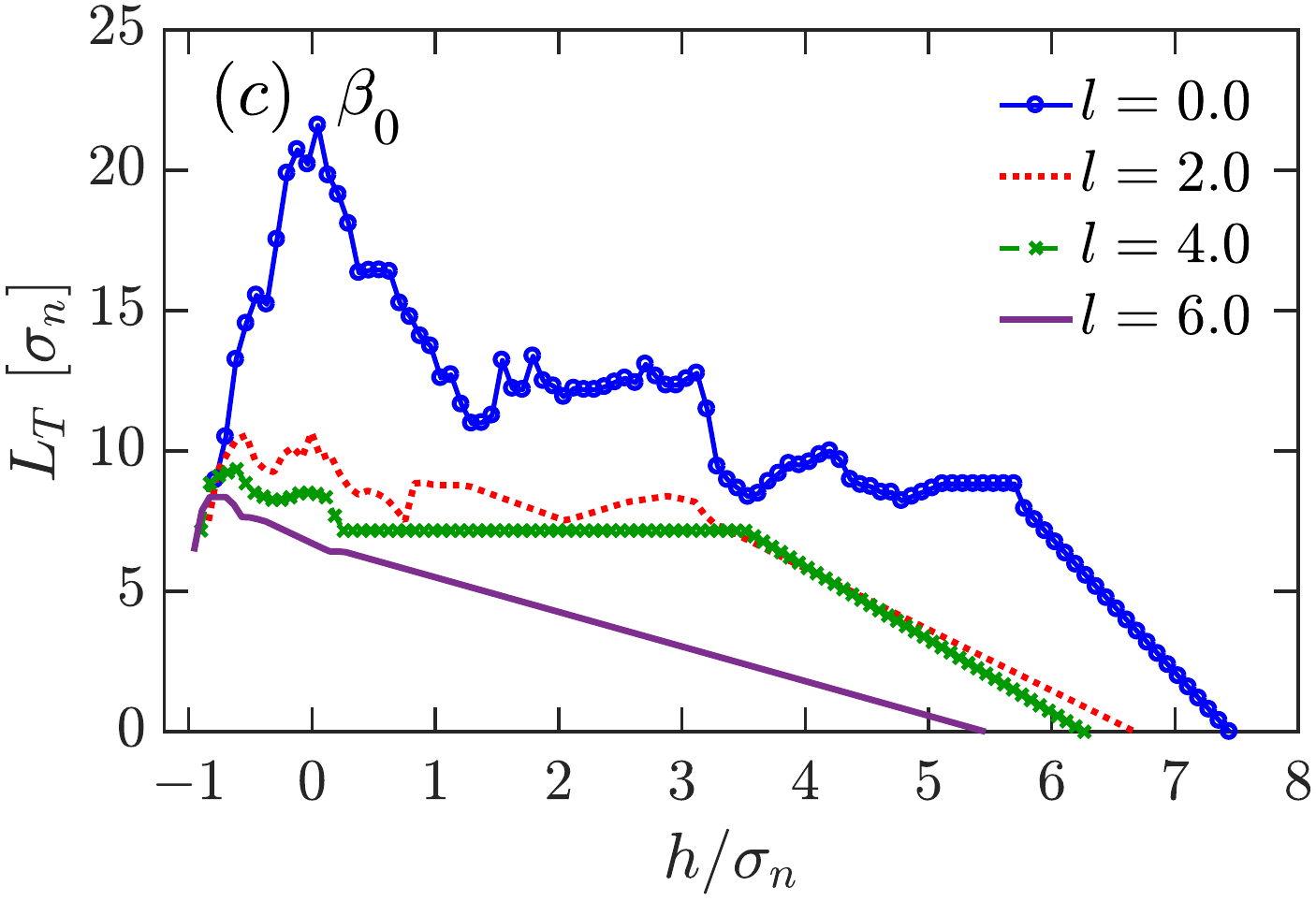} \hspace{2mm} 
\includegraphics[width=0.45\textwidth]{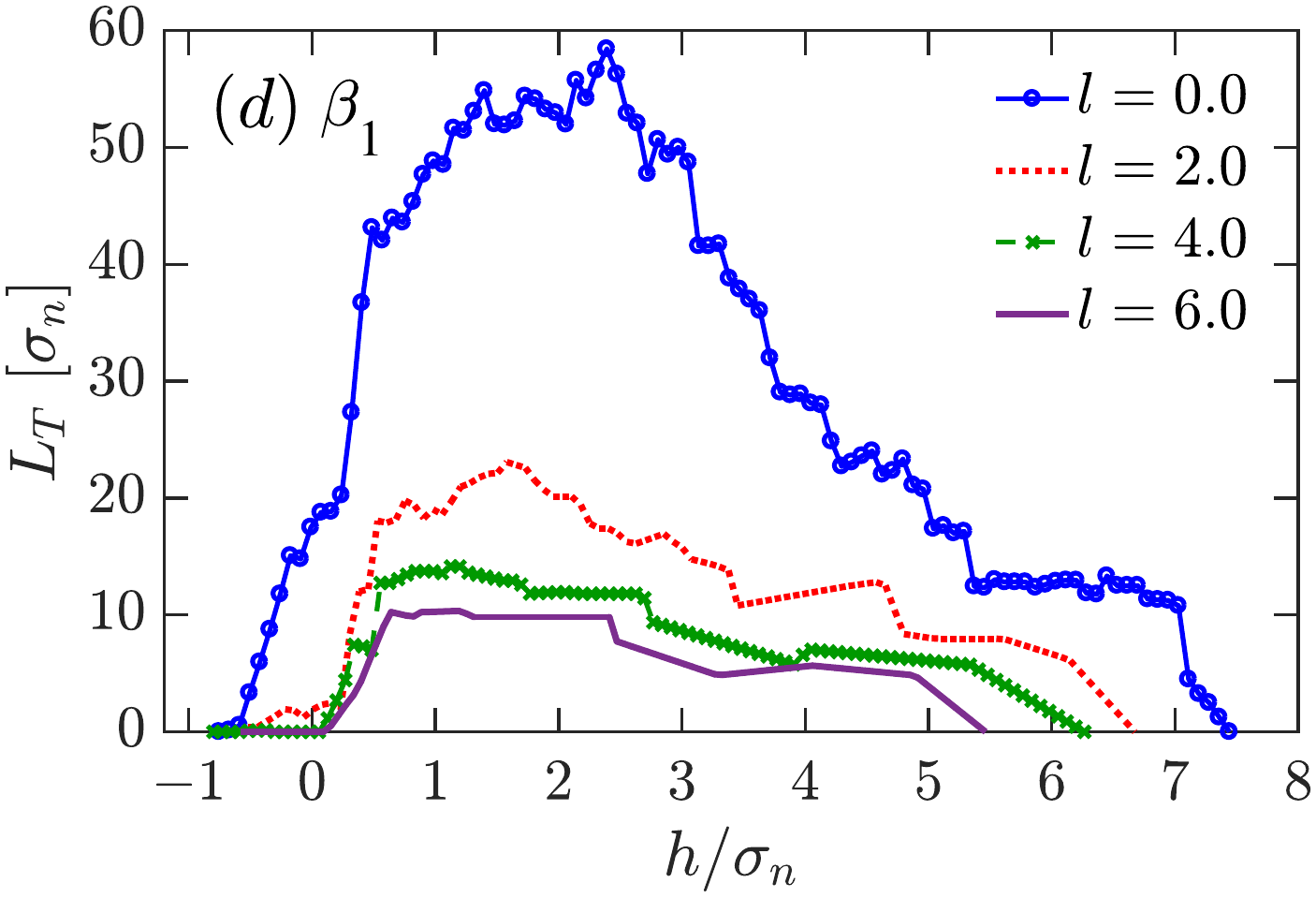}}             
\caption{Topological characteristics of the observed H\,\textsc{i} number 
density $n$ of figure~\ref{fig:gs1} under Gaussian smoothing. The dependence of 
the Betti numbers per unit area, (\textit{a})~$\beta_0$ and 
(\textit{b})~$\beta_1$ versus the of isocontour level $h$ (given in the units 
of $\sigma_n$, the standard deviation of $n$). The total length of barcodes (in 
the units of $\sigma_n$) is shown as a function of $h$ for 
(\textit{c})~$\beta_0$ and (\textit{d}) $\beta_1$. Solid blue lines are for the 
density field at the original resolution, the other lines represent fields 
smoothed using the Gaussian kernel (\ref{nGn}) with $l = 2$, 4 and 6.} 
\label{fig:gs2}
\end{figure}
\begin{figure} 
\centerline{
\includegraphics[width=0.45\textwidth]{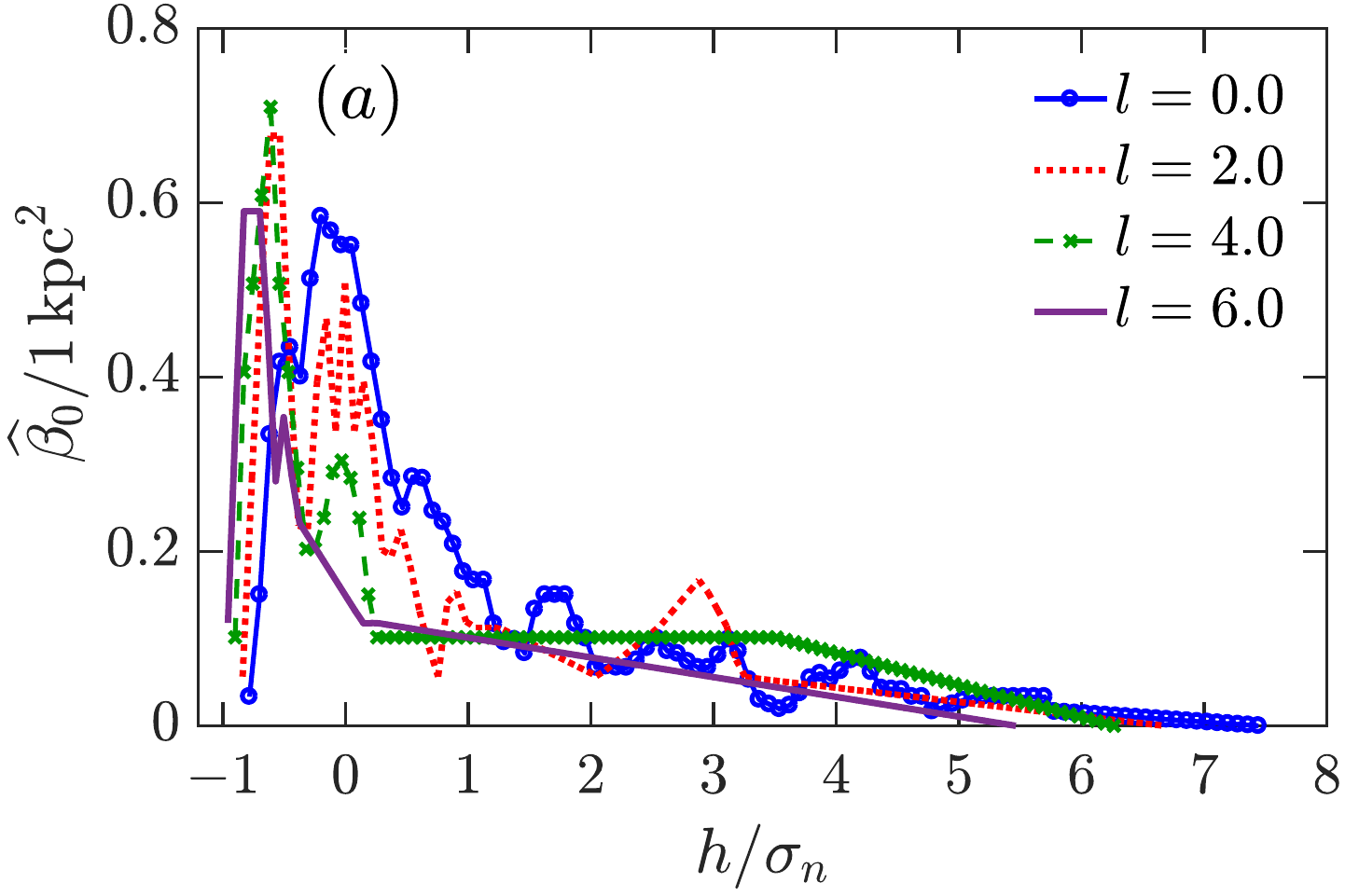} \hspace{2mm} 
\includegraphics[width=0.45\textwidth]{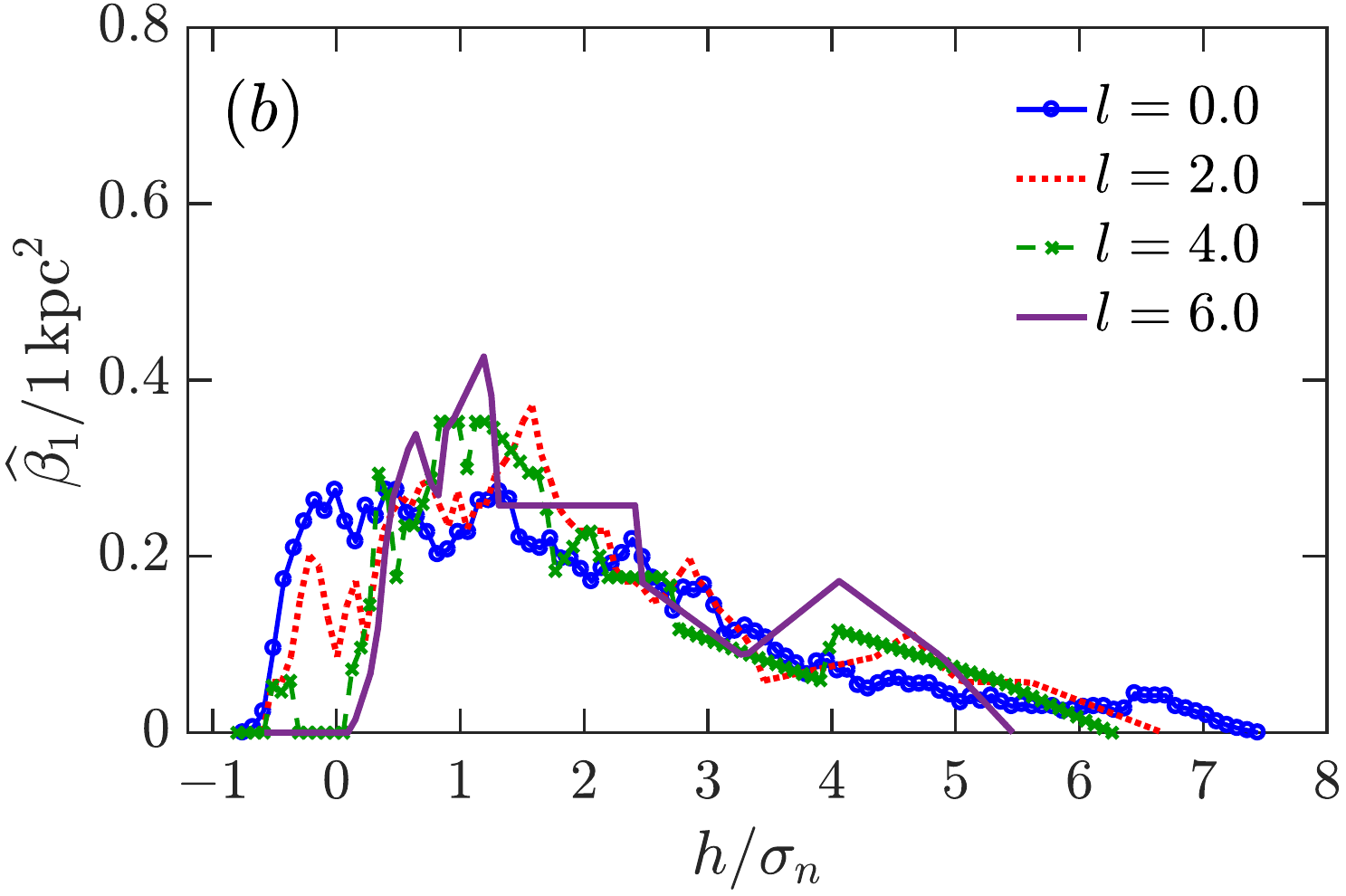}}             
\centerline{
\includegraphics[width=0.45\textwidth]{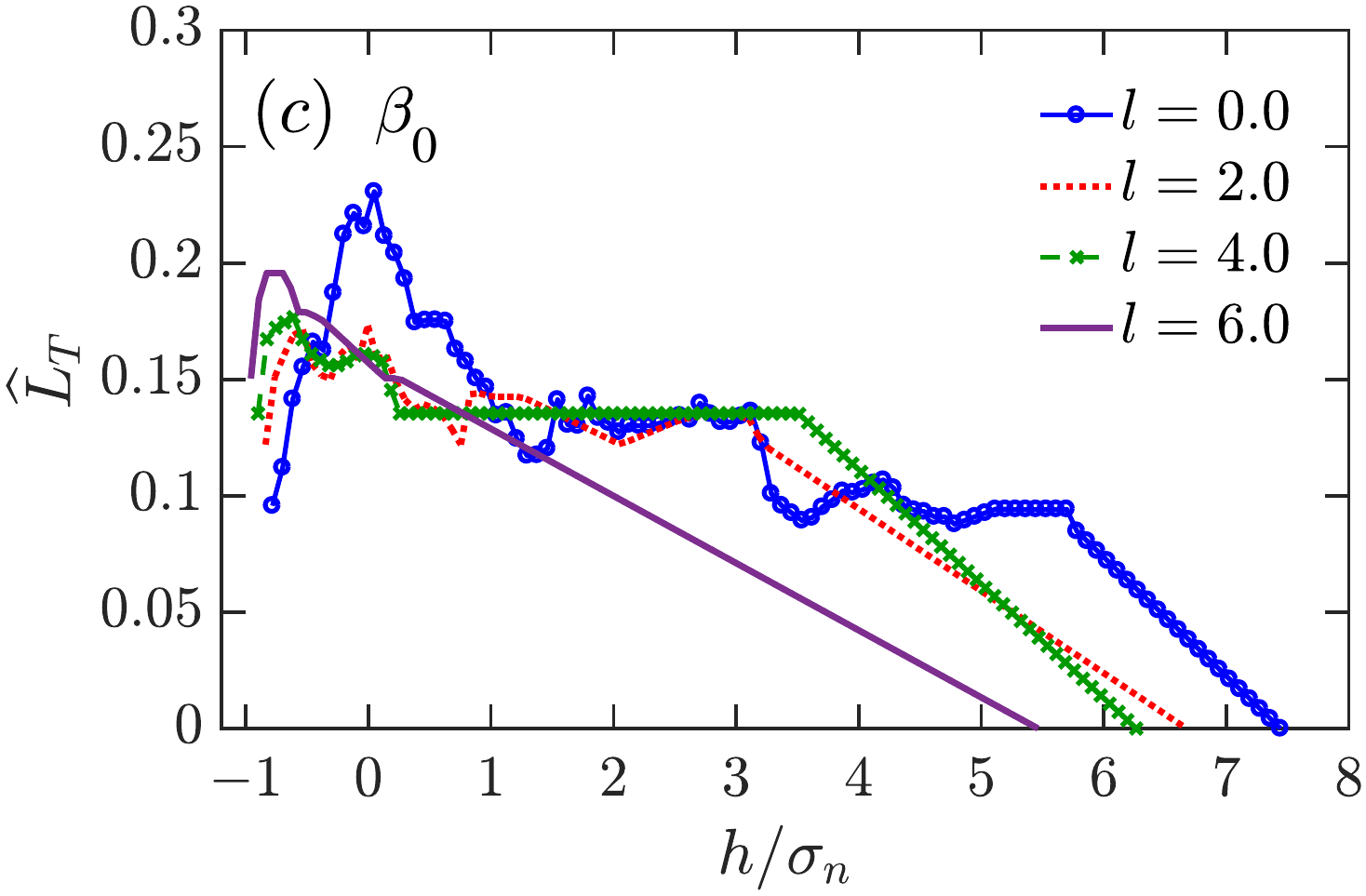} \hspace{2mm}  
\includegraphics[width=0.45\textwidth]{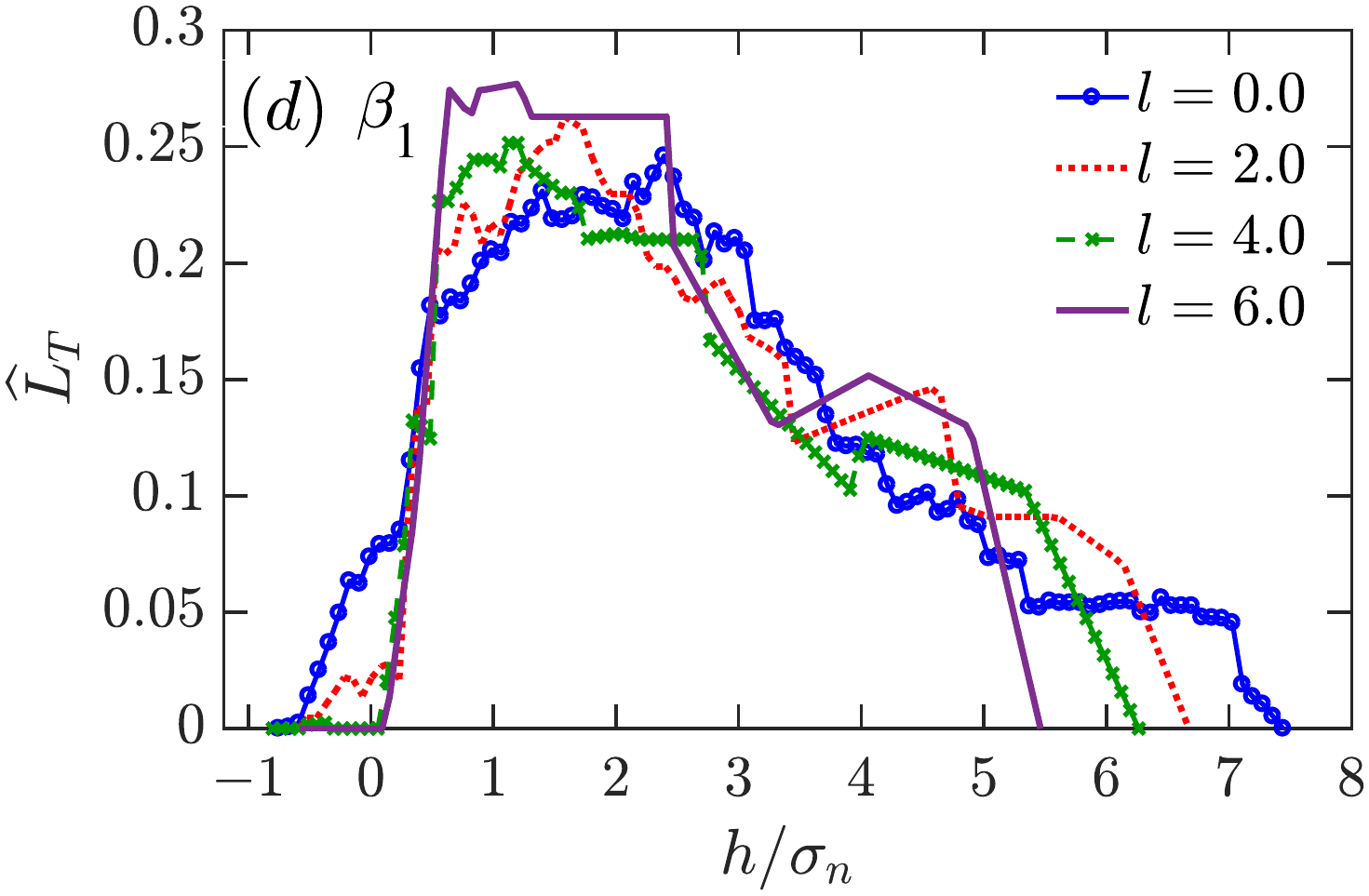}}              
\caption{As figure~\ref{fig:gs2}, but with all curves normalized to unit 
underlying area.} 
\label{fig:gs3}
\end{figure}

\section{Topological characteristics and the image resolution}
\label{sec:res}
In this section we will show that topological statistics, introduced above, 
remain stable when the image resolution decreases.

Figure~\ref{fig:gs1}(\textit{a}) shows an original observation of the hydrogen 
gas density distribution, $n(\phi, z)$,  in a region of the Milky Way, from the 
same data as figure~\ref{fig:f1}({\it a}), after data standardization using 
equation~(\ref{eq:stand}). The image is of 131 pixels height by 349 pixels 
wide. The horizontal extent of the region shown is about $5.77\kpc$, so the 
pixel width is about $17\pc$. The resolution of the data was coarsened by 
convolving the image with a 2D Gaussian filter,
\begin{equation}\label{nGn}
\langle n \rangle(\vect{x}) = \frac{1}{2\upi l^2} \int n(\vect{x}')
\text{e}^{-(\vect{x}-\vect{x}')^2/(2l^2)}\,\text{d}^2 \vect{x}'\,.
\end{equation} 
We varied the width of the Gaussian kernel as $l = 2, 4, 6$ pixels and the 
resulting smoothed images are shown in 
figure~\ref{fig:gs1}(\textit{b}--\textit{d}). The smoothed images were 
standardized and a topological filtration applied. Degrading the image 
resolution using bi-cubic interpolation, where the output pixel value is a 
weighted average of pixels in the $4\times 4$ neighbourhood, has the same 
effect as Gaussian smoothing.

The topological filtration of a 2D image is quantified in terms of two 
topological variables, the number of components $\beta_0$ and the number of 
holes $\beta_1$, both functions of the threshold level $h$. For a statistically 
stationary random field in 2D, the magnitudes of the Betti numbers are 
proportional to the area of the image and they are therefore usefully presented 
per unit area. The normalization of the Betti numbers per correlation cell 
(area or volume) was suggested by \citet{Makar2018MNRAS} to be more physically 
informative and sensitive to the topological properties of the random fields 
rather than its spatial scale. Here we follow the simple approach, calculating 
the Betti numbers per $1\kpc^2$.

The Betti numbers and the total length of the barcodes obtained from filtering 
the gas density fields of figure~\ref{fig:gs1} are shown in 
figure~\ref{fig:gs2}. It is clear that smoothing with $l=4\text{--}6$ leads to 
a significant loss of detail that affects the topological structure of the 
data, especially for $\beta_0$. It is understandable that smoothing reduces the 
numbers of both components  and holes (i.e., $\beta_0$ and $\beta_1$), and 
contracts the barcodes. However, further normalization of the dependencies 
$\beta_0(h)$, $\beta_1(h)$ and $L_T(h)$ alleviates the problem of image 
resolution. Figure~\ref{fig:gs3} presents the same results as 
figure~\ref{fig:gs2} but normalized to unit area under each curve. Thus, images 
of different resolutions can be compared directly provided both the observable 
quantities and the results of their topological filtration have been properly
standardized.  

Thus, topological filtration helps to identify a range of resolutions where 
statistical properties of the random field remain stable, extending this beyond 
second-order moments or power spectra. For the specific field of the 
interstellar gas number density, smoothing affects $\beta_0$ more than 
$\beta_1$. For example, $\beta_1(h)$ still contains some details for $l=4$ 
whereas $\beta_0(h)$ loses them for $h\gtrsim0.5\sigma_n$. This suggests that 
$n(\vect{x})$ has narrow minima and broad maxima, so that the minima are lost 
first as the resolution is being reduced. 

\section{Comparing the observed and simulated gas density fields}
\label{sec:app}

\begin{figure} 
\centerline{\includegraphics[width=0.95\textwidth]{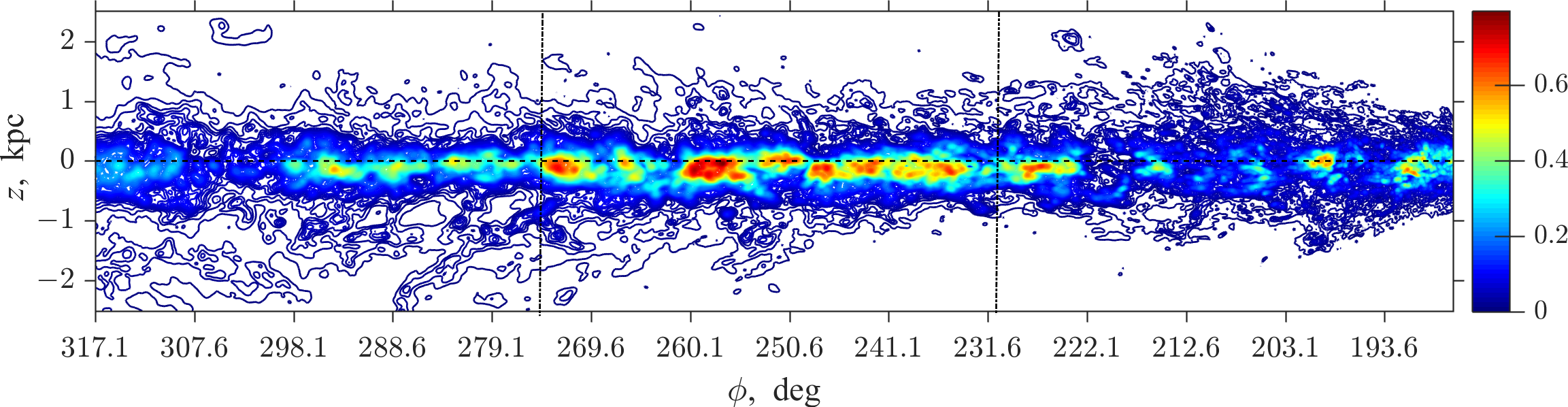}} \vspace{2mm}
\centerline{\includegraphics[width=0.95\textwidth]{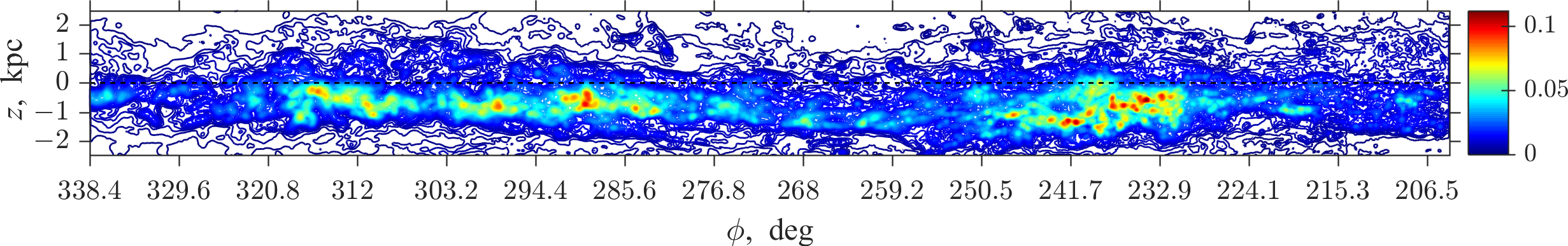}}
\caption{The H\,{\sc i} number density $n\,(\mathrm{atom\, cm}^{-3})$ from the 
GASS survey  \citep{GASS10}  in $5\kpc$-wide  strips along the Galactic 
mid-plane ($|z|\leq2.5\kpc$)  at $R = 10\kpc$ (upper panel) and $20\kpc$ 
(bottom panel). The vertical lines in the upper panel separate three regions 
where gas density apparently gas distinct properties and which are analysed 
separately.} 
\label{fig:ot1}
\end{figure}

\begin{figure} 
\centerline{
\includegraphics[width=0.45\textwidth]{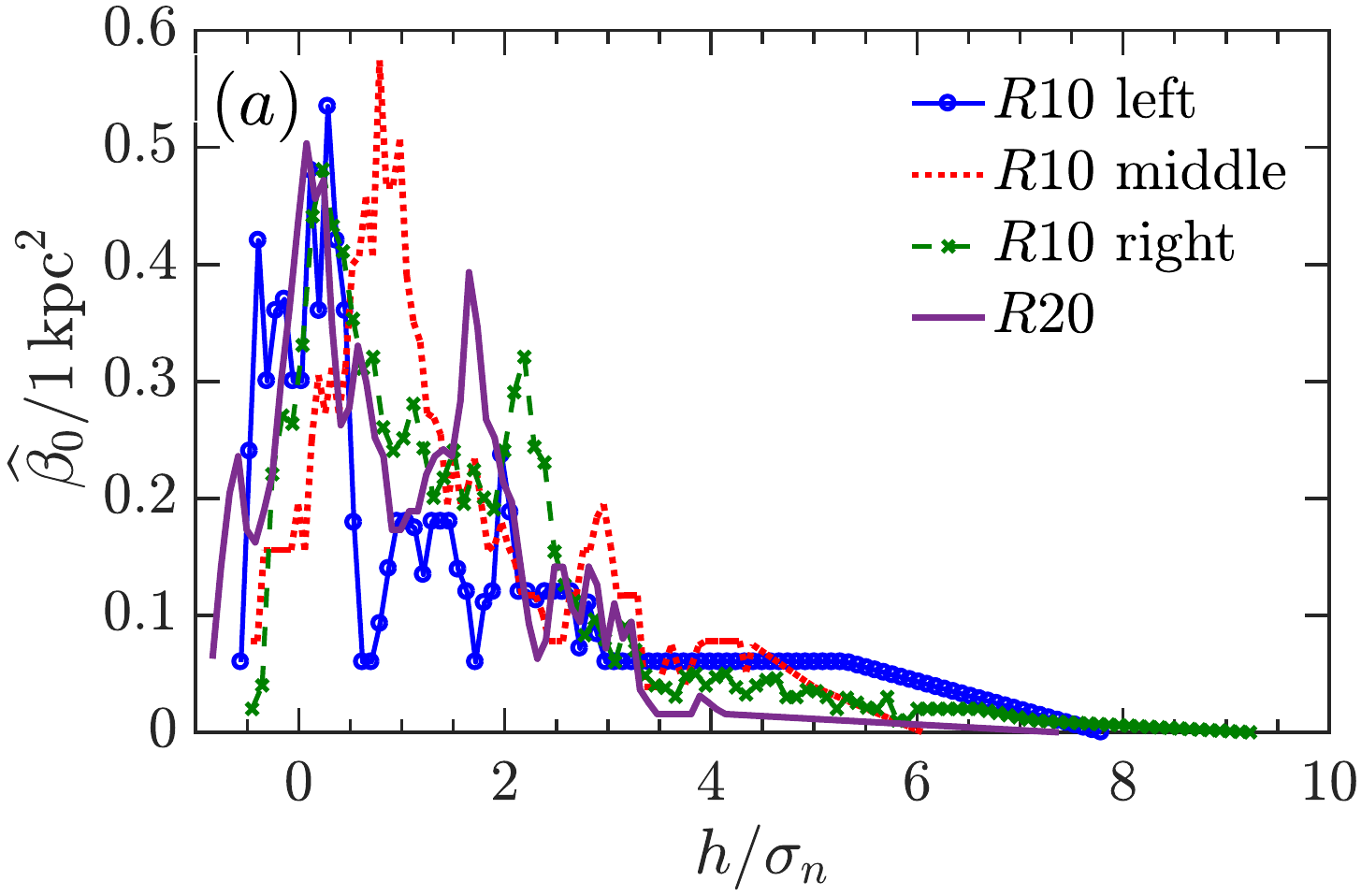} \hspace{2mm} 
\includegraphics[width=0.45\textwidth]{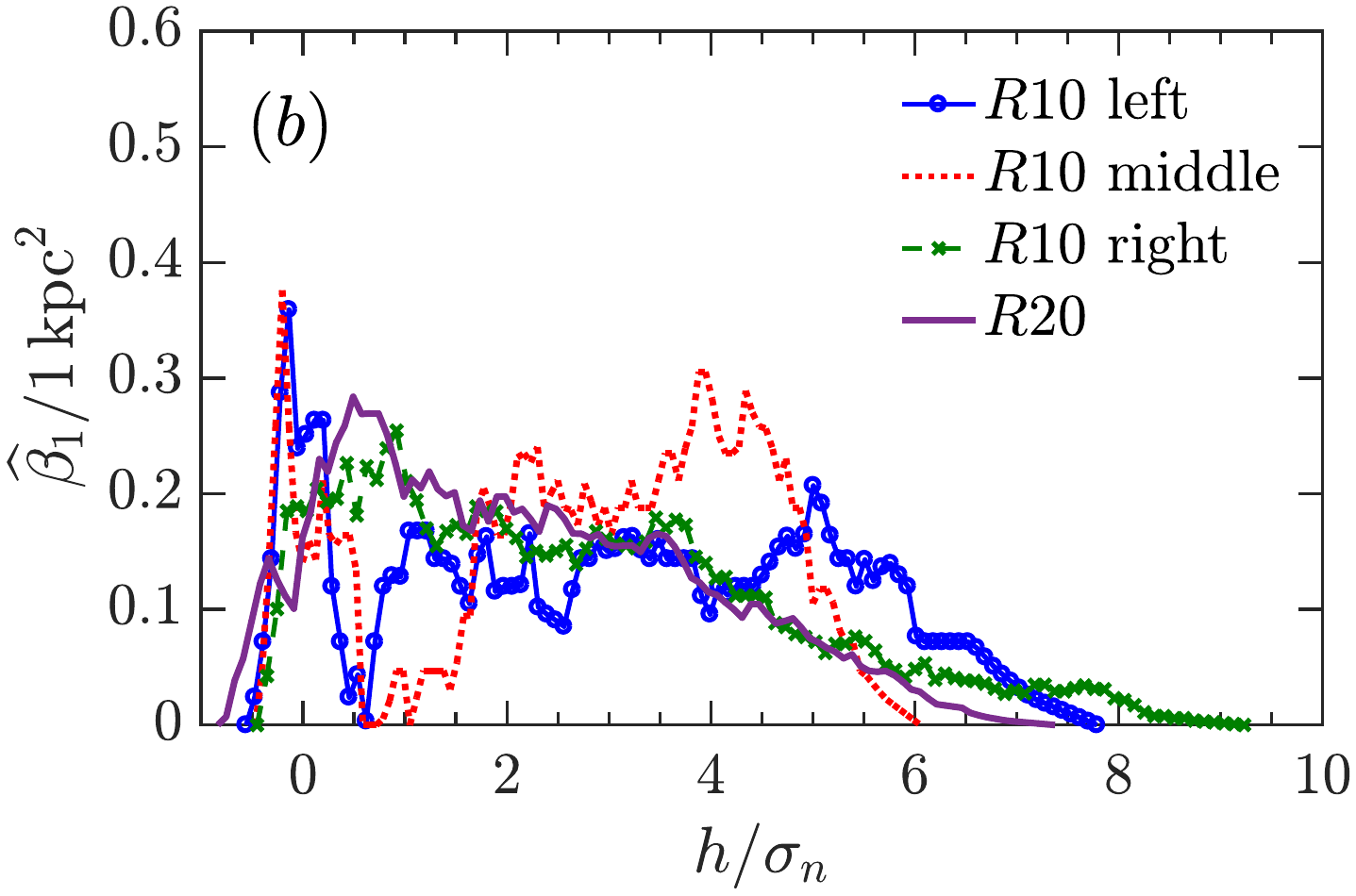}}				
\centerline{
\includegraphics[width=0.45\textwidth]{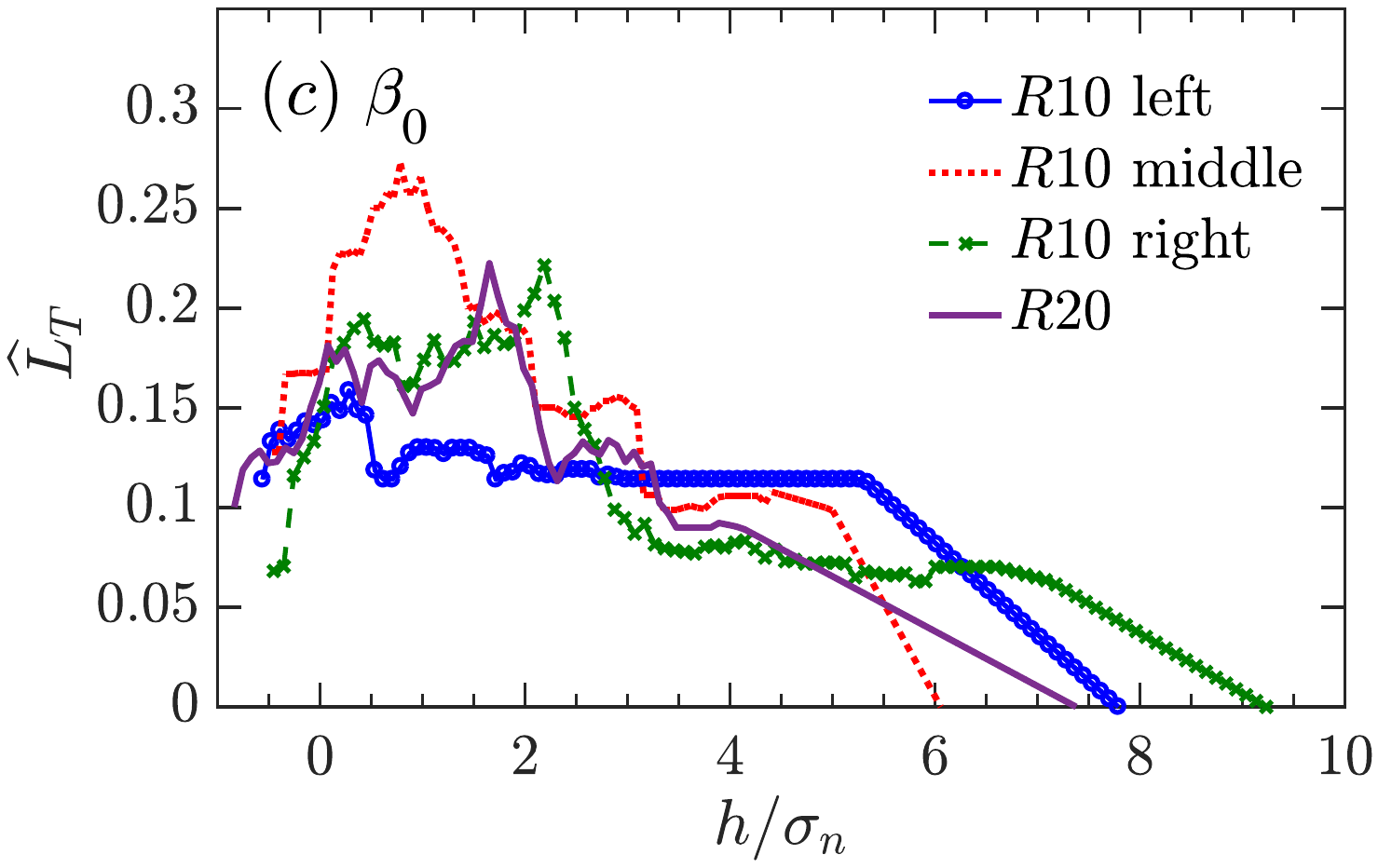} \hspace{2mm} 
\includegraphics[width=0.45\textwidth]{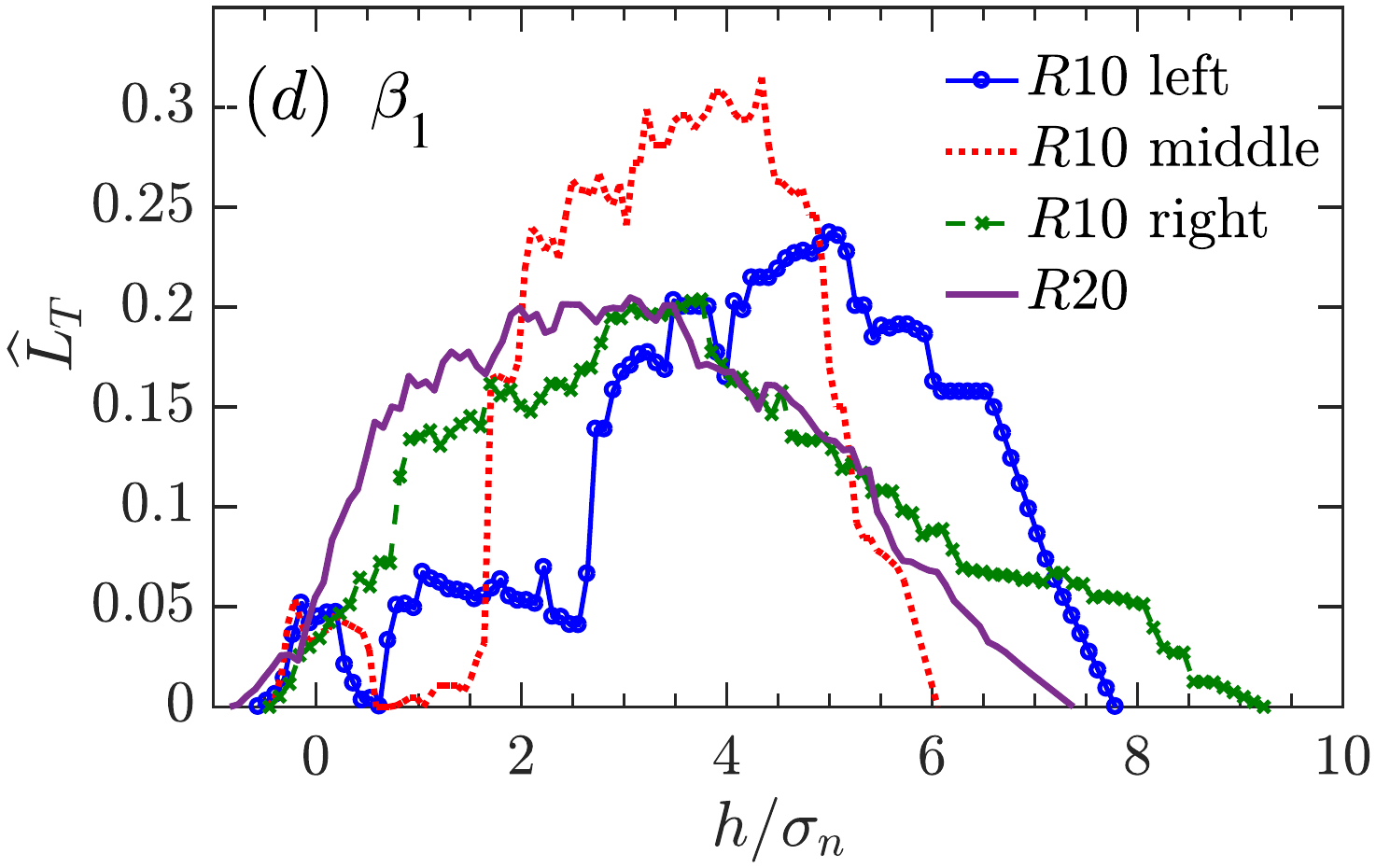}}			
\caption{Results of topological analysis of the H\,{\sc i} number density in 
four regions in the Milky Way (as specified in the legends) shown in 
figure~\ref{fig:ot1} and described in the text. (\textit{a}) and (\textit{b}): 
The Betti numbers per unit area, $\beta_0$ and  $\beta_1$, respectively. 
(\textit{c})  and (\textit{d}): Total barcode lengths for $\beta_0$ $\beta_1$, 
respectively. All variables are presented as functions of the level height $h$ 
in the units of the standard deviation of the density variations $\sigma_n$. 
All curves are normalized to the unit underlying area. } 
\label{fig:ot4}
\end{figure}

\begin{figure} 
\centerline{
\includegraphics[width=0.45\textwidth]{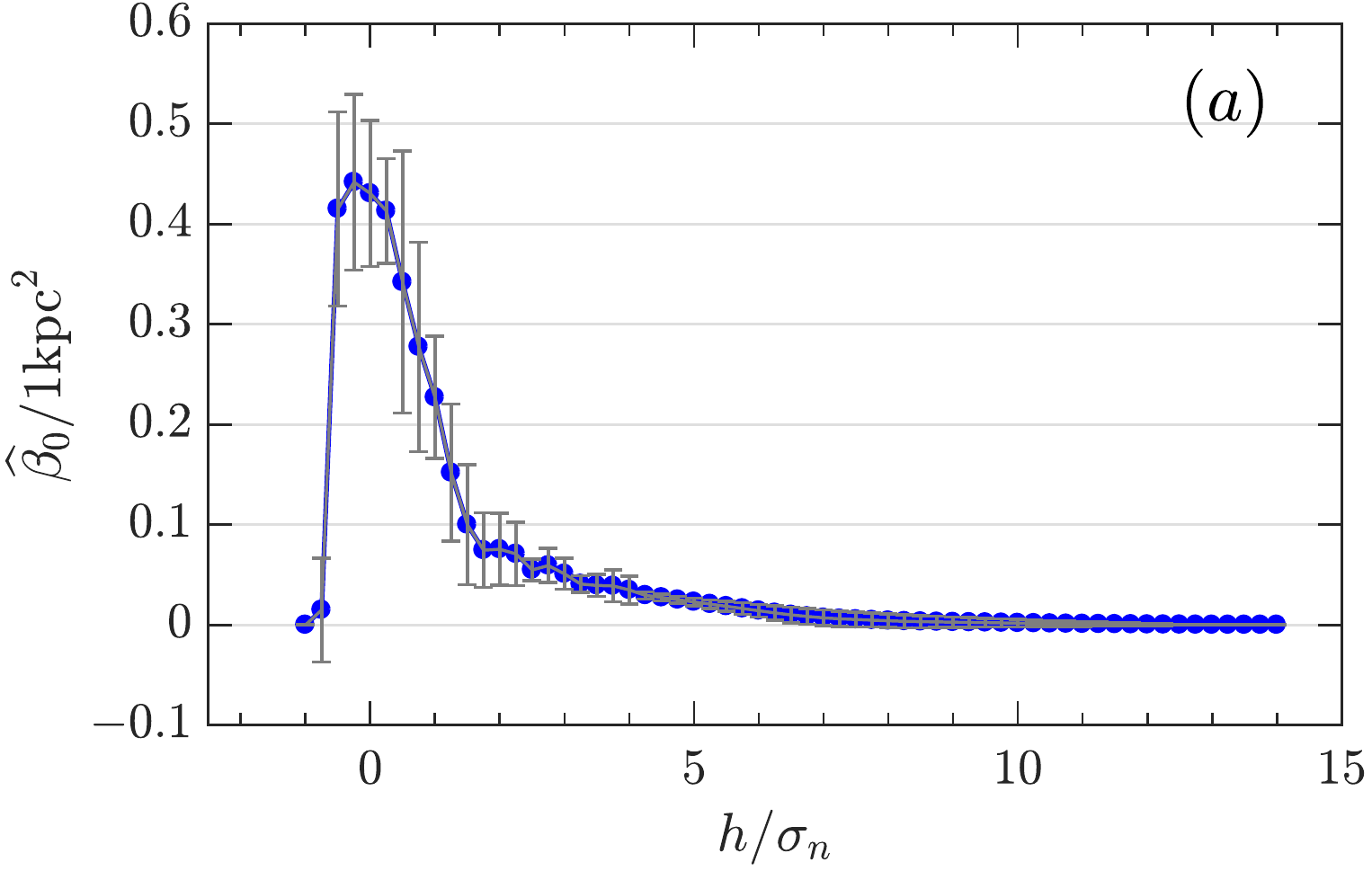} \hspace{2mm}
\includegraphics[width=0.45\textwidth]{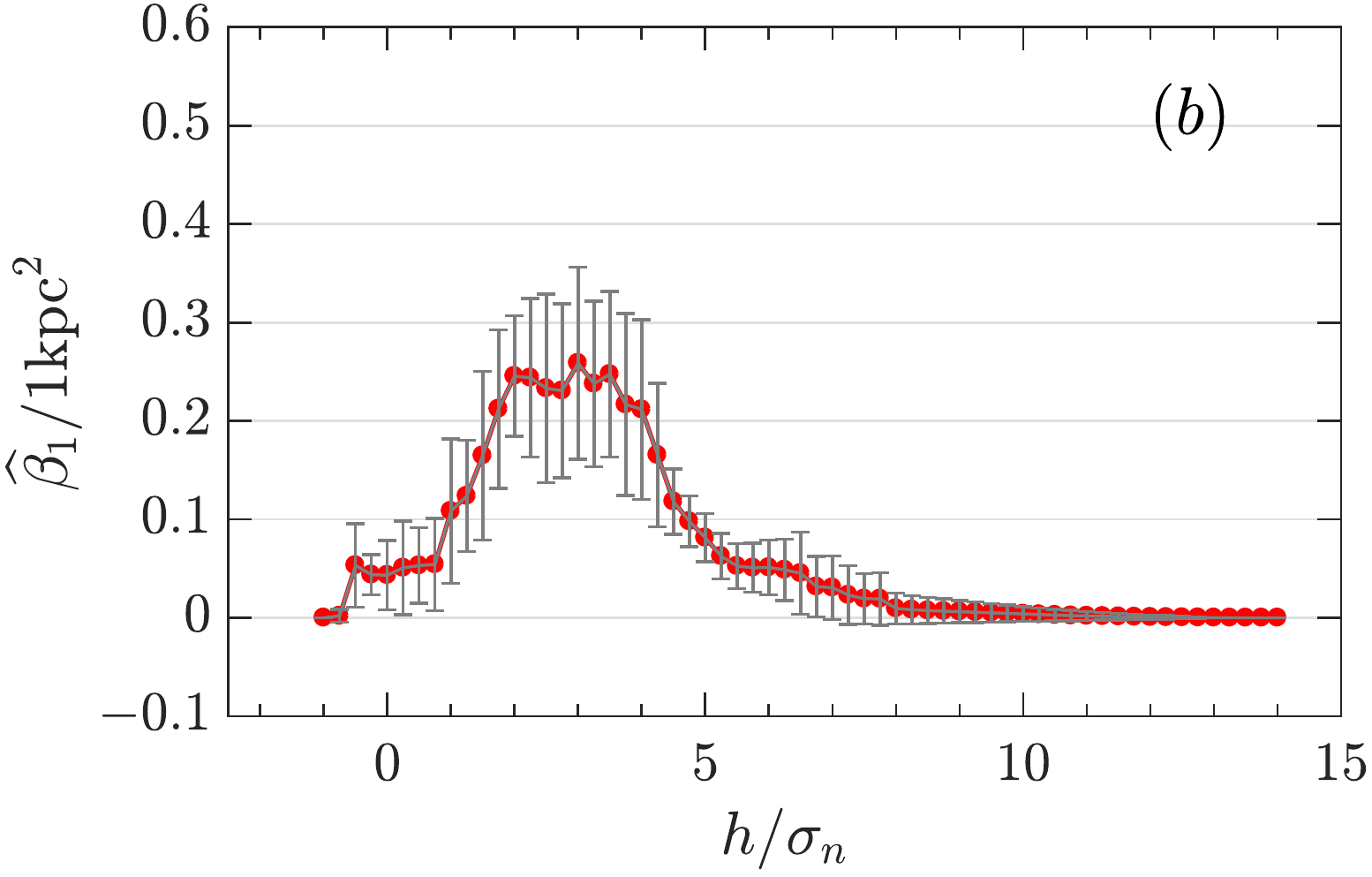}}
\vspace{2mm}
\centerline{
\includegraphics[width=0.45\textwidth]{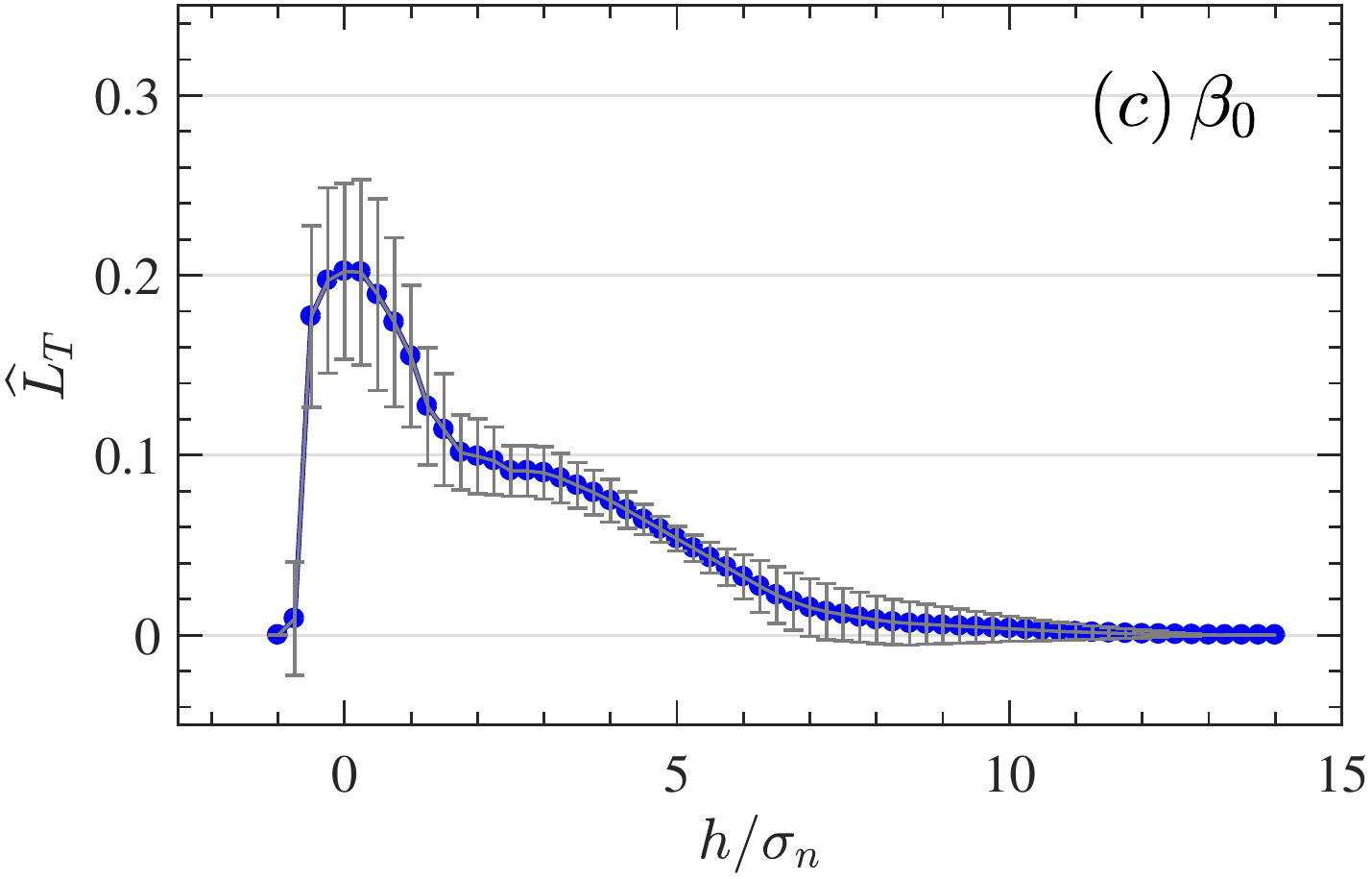} \hspace{2mm}
\includegraphics[width=0.45\textwidth]{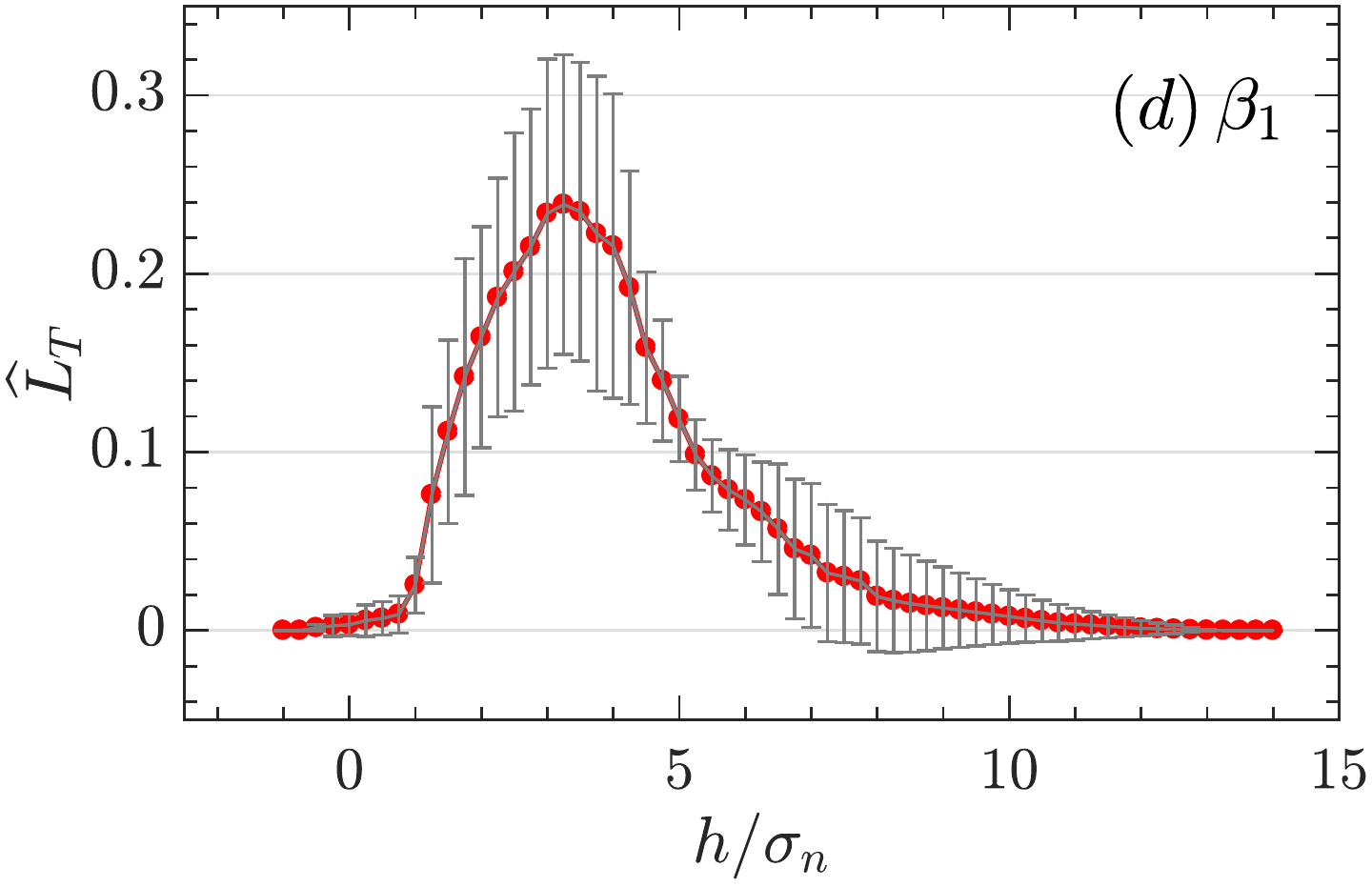}}
\caption{As figure~\ref{fig:ot4} but for the simulated interstellar gas. The 
error bars represent the standard deviation of the scatter of the corresponding 
variable between the twelve snapshots used for the analysis.} 
\label{fig:simulPDF}
\end{figure}

In this section we compare 2D gas density fields observed in different parts of 
the Milky Way \citep[from the GASS survey --][]{GASS10} and taken from 3D MHD 
simulations of the ISM \citep{Gent2013p1,Gent2013p2}. 

The GASS data, shown in figure~\ref{fig:ot1}, are from cylindrical surfaces at 
$R=10$ and $20\kpc$. We consider separately three regions at $R=10\kpc$, 
separated by the vertical lines, that have equal areas. The average values of 
$n$ are quite different in the three regions because of the spiral arm pattern 
of a significant strength at $R=10\kpc$.  Its strength decreases with $R$ and 
variations of $n$ with azimuth $\phi$ are much weaker at $R=20\kpc$ (note the 
difference in the colour schemes in the two panels of figure~\ref{fig:ot1}). 
Therefore, we do not divide the region at $R=20\kpc$ into separate parts.
 
We also analysed, using the same tools as for the observed gas distribution,  
gas density obtained from numerical simulations of the interstellar medium in a 
$1\kpc\times 1\kpc\times 2\kpc$ domain. Parameters of the simulation are close 
to those in the Solar neighbourhood, $R=8.5\kpc$. Random flows in the simulated 
ISM are driven by explosions of supernova stars that inject into the system 
large amounts of thermal and kinetic energy. The model includes, among other 
effects, large-scale velocity shear due to the Galactic differential rotation, 
stratification in a gravitational field of stars, heating by the supernovae, 
optically thin radiative cooling and magnetic fields which are due to a 
mean-field dynamo action. We took a vertical cross-section through the middle 
of the computational domain at twelve moments separated by $2.5\tau$ where 
$\tau$ is the correlation length of the random flow in the time domain. The 
time lag is large enough for the snapshots to be statistically independent to a 
reasonable approximation. Figure~\ref{fig:f1}(\textit{b}) shows one such 
cross-section. 

We standardized the data using equation~(\ref{eq:stand}), applied the 
topological filtration described in \S\ref{sec:filt}, and computed the Betti 
numbers and barcode lengths for each level $h$. The results were normalized to 
eliminate, as much as possible, the effects of the image resolution and 
systematic trends as described in \S\ref{sec:trends} and \S\ref{sec:res}. 
The observations at $R = 20\kpc$ have a lower linear resolution than at 
$R = 10\kpc$ because the angular width of the telescope beam was the same. 
Moreover, the resolution in each panel of figure~\ref{fig:ot1} decreases from 
right to left, since the right end of the Galactocentric azimuthal range shown 
is much closer to the Sun. The numerical simulations have a resolution higher 
than any of the observational data. 

The results shown in figure~\ref{fig:ot4} suggest that the gas distributions in 
the right-hand part of figure~\ref{fig:ot1}(\textit{a}) and in 
figure~\ref{fig:ot1}(\textit{b}) are topologically similar, and both are 
different from the other two regions at $R=10\kpc$. This can be explained
by the fact that the left and middle zones in figure~\ref{fig:ot1} mostly lie 
within the Carina spiral arm (roughly, the azimuthal range $300^\circ < \phi < 
250^\circ$) whereas the right zone mostly comprises the inter-arm region close 
to the Sun (roughly, the azimuthal range $230^\circ < \phi < 200^\circ$). 
Spiral arms are the sites of higher star formation rate. Star formation is very 
likely to produce a more energetic forcing of the interstellar turbulence, via 
supernova explosions and expanding supernova remnants as well as winds from 
massive stars. The outer Galaxy $R\gtrsim 16\kpc$ hosts only weak star formation
and spiral arms are expected to be less pronounced there; condition of the 
interstellar gas may be expected to be similar to that in the inter-arm regions 
near the Sun. Different properties of turbulence in the arms and inter-arm 
regions are often suspected but there is little observational support for this
assumption, mainly because the difference is rather subtle. It is remarkable 
that the topological data analysis appears to be more sensitive to the expected 
difference than other methods, but this suggestion needs to be carefully 
verified.  

We show in figure~\ref{fig:simulPDF} the result of a topological filtration of 
the simulated gas density. The curves represent means and the error bars the 
standard deviations across the twelve snapshots. The pronounced,  rather 
symmetric maximum of  $\beta_1(h)$ appears to be closest to that for the middle 
zone of the observational data at $R=10\kpc$ (the dotted red lines in 
figure~\ref{fig:ot4}). That region in the observations shows highly skewed, 
long-tailed dependence on $h$ for $\beta_0$, which is also the case for the 
simulated data. The middle zone of figure~\ref{fig:ot1} is dominated by the 
Carina spiral arm, and this suggests that the parameters selected for the 
numerical simulation are more appropriate for the spiral arms of the Milky
Way. The promise of topological data analysis is quite 
evident but more work with both observational and numerical data is required to 
verify and substantiate these suggestions.

\section{Conclusions and discussion}\label{Con}
We have outlined an approach to topological data analysis that allows for 
comparisons of random fluctuations in different datasets that does not rely on 
assumptions of Gaussian statistics. The analysis follows the following steps:
\begin{enumerate}
\item Standardize the data by subtracting the mean and normalizing by the 
standard deviation. This makes the topological measures insensitive to the 
magnitude of the data.

\item Compute the Betti numbers $\beta_0$ and $\beta_1$ (for two dimensional 
data), and the total barcode length, at each level $h$ of a topological 
filtration.

\item Normalise the Betti numbers and barcode lengths to the unit length or 
area as appropriate.

\item Normalize the results further to unit area under each curve. 

\item Working with the normalized curves removes the sensitivity of the 
topological measures to the presence of large-scale trends and to the 
resolution of the data.

\item Topologically similar data will have similar normalized variations of 
$\beta_0$, $\beta_1$ and $L_T$ with $h$. 
\end{enumerate}

The effects of the standardization and systematic trends depend on the 
scale separation between the mean values and fluctuations. For example, for a 
weaker scale (or frequency) separation between the	two terms in 
equation~(\ref{fxsin}), the points in the standardized persistence diagram 
similar to figure~\ref{fig:F7}(\textit{c}) no longer coincide. For $f(x)$ of 
this form, 	the standardization is still useful, when the frequency of the 
second term is changed from 0.2 to 0.5, but not for frequency ratios that are 
still closer to unity. The marginal frequency ratio (or scale separation) is, 
of course, model-dependent.

By applying this approach to observations and numerical simulations of the gas 
density distribution in the turbulent interstellar gas of the Milky Way, we 
were able to distinguish observations of the spiral arms from those of 
inter-arm regions and to identify which regions of the Galaxy most closely 
resemble the domain of the simulations.

The kinetic and magnetic Reynolds numbers of the simulations used here do not 
exceed 100 \citep{HSSFG17} and are by far smaller than those in the 
interstellar gas of the Milky Way. This means that the simulated fluctuations 
have a much narrower range of scales than the observational data. However, 
comparison of figures~\ref{fig:ot4} and \ref{fig:simulPDF} suggests that this 
does not undermine the topological comparison and its discriminatory power. 
This is understandable since the whole approach of topological filtration 
isolates the most significant, `persistent' features of random fields, and 
these are dominated by fluctuations of larger scales. Robustness of the 
topological filtration under changing image resolution has a similar reason. 

\section*{Acknowledgments}
We gratefully acknowledge financial support of the Leverhulme Trust (Research 
Grant RPG-2014-427) and Institute of Information and Computational Technologies 
(Grant AR05134227).

\bibliographystyle{jpp}
\bibliography{tdaturb}
\end{document}